\documentclass[11pt,a4paper]{article}

\usepackage{lmodern}
\usepackage[T1]{fontenc}
\usepackage[utf8]{inputenc}
\usepackage{amsmath,amssymb}
\usepackage{mathtools}
\usepackage{amsthm}
\usepackage{bm}
\usepackage{geometry}
\usepackage{tikz}
\usetikzlibrary{arrows.meta,positioning,calc,decorations.pathreplacing}
\definecolor{cPurple}{RGB}{120,110,200}
\definecolor{cBlue}{RGB}{40,110,180}
\definecolor{cTeal}{RGB}{30,150,120}
\definecolor{cCoral}{RGB}{200,90,50}
\usepackage{graphicx}
\usepackage[colorlinks=true,linkcolor=blue,citecolor=blue,urlcolor=blue]{hyperref}
\hypersetup{pdftitle={Left-right symmetry breaking in E6L x E6R occurs only in spacetime - with possible implications for strong CP},pdfauthor={Tejinder P. Singh}}

\newcommand{\ket}[1]{\lvert #1\rangle}

\newcommand{\rep}[1]{\mathbf{#1}}
\newcommand{\arep}[1]{\overline{\mathbf{#1}}}
\newtheorem{proposition}{Proposition}
\newtheorem{corollary}{Corollary}
\newtheorem{remark}{Remark}
\allowdisplaybreaks

\title{Left--right symmetry breaking in $E_6^L\times E_6^R$ occurs only in spacetime ---
with possible implications for strong $CP$}

\author{Tejinder P. Singh\\[4pt]
\small Tata Institute of Fundamental Research, Homi Bhabha Road, Mumbai 400005, India\\
\small \texttt{tpsingh@tifr.res.in}}

\date{July 7, 2026; revised November 16, 2026}

\begin{document}
\maketitle

\begin{abstract}
The octonionic $E_6^L\times E_6^R$ program aims to unify the Standard Model with gravitation by
trinifying each exceptional factor, $E_6\to SU(3)\times SU(3)\times SU(3)$. The two-sided
structure then carries \emph{two} colour groups: the visible $SU(3)_c$ from the left factor, and a
second $SU(3)_{c'}$ from the right. Gauged like the first, this second colour would give the
charged lepton a colour index, and so must be suppressed by hand --- for instance as a global
symmetry, broken at the electroweak scale. We argue that no such device is needed: $SU(3)_{c'}$ is
not a second colour at all, but the \emph{same} group as the visible $SU(3)_c$.

The reason is that the left--right symmetry relating the two factors lives in \emph{spacetime},
not in the internal sector. In the gravi-weak reading the right-sector $SU(2)_R$ is the
gravitational frame group, not a second weak force, so the left--right exchange acts on the Dirac
field as ordinary spacetime parity. A spacetime operation cannot reach into the internal sector
and double the colour that lives there; it merely relates the one internal colour as seen in the
two octonionic frames. Colour is therefore single and vector-like, $SU(3)_{c'}=SU(3)_c$, and the
charged lepton is a colour singlet --- never a coloured triplet, so there is nothing to suppress.
What the right sector \emph{does} contribute is a single colour-blind datum: the square-root of
mass $\sqrt m$, whose values are an exceptional-Jordan input (the colour-counting operator $N_R$
fixes only the colour representation, never the value of $\sqrt m$). The two-step right-sector
breaking $SU(3)_R\to SU(2)_R\times U(1)_{Y_{\mathrm{dem}}}\to U(1)_{\mathrm{dem}}$, the exact
mirror of $SU(3)_L\to SU(2)_L\times U(1)_Y\to U(1)_{\mathrm{em}}$, does leave an unbroken
dark-electromagnetic gauge factor; but an anomaly no-go proven here (Appendix~A) shows that on the
visible fermions every family-universal, vectorial anomaly-free abelian charge lies in the span of
electric charge and $B-L$,
which the $\sqrt m$ pattern does not --- no sign convention, chirality assignment, or redefinition
rescues it. The gauged $U(1)_{\mathrm{dem}}$ therefore carries the parity-mirror of electric
charge, its charged matter is the dark sector, and the visible fermions are dem-neutral, coupling
at most through kinetic mixing; no visible fifth force is predicted. The $\sqrt m$ values enter
visible physics only as the spectral labels of the exceptional-Jordan mass operator --- the
Dynkin-swap image of the dark charge lattice --- which settles, in the mirror-canonical direction,
the question of which pattern the gauged charge carries. The same recognition also fixes the
Standard-Model hypercharge as $Y=Q-T^3_L$, with electric charge $Q=N/3$ from the $\mathrm{Cl}(6)$
number operator and no right-sector generator --- a consistency relation recovered in a
sector where the answer is independently known.

We then draw out the consequence for the strong-$CP$ problem. A single vector-like colour together
with the spontaneous parity protects the QCD vacuum angle --- parity restricts it to the
$CP$-conserving values $\{0,\pi\}$, and $\theta_{\mathrm{QCD}}=0$ is the branch matched to the sign
content of the mass determinant (Sec.~\ref{sec:strongcp}) --- inherited
from the gravi-weak reading of $SU(2)_R$ as gravitational, not from QCD itself. The quark
mass-determinant phase vanishes too, $\arg\det M=0$: the flavour rotors that diagonalize the mass
matrices have \emph{real determinant} (shown for the Cabibbo rung, and for the full
three-generation matrix within the flavour texture of Ref.~\cite{TeliSinghMixing}), while the
exceptional-Jordan spectra are real, with matching sign parities in the up and down sectors. The
two add to
the physical $\bar\theta=\theta_{\mathrm{QCD}}+\arg\det M=0$ at tree level, and this coexists with a
nonzero CKM phase --- a \emph{different}, off-diagonal invariant of the same rotors. The appeal of
the mechanism is that both ingredients are already present in the octonionic mass construction,
rather than being added to solve strong $CP$.

We present strong $CP$ as an \emph{application} of this structure, not as a solved problem. The
restriction of $\theta_{\mathrm{QCD}}$ rests on the gravi-weak identification, a working hypothesis
of the program, and the choice of $0$ over $\pi$ is a selection, not a derivation; the two colours
are merged dynamically, by a BF$\to$Yang--Mills reduction to a
single colour connection that the same hypothesis forces (a spacetime breaking cannot touch the
internal colour). The decisive open question is radiative: parity solutions are notoriously undone
by loops. In the minimal universal-seesaw parity model a two-loop
$\bar\theta\sim10^{-3}$--$10^{-2}$, not suppressed by the weak scale, arises through scalar
exchange once parity breaking enters the heavy-quark masses~\cite{deVriesDraperPatel}. That
computation does not apply here as it stands --- the framework has no such heavy-quark sector, and
no gauged $W_R$ --- but the absence of $W_R$ does not remove its mechanism, which the Higgs-bridge
sector of this framework shares, and the model-specific estimate is not yet in hand. Throughout, we
mark carefully which statements are derived, which are program-level inputs, and which remain open.
\end{abstract}

\tableofcontents

\section{Introduction}
\label{sec:intro}

The strong-$CP$ problem --- why the QCD vacuum angle
$\bar\theta=\theta_{\mathrm{QCD}}+\arg\det M$, rendered physical by the chiral anomaly
and instantons~\cite{tHooft1976}, is experimentally consistent with zero,
$\bar\theta\lesssim10^{-10}$ (from the neutron electric-dipole-moment bound~\cite{nEDM}
via the QCD estimate of $d_n(\bar\theta)$~\cite{CrewtherEtAl,PospelovRitz}), despite
there being no Standard-Model reason for it to be small --- has three broad classes of
resolution. A Peccei--Quinn symmetry~\cite{PecceiQuinn} promotes $\bar\theta$ to a
dynamical field, at the cost of an axion~\cite{Weinberg1978,Wilczek1978}. Spontaneous
$CP$ violation of Nelson--Barr type~\cite{Nelson1984,Barr1984} makes $CP$ a symmetry of
the Lagrangian, broken so as to leave $\arg\det M$ zero at tree level. Spontaneous
\emph{parity} breaking, finally, uses an exact parity to restrict the $\theta$-term to
its $CP$-conserving values, while the determinant of the quark mass matrix is kept real ---
by Hermitian mass matrices in the bidoublet left--right models (for real vacuum values), or
by the seesaw structure of the universal-seesaw models~\cite{MohapatraBabu}. The parity
route needs no new light particle, but it requires two structural ingredients to be
supplied: an exact parity in the unbroken theory, and a mechanism rendering the
determinant of the quark mass matrix real.

This paper concerns a setting in which both ingredients are not imposed but
\emph{inherited}. In the octonionic unification program built on the complexified
exceptional Jordan algebra $J_3(\mathbb{O}_\mathbb{C})$, one fermion generation is
realized as a Clifford module over the complex octonions, electric charge is the
central $U(1)$ of that module, and fermion masses arise as the Jordan eigenvalues of
an octonionic-Hermitian operator~\cite{SinghMassRatios,TeliSinghMixing}. The
construction places the visible and mass sectors in two factors of an
$E_6^L\times E_6^R$ product, related by a left--right symmetry. We will see that the
parity needed for strong $CP$ \emph{is} that left--right symmetry, correctly
interpreted, and that the reality needed for $\arg\det M=0$ --- real quark spectra with
matched sign parities, together with flavour rotors of real determinant --- is supplied by the same
octonionic and flavour structure already in place for the masses and their mixings.

That interpretation is the subject of the paper, and it resolves two structural tensions in the
two-sided construction. The right-sector trinification produces a factor $SU(3)_{c'}$ --- a
nominal ``second colour'' --- which, taken as an ordinary gauged interaction, would give the
charged lepton a colour index unless suppressed by hand: a group defined by what it must not do.
Separately, a Standard-Model hypercharge built from right-sector Cartan generators requires an
extra interface assumption. We argue that both tensions trace to a single over-extension ---
treating the left--right structure as acting, and breaking, in the \emph{internal} sector --- and
that one recognition removes both.

The recognition is this: left--right is a \emph{spacetime} symmetry. It is
realized on the Dirac field as parity, so its breaking is confined to the spacetime
(gravi-weak) frame; on the internal sector it acts trivially --- identifying the two
would-be colour groups as one --- and so is never broken there. The right-sector
$SU(2)_R$ is the gravitational frame group, not a second electroweak interaction.
Left--right is thus \emph{broken in the spacetime sector and preserved in the internal
sector}, and from this single asymmetry several consequences follow. Hypercharge is built without any right-sector
generator, $Y=Q-T^3_L$, and reproduces the Standard-Model assignments exactly
(Sec.~\ref{sec:hypercharge}). The ``second colour'' dissolves: its non-abelian part
is identified with the one physical colour, while $\sqrt m$ survives as a \emph{distinct},
colour-blind grading --- not the colour-counting $N_R$, and, by an anomaly no-go proven in
Appendix~\ref{app:rhs}, not a gauged charge on the visible fermions either: the gauged
$U(1)_{\mathrm{dem}}$ of the right electroweak chain carries the mirror of electric charge, on the
dark sector, and the visible $\sqrt m$ is the spectral label of the Jordan mass operator
(Sec.~\ref{sec:onecolour}, Appendix~\ref{app:rhs}); the electron
is then a colour singlet bearing that grade, never a coloured triplet (Sec.~\ref{sec:electron}).
The flavour-sector Dynkin swap that relates the down and lepton mass ladders is a
\emph{distinct} operation and is untouched, so the mass-ratio program is unaffected
(Sec.~\ref{sec:dynkin}). And from a single vector-like colour together with the
spontaneous parity follows the conservation of strong $CP$
(Sec.~\ref{sec:strongcp}). We present the hypercharge case first, as a check in a
sector where the answer is independently known, and then the colour case, where the
same move yields strong $CP$ as its payoff. Figure~\ref{fig:decomposition} gathers this anatomy in a single picture --- the split-bioctonionic dimensions dividing into a spacetime block (which breaks, yielding the two Higgs, the $(3,3)$ spacetime and the two $4$D worlds) and a breaking-blind internal block, whose single colour $SU(3)_c=SU(3)_{c'}$ is the premise on which the strong-$CP$ argument rests --- and may be read now as a roadmap for the sections that follow.

\begin{figure}[tp]
\centering
\scalebox{0.90}{%
\begin{tikzpicture}[
  >={Stealth[length=2mm]},
  topbox/.style={draw=cPurple!80, fill=cPurple!12, rounded corners=3pt, align=center, inner sep=5pt, font=\small},
  sblue/.style={draw=cBlue!75, fill=cBlue!8, rounded corners=4pt, align=left, inner sep=6pt, font=\footnotesize, text width=5.6cm},
  steal/.style={draw=cTeal!75, fill=cTeal!9, rounded corners=4pt, align=left, inner sep=6pt, font=\footnotesize, text width=5.3cm},
  keyb/.style={draw=cCoral!85, fill=cCoral!14, rounded corners=4pt, align=center, inner sep=5pt, font=\small, line width=0.7pt},
  note/.style={draw=gray!45, dashed, rounded corners=3pt, align=left, inner sep=4pt, font=\scriptsize, text width=3.2cm},
  forkarr/.style={->, draw=gray!70, line width=0.8pt},
  relay/.style={->, draw=gray!65, line width=0.8pt},
  basearr/.style={->, draw=cBlue!85, line width=0.9pt},
  fibarr/.style={->, draw=cCoral!85, line width=0.9pt},
  lbl/.style={font=\footnotesize, align=center},
  slbl/.style={font=\scriptsize, align=center},
]

\node[topbox] (top) at (0,6.4) {$E_6^L \times E_6^R$\\[2pt]\footnotesize on 16D split-bioctonionic space};

\node[note] (pre) at (-5.05,6.3)
  {\textbf{Before breaking:} near-exact L--R symmetry, only a Dynkin swap --- surviving in the flavour sector};
\draw[gray!45, dashed, line width=0.3pt] (pre.east) -- (top.west);

\node[sblue] (sp) at (-3.8,4.2)
  {\textbf{Spacetime sector}\\ 8D split-biquaternion $=4+4$\\[2pt]
   $2$ real directions $\to$ two Higgs\\
   $6$ imaginaries $\to$ 6D spacetime $(3,3)$\\
   $\to$ two 4D spacetimes (gravi-weak)};
\node[steal] (in) at (3.8,4.2)
  {\textbf{Internal sector}\\ 8D $=4+4$, breaking-blind\\[2pt]
   untouched by the spacetime breaking\\
   $\to$ the two copies coincide\\
   $\to$ one 4D internal space};

\draw[forkarr] (top.south) to[bend right=8] (sp.north);
\draw[forkarr] (top.south) to[bend left=8] (in.north);
\node[slbl] at (0,5.55) {symmetry breaking};

\node[keyb] (key) at (0,1.25) {$SU(3)_c = SU(3)_{c'}$\\[2pt]\footnotesize one vector-like colour};
\node[slbl] at (0,0.45) {shared colour fibre};

\begin{scope}[shift={(-3.8,-1.8)}]
  \draw[gray!55, fill=gray!8, line width=0.4pt]
     (-1.55,-0.2) -- (1.45,0.18) -- (1.75,-0.28) -- (-1.25,-0.66) -- cycle;
  \foreach \x/\xt in {-1.1/-1.0, -0.45/-0.32, 0.2/0.36, 0.85/1.04} {
     \draw[gray!65, line width=0.5pt] (\x,0.0) -- (\xt,1.35);
     \fill[gray!75] ($(\x,0.0)!0.45!(\xt,1.35)$) circle (1.5pt);
  }
  \coordinate (Lfib) at (0.1,1.95);
  \coordinate (Lbase) at (-1.3,-0.45);
  \node[lbl] at (0.1,1.62) {$SU(3)_c \otimes U(1)_{\mathrm{em}}$};
  \node[slbl] at (0.1,-1.0) {2nd (anti-)spacetime $(-,-,-,+)$};
  \node[slbl] at (0.1,-1.32) {$SU(2)_L\otimes U(1)_Y$ broken};
  \node[slbl] at (0.1,-1.64) {weak $=$ its geometry};
\end{scope}

\begin{scope}[shift={(3.8,-1.8)}]
  \draw[gray!55, fill=gray!8, line width=0.4pt]
     (-1.55,-0.2) -- (1.45,0.18) -- (1.75,-0.28) -- (-1.25,-0.66) -- cycle;
  \foreach \x/\xt in {-1.1/-1.0, -0.45/-0.32, 0.2/0.36, 0.85/1.04} {
     \draw[gray!65, line width=0.5pt] (\x,0.0) -- (\xt,1.35);
     \fill[gray!75] ($(\x,0.0)!0.45!(\xt,1.35)$) circle (1.5pt);
  }
  \coordinate (Rfib) at (0.1,1.95);
  \coordinate (Rbase) at (-1.3,-0.45);
  \node[lbl] at (0.1,1.62) {$SU(3)_{c'} \otimes U(1)_{\mathrm{DEM}}$};
  \node[slbl] at (0.1,-1.0) {our 4D spacetime $(-,+,+,+)$};
  \node[slbl] at (0.1,-1.32) {$SU(2)_R\otimes U(1)_{Y\mathrm{DEM}}$ broken};
  \node[slbl] at (0.1,-1.64) {GR $=$ its geometry};
\end{scope}

\draw[relay] (in.south) to[bend left=15] (key.north east);
\draw[basearr] (sp.south west) to[out=-90,in=150] (Lbase);
\draw[basearr] (sp.south east) .. controls (-2.5,1.0) and (1.2,-2.7) .. (Rbase);
\draw[fibarr] (key.south west) to[out=-150,in=60] (Lfib);
\draw[fibarr] (key.south east) to[out=-30,in=120] (Rfib);

\draw[decorate, decoration={brace, amplitude=7pt, mirror}, gray!70] (-5.9,-3.7) -- (5.9,-3.7);
\node[lbl] at (0,-4.3) {\textbf{Our Universe}};

\end{tikzpicture}}%
\caption{\textbf{Dimensional anatomy of the $E_6^L\times E_6^R$ construction, and why there is
only one colour.} The unification lives on a $16$-(real-)dimensional split-bioctonionic space.
Before symmetry breaking the two $E_6$ factors are related by a near-exact left--right symmetry;
the sole exception is a Dynkin swap, which is \emph{not} the parity and survives, after breaking,
in the flavour sector. The breaking itself is a single event wearing three hats: it is at once the
left--right breaking, an electroweak breaking ($SU(2)_L\otimes U(1)_Y$, visible side) and a
gravi-dark breaking ($SU(2)_R\otimes U(1)_{Y_{\mathrm{dem}}}$, gravidem side). It separates the
$16$ dimensions into two $8$-dimensional blocks but breaks only one of them, the spacetime block. The \emph{spacetime} block is a
split-biquaternion ($4+4$): its two real directions supply the two Higgs fields, while its six
quaternionic imaginaries furnish a $6$D spacetime of signature $(3,3)$, inside which two $4$D
spacetimes become embedded once the gravi-weak symmetry breaks --- our Lorentzian world
$(-,+,+,+)$, whose geometry is general relativity (the $SU(2)_R$ frame connection), and a second,
``anti''-spacetime $(-,-,-,+)$, whose geometry is the weak force (the $SU(2)_L$ connection). The
\emph{internal} block ($4+4$) is untouched by this spacetime breaking, so the two would-be colour
copies do not separate but collapse to a single $4$D internal space. This is the geometric content
of the identification $SU(3)_{c'}=SU(3)_c$. The identification does not touch the surviving
abelian Cartans of the two \emph{electroweak} factors, which appear in the two bundles ---
$U(1)_{\mathrm{em}}$ (left) and $U(1)_{\mathrm{DEM}}$ (right, whose charged matter is the dark
sector; Appendix~\ref{app:rhs}). Here the two
sides are \emph{not} mirror images: on the left the electroweak Cartan coincides with the colour
trace, so electric charge is colour-welded ($Q=N/3$); on the right it does not, and $\sqrt m$ is a
colour-blind, unwelded \emph{spectral} grading whose lattice mirrors the surviving dark charge
$T^3_R+Y_{\mathrm{dem}}$ --- a charge carried by the dark sector, not by the visible fermions
(anomaly no-go, Appendix~\ref{app:rhs}) --- while the right colour trace $N_R$
serves only to label the colour representation (Sec.~\ref{sec:onecolour}, Appendix~\ref{app:rhs}).
The two emergent spacetimes thus carry bundles whose bases differ but whose colour fibre is
one and the same, identical precisely because the internal directions never felt the breaking. The
dimensional assignments are structural inputs of the framework; what the paper adds is the reading
that the left--right breaking is confined to the spacetime block, so colour is single and
vector-like --- the premise the strong-$CP$ argument rests on.}
\label{fig:decomposition}
\end{figure}

We are explicit throughout about what is established and what is assumed, and adopt
the program's epistemic tags: \textbf{[D]} for a result derived here or in cited
work; \textbf{[P]} for a program-level working hypothesis (chiefly the gravi-weak
identification, which the whole construction shares); \textbf{[C]} for a conditional
claim --- one forced by stated premises but carrying a named, checkable residual, so
that it sits between \textbf{[D]} and \textbf{[O]} (we write ``\textbf{[C]} leaning
\textbf{[D]}'' when the residual is a convention rather than a computation); and
\textbf{[O]} for a genuinely open point. So that the bookkeeping does not have to be repeated
at every claim, the dependency structure is stated \emph{once in full}, in the ledger of
Sec.~\ref{sec:discussion}, which the reader should treat as the authoritative accounting;
local status remarks elsewhere are pointers to it. In one sentence: the $\arg\det M=0$ half of
the strong-$CP$ result is the more secure (\textbf{[D]} for the Cabibbo rung,
texture-contingent for the full three-generation matrix); the $\theta_{\mathrm{QCD}}=0$ half is
conditional on the gravi-weak identification (\textbf{[P]}) and on a branch selection --- parity
alone gives $\theta_{\mathrm{QCD}}\in\{0,\pi\}$ (\textbf{[C]}) --- together with an emergence-map
caveat logged in Sec.~\ref{sec:strongcp}; and loop-level safety is open
(\textbf{[O]}) --- strong $CP$ entering as an \emph{application} of structural results that are
themselves independent of it.

The paper is organized as follows. Section~\ref{sec:framework} sets out the
framework --- the $\mathrm{Cl}(6)$ one-generation module, the gravi-weak character of
the right sector, and the Jordan mass operator --- at a level sufficient to follow the
argument without the companion works. Section~\ref{sec:hypercharge} derives
hypercharge as the principle's first, checkable instance. Section~\ref{sec:parity}
states and proves the central principle, that the left--right symmetry acts as
spacetime parity. Sections~\ref{sec:onecolour}--\ref{sec:dynkin} work out the colour
sector: the dissolution of $SU(3)_{c'}$, the electron's grading-only relation to it,
and the distinctness of the Dynkin swap. Section~\ref{sec:strongcp} applies the structural
results to strong $CP$; Section~\ref{sec:loops} addresses loop regeneration honestly; and
Section~\ref{sec:discussion} collects the dependency structure and the outlook, including the
cosmology of the spontaneous parity. Appendix~\ref{app:rhs} establishes the right-sector
representation content and proves the anomaly no-go (Prop.~\ref{prop:nogo}) that fixes the dark
charge and renders the visible fermions dem-neutral.
Figure~\ref{fig:roadmap} summarizes the claim architecture and the epistemic status of each
step.

\begin{figure}[t]
\centering
\begin{tikzpicture}[
  >={Stealth[length=2.2mm]},
  box/.style={draw, rounded corners, align=center, inner sep=4pt, font=\footnotesize, text width=40mm},
  side/.style={draw, rounded corners, align=center, inner sep=4pt, font=\footnotesize, text width=42mm},
  open/.style={draw, dashed, rounded corners, align=center, inner sep=4pt, font=\footnotesize, text width=40mm},
  link/.style={->, thick},
  node distance=9mm and 12mm,
]
\node[box] (grav) {\textbf{[P]} Gravi-weak input: $SU(2)_R$ gravitational $\Rightarrow$ $L$--$R$ exchange $=$ spacetime parity};
\node[side, right=of grav] (side) {also from parity:\\ hypercharge $Y=Q-T^3_L$ (consistency relation);\\ single vector-like $SU(3)_c$ \textbf{[C]}};
\node[box, below=of grav] (theta) {\textbf{[P]} $\theta_{\mathrm{QCD}}\in\{0,\pi\}$;\\ $0$ on the matched branch \textbf{[C]}};
\node[box, below=of theta] (tbar) {\emph{Application:} $\bar\theta=\theta_{\mathrm{QCD}}+\arg\det M=0$ at tree level (CKM $CP$ preserved)};
\node[box, left=of tbar] (rotor) {\textbf{[D]} $SU(3)_F$ rotors, real generator: $\arg\det M=0$};
\node[open, below=of tbar] (loop) {\textbf{[O]} Higgs bridge $B_H$ $\Rightarrow$ radiative stability (open)};
\draw[link] (grav) -- (side);
\draw[link] (grav) -- (theta);
\draw[link] (theta) -- (tbar);
\draw[link] (rotor) -- (tbar);
\draw[link, dashed] (tbar) -- (loop);
\end{tikzpicture}
\caption{Claim architecture and epistemic status. The structurally real mass-determinant
phase $\arg\det M=0$ \textbf{[D]} follows from the special-unitary flavour rotors and is
independent of the wider program. The gravi-weak identification of $SU(2)_R$ \textbf{[P]} yields
a spacetime-parity reading of the left--right exchange, hence $\theta_{\mathrm{QCD}}\in\{0,\pi\}$
(with $0$ selected on the matched branch, Sec.~\ref{sec:strongcp}), and ---
at the level of frame-conjugacy --- a single vector-like colour \textbf{[C]} together with the
hypercharge consistency relation. Strong $CP$ ($\bar\theta=0$ at tree level) is their
\emph{application}; its radiative stability \textbf{[O]} awaits the Higgs-bridge matrix
elements; the minimal universal-seesaw parity model is known not to be generically safe at
two loops~\cite{deVriesDraperPatel}. Not drawn: the anomaly no-go of Appendix~\ref{app:rhs}
(Prop.~\ref{prop:nogo}, \textbf{[D]}), which fixes the gauged dark charge to the mirror-canonical
pattern and renders the visible fermions dem-neutral --- it feeds the force inventory of
Sec.~\ref{sec:discussion} but no arrow of the strong-$CP$ chain above.}
\label{fig:roadmap}
\end{figure}

\section{The framework}
\label{sec:framework}

We collect the ingredients the argument uses, at a level sufficient to follow the
paper without the companion works, while citing those works for the derivations.
Three pieces are needed: the construction of one fermion generation as a Clifford
module over the complex octonions, which fixes electric charge algebraically
(Sec.~\ref{sec:cl6}); the two-sided $E_6^L\times E_6^R$ trinification and the
gravi-weak character of the right sector (Sec.~\ref{sec:graviweak}); and the
exceptional-Jordan mass operator, whose spectrum is real (Sec.~\ref{sec:jordan}). All
three are inputs to the present paper, tagged \textbf{[P]}, and are developed in
Refs.~\cite{SinghMassRatios,TeliSinghMixing,SinghEmergence,SinghScaffolding}. The
division-algebraic reading of Standard-Model structure on which they build has a long
lineage --- octonions and quarks~\cite{GunaydinGursey}, the octonion algebra in
particle physics~\cite{BaezOctonions}, and ladder-operator constructions of the gauge
group and electric charge~\cite{Furey1603,Furey2018} --- as does the use of the
exceptional groups: $E_6$ unification~\cite{GurseyRamondSikivie}, its quark--lepton and
trinification structure~\cite{AchimanStech}, and recent exceptional-Jordan derivations
of the Standard-Model gauge group from $F_4$/$E_6$
automorphisms~\cite{TodorovDuboisViolette,Boyle}.

\subsection{One generation as a Clifford module over the complex octonions}
\label{sec:cl6}

We work with the complex octonions $\mathbb{O}_\mathbb{C} = \mathbb{C}\otimes\mathbb{O}$,
i.e.\ the (real, non-associative, division) octonion algebra $\mathbb{O}$ with basis
$\{e_0\equiv 1, e_1,\dots,e_7\}$, $e_a e_b = -\delta_{ab} + f_{abc}\,e_c$, tensored with a
commuting complex unit $i$ ($i^2=-1$, $[\,i,e_a\,]=0$). The complex unit $i$ is not one of
the octonionic imaginary units; it is the field scalar of the complexification.
The right sector of the two-sided construction (Sec.~\ref{sec:graviweak}) lives in a
split-bioctonionic doubling, in which a distinguished eighth imaginary unit $e_8$ appears
alongside $e_7$ as the unit singled out by that sector; we use $e_8$ in this sense
throughout, and refer to Ref.~\cite{SinghScaffolding} for the split-bioctonionic
construction.

\paragraph{Clifford algebra from left multiplication.}
Each $x\in\mathbb{O}_\mathbb{C}$ defines a left-multiplication operator
$L_x:\mathbb{O}_\mathbb{C}\to\mathbb{O}_\mathbb{C}$, $L_x(y)=xy$. Because $\mathbb{O}$ is
not associative, $L_xL_y\neq L_{xy}$ in general; the associative algebra generated by the
$\{L_{e_a}\}$ under composition is, however, well defined, and is isomorphic to the complex
Clifford algebra
\begin{equation}
  \mathrm{Cl}(6,\mathbb{C}) \;\cong\; \mathbb{C}(8),
  \label{eq:cl6}
\end{equation}
the algebra of $8\times8$ complex matrices~\cite{Furey2018,DixonBook}. All operator
identities below are understood in this generated associative algebra; they are statements
about composed left-multiplications, not about products taken in $\mathbb{O}_\mathbb{C}$.

\paragraph{Complex structure and minimal ideal.}
Single out $e_7$ and form
\begin{equation}
  I \;:=\; i\,e_7, \qquad I^2 = i^2\,L_{e_7}^2 = (-1)(-1) = +1,
\end{equation}
where $L_{e_7}^2=-1$ by the alternative law $e_7(e_7 y)=e_7^2\,y=-y$. Thus $I$ is an
involution, and
\begin{equation}
  \omega_\pm \;:=\; \tfrac12\,(1\pm I), \qquad
  \omega_\pm^2=\omega_\pm,\quad \omega_+\omega_-=0,\quad \omega_++\omega_-=1,
\end{equation}
are complementary primitive idempotents. The space
$S:=\mathrm{Cl}(6,\mathbb{C})\,\omega_+$ is a minimal left ideal, an $8$-complex-dimensional
irreducible module --- one Weyl fermion's worth of internal states.

\paragraph{Fermionic (Witt) basis.}
Pair the remaining six imaginary units into three complex combinations
\begin{equation}
  a_1^\dagger=\tfrac12(e_5+i e_4),\quad
  a_2^\dagger=\tfrac12(e_3+i e_1),\quad
  a_3^\dagger=\tfrac12(e_6+i e_2),
  \qquad a_i = (a_i^\dagger)^\ast,
\end{equation}
i.e.\ $a_i^\dagger=\tfrac12(e_{\rho(i)}+i e_{\sigma(i)})$ with
$(\rho,\sigma)=(5,4),(3,1),(6,2)$. As operators on $S$ they are nilpotent and obey the
canonical anticommutation relations of three fermionic modes,
\begin{equation}
  \{a_i,a_j^\dagger\}=\delta_{ij},\qquad
  \{a_i,a_j\}=\{a_i^\dagger,a_j^\dagger\}=0,\qquad
  a_i\,\omega_+ = 0 ,
  \label{eq:CAR}
\end{equation}
so $\omega_+$ is the Fock vacuum and $S$ is the three-mode Fock space generated by the
$a_i^\dagger$~\cite{Furey2018}.

\paragraph{The eight states.}
Acting on $\omega_+$ with $0,1,2,3$ creation operators produces the $1+3+3+1$ states
\begin{equation}
  \begin{aligned}
    &\ket{\nu}=\omega_+, \\
    &\ket{\bar d_i}=a_i^\dagger\,\omega_+, \\
    &\ket{u_i}=a_j^\dagger a_k^\dagger\,\omega_+ \ \ (\text{$ijk$ cyclic}), \\
    &\ket{e^+}=a_1^\dagger a_2^\dagger a_3^\dagger\,\omega_+ .
  \end{aligned}
  \label{eq:states}
\end{equation}
The conjugate ideal $\bar S=\mathrm{Cl}(6,\mathbb{C})\,\omega_-$ supplies the
charge-conjugate octet $\{\bar\nu, d_i, \bar u_i, e^-\}$; together the two ideals carry the
$16$ Weyl states of one Standard-Model generation (the $\mathbf{16}$ of $\mathrm{Spin}(10)$,
including a sterile $\nu_R$).

\paragraph{Number operator and electric charge.}
Define the number operator and the electric-charge operator
\begin{equation}
  N \;=\; \sum_{i=1}^{3} a_i^\dagger a_i \;\in\;\{0,1,2,3\},
  \qquad
  Q \;=\; \frac{N}{3}.
  \label{eq:QisN3}
\end{equation}
$Q$ is the central (trace) $\mathfrak{u}(1)$ of $\mathrm{Cl}(6,\mathbb{C})\supset
\mathfrak{u}(3)$. Evaluating \eqref{eq:QisN3} on \eqref{eq:states} returns the physical
charges with no further input:
\begin{equation}
  Q(\nu)=0,\quad Q(\bar d_i)=\tfrac13,\quad Q(u_i)=\tfrac23,\quad Q(e^+)=1 .
\end{equation}
The fractional quark charges and the quark--lepton offset are thus \emph{algebraic}: they
are the integer eigenvalues of $N$ read in units of $\tfrac13$, not an embedding choice. We
return to this point in Sec.~\ref{sec:hypercharge}, where hypercharge is built from $Q$.

\paragraph{Colour $SU(3)_c$ as the stabilizer of $e_7$.}
The octonion automorphism group is $G_2=\mathrm{Aut}(\mathbb{O})$. The subgroup fixing the
distinguished unit $e_7$ --- equivalently commuting with $I$ and preserving $\omega_\pm$ ---
is
\begin{equation}
  SU(3)_c \;=\; \mathrm{Stab}_{G_2}(e_7)\;\subset\; G_2,
  \label{eq:su3c}
\end{equation}
acting on the triple $(a_1,a_2,a_3)$ as the fundamental $\mathbf 3$ (and on
$(a_1^\dagger,a_2^\dagger,a_3^\dagger)$ as $\bar{\mathbf 3}$). The Fock grading
\eqref{eq:states} therefore organizes into colour representations as
\begin{equation}
  \underbrace{\nu}_{\mathbf 1,\,N=0}\ \oplus\
  \underbrace{\bar d_i}_{\bar{\mathbf 3},\,N=1}\ \oplus\
  \underbrace{u_i}_{\mathbf 3,\,N=2}\ \oplus\
  \underbrace{e^+}_{\mathbf 1,\,N=3},
\end{equation}
using $\bar{\mathbf 3}\wedge\bar{\mathbf 3}=\mathbf 3$ and
$\bar{\mathbf 3}\wedge\bar{\mathbf 3}\wedge\bar{\mathbf 3}=\mathbf 1$. Leptons are precisely
the colour singlets ($N=0,3$) and quarks the colour triplets ($N=1,2$): ``lepton number as a
fourth colour'' is automatic, and the same datum $N$ that fixes the colour representation
also fixes the charge through \eqref{eq:QisN3}.

\medskip\noindent\emph{Two features will be load-bearing later.} First, $Q=N/3$ is fixed by
the number operator alone \textbf{[D]}; no $U(1)$ normalization is chosen by hand, and the
quark--lepton ``factor of $3$'' lives in the $N$-spectrum. Second, there is here a
\emph{single} colour group, $SU(3)_c=\mathrm{Stab}_{G_2}(e_7)$, acting on the $a_i$ of this
one ideal; no second colour group appears in the one-generation module. The status of a
putative right-sector ``$SU(3)_{c'}$'' --- and its dissolution --- is taken up in
Sec.~\ref{sec:onecolour}.

\subsection{Trinification and the gravi-weak character of the right sector}
\label{sec:graviweak}

\paragraph{Two-sided trinification.}
One generation's ideal is placed inside a product of two exceptional factors,
$E_6^L\times E_6^R$, each trinifying at the electroweak scale,
\begin{align}
  E_6^L &\;\longrightarrow\; SU(3)_c \times SU(3)_{F,L}\times SU(3)_L, \label{eq:trinL}\\
  E_6^R &\;\longrightarrow\; SU(3)_{c'}\times SU(3)_{F,R}\times SU(3)_R . \label{eq:trinR}
\end{align}
The left electroweak factor descends in the familiar way,
$SU(3)_L\to SU(2)_L\times U(1)\to U(1)_{em}$, and $SU(2)_L$ is the visible weak gauge
group. The two flavour factors $SU(3)_{F,L},\,SU(3)_{F,R}$ are \emph{global}: they
supply the $\mathrm{Sym}^3(\mathbf 3)$ ladder geometry that organizes the three
generations and the Jordan mass operator, but are not gauged~\cite{SinghMassRatios}. The
right-sector factor $SU(3)_{c'}$ is the nominal ``second colour''; its status is the
subject of Sec.~\ref{sec:onecolour} and we defer it. Here we examine the remaining right
electroweak factor $SU(3)_R$, whose character is the one structural input the rest of the
paper leans on.

\paragraph{Why the right $SU(2)$ is gravitational.}
The right electroweak factor does not descend to a second \emph{visible} electroweak
interaction. It breaks as
\begin{equation}
  SU(3)_R \;\longrightarrow\; SU(2)_R \times U(1)_{Y_{\mathrm{dem}}}
  \;\longrightarrow\; U(1)_{\mathrm{dem}},
  \label{eq:su3R}
\end{equation}
with $SU(2)_R$ identified as a \emph{gravitational} gauge group rather than an internal
one. The identification is the gravi-weak one~\cite{Percacci,NestiPercacci,AlexanderMarcianoSmolin,WSI_BF}, and the
mechanism that makes it natural is the self-dual (first-order) formulation of gravity.
In four dimensions gravity is a gauge theory of the local frame group, which in Euclidean
signature factorizes chirally,
\begin{equation}
  \mathrm{Spin}(4)\;\cong\;SU(2)_+\times SU(2)_- .
\end{equation}
The spin connection splits accordingly into self-dual and anti-self-dual parts,
$\omega^{\pm}=\tfrac12\big(\omega\pm\star\,\omega\big)$, which are the connections of the two
chiral $SU(2)$ factors. In the Plebanski/Ashtekar formulation the Einstein--Hilbert action
is reproduced from a \emph{single} chiral half of the connection, so the gravitational
degrees of freedom are carried by one $SU(2)$ and the complementary chiral $SU(2)$ is free
to be gauged as an internal symmetry.\footnote{The chiral split is sharp in the Euclidean
base; the Lorentzian continuation inherits it through the leafwise Wick rotation
of~\cite{WSI_BF}, where the Ashtekar reality conditions relate the two chiral halves.}
The gravi-weak identification assigns the visible weak group to the latter and the
gravitational connection to the former --- here $SU(2)_R$. In the present program this split
is realized dynamically by the six-dimensional $SO(3,3)$ BF reduction of Wesley, Singh and
Isidro~\cite{WSI_BF}, which resolves the six-dimensional base into two four-dimensional
leaves, each carrying a tetrad and a spin connection: $SU(2)_R$ is the (self-dual)
gravitational connection on its leaf, while $SU(2)_L$ acts as the weak gauge field on the
visible leaf.

\paragraph{Dark electromagnetism and the structural conclusion.}
The chain \eqref{eq:su3R} runs in two steps, mirroring the visible
$SU(3)_L\to SU(2)_L\times U(1)_Y\to U(1)_{\mathrm{em}}$: the first breaking leaves the dark
hypercharge $U(1)_{Y_{\mathrm{dem}}}$ (generated by the right electroweak Cartan $\lambda_8^R$), and
a second Higgs~\cite{WSI_BF} then breaks $SU(2)_R\times U(1)_{Y_{\mathrm{dem}}}\to
U(1)_{\mathrm{dem}}$, leaving the unbroken \emph{dark electromagnetism} $U(1)_{\mathrm{dem}}$. Just
as electric charge is the Gell-Mann--Nishijima combination $Q=T^3_L+Y$, the surviving dark charge is
$Q_{\mathrm{dem}}=T^3_R+Y_{\mathrm{dem}}$, the exact mirror of $Q$. \emph{Which} values that gauged
charge carries, and on \emph{what} matter, is settled in Appendix~\ref{app:rhs} by an anomaly
computation (Prop.~\ref{prop:nogo}): on the visible fermions no anomaly-free $U(1)$ can realize
the $\sqrt m$ pattern, so the gauged $Q_{\mathrm{dem}}$ carries the mirror of electric charge and
its charged matter is the \emph{dark} sector, while the $\sqrt m$ values label the Jordan spectrum
of Sec.~\ref{sec:jordan} as the Dynkin-swap image of that charge lattice --- a spectral grading of
the mass operator, not a gauge charge of visible matter. $U(1)_{\mathrm{dem}}$ is colour-blind (an
electroweak-right Cartan combination, commuting with colour) and is \emph{not} the Standard-Model
hypercharge direction. It
must not be confused with the trace of the right \emph{colour} group $SU(3)_{c'}$ --- the number
operator $N_R$ of Sec.~\ref{sec:onecolour}, a distinct, colour-\emph{welded} $U(1)$ that fixes only
the colour representation and never the value of $\sqrt m$; Appendix~\ref{app:rhs} keeps the two
apart. We take the gravi-weak identification as a working hypothesis of the program,
to be tagged \textbf{[P]} throughout: the explicit leaf-by-leaf map and the verification
that it is normalization-faithful (that the relative scale of $SU(2)_R$ and the visible
generators is fixed by the construction rather than chosen) are not established here, and we
do not claim them. We also note that gravi-weak unification is an active and non-standard
proposal, with its own open problems; we use it as a stated input rather than defend it. One such
problem should be surfaced rather than left implicit, since an exact parity relates the couplings
of the pair it exchanges: in the $SO(3,3)$ construction of Ref.~\cite{WSI_BF} this is realized
through a single master coupling, with $G_N=G_W$ at the unbroken point, and how the enormous
infrared disparity between the weak and gravitational couplings descends from that single input is
an open problem of the gravi-weak program which the present paper inherits within \textbf{[P]}.
What the arguments below require is only the qualitative conclusion, which we state for
emphasis:
\begin{quote}
\itshape
the right-sector electroweak generators --- the Cartans $T^3_R,\,T^8_R$ of $SU(2)_R$ and
$U(1)_{\mathrm{dem}}$ --- are frame, gravitational, and dark objects, not internal
generators of the visible Standard Model.
\end{quote}
Appendix~\ref{app:rhs} sharpens the abelian half of this statement from interpretation to
consistency: on the visible content every anomaly-free $U(1)$ lies in
$\mathrm{span}\{Y,\,B{-}L\}$, so the right-sector Cartans --- whatever their geometric meaning ---
can contribute no independent visible charge (Prop.~\ref{prop:nogo}).
This single input is what Secs.~\ref{sec:hypercharge} and \ref{sec:parity} turn on. In
Sec.~\ref{sec:hypercharge} it forces hypercharge to be built \emph{without} any right-sector
generator, $Y=Q-T^3_L$; in Sec.~\ref{sec:parity} it is what makes the left--right exchange
$SU(2)_L\leftrightarrow SU(2)_R$ act on the Dirac field as a spacetime operation --- parity,
which exchanges the two chiral halves of the frame group --- rather than as an internal
relabelling. The dissolution of $SU(3)_{c'}$ in Sec.~\ref{sec:onecolour} is the same lesson
in the colour sector.

\subsection{The Jordan mass operator and the reality of its spectrum}
\label{sec:jordan}

\paragraph{The mass operator.}
Masses enter through an element of the complexified exceptional Jordan
algebra~\cite{JordanVNW,DrayManogue} $J_3(\mathbb{O}_\mathbb{C})$ --- the
$27$-dimensional Albert algebra of $3\times3$
octonionic-Hermitian matrices
\begin{equation}
  X \;=\;
  \begin{pmatrix}
    \xi_1 & \bar z_3 & z_2 \\
    z_3 & \xi_2 & \bar z_1 \\
    \bar z_2 & z_1 & \xi_3
  \end{pmatrix},
  \qquad \xi_a\in\mathbb{C},\quad z_a\in\mathbb{O}_\mathbb{C},
  \label{eq:Xform}
\end{equation}
with $X=X^\star$ under octonionic conjugation of the off-diagonal entries (here
$\bar z$ denotes the octonionic conjugate, which reverses the seven octonionic imaginary
units but \emph{not} the field scalar $i$). The reduced structure group preserving the cubic
norm $N(X)$ is $E_6$, and $F_4=\mathrm{Aut}\,J_3(\mathbb{O})$~\cite{SinghMassRatios}. A
scalar order parameter $X(x)\in J_3(\mathbb{O}_\mathbb{C})$ acquires a vacuum value
$\langle X\rangle$, which the $E_6$-invariant cubic Yukawa converts into an internal mass
operator acting on the one-generation states of Sec.~\ref{sec:cl6}; the physical masses are
the Jordan eigenvalues of $\langle X\rangle$~\cite[Sec.~V]{SinghMassRatios}.

\paragraph{Cubic spectrum and the universal spread.}
Every $X\in J_3(\mathbb{O}_\mathbb{C})$ satisfies a cubic minimal polynomial
\begin{equation}
  \lambda^3 - T\,\lambda^2 + S\,\lambda - D = 0,
  \qquad
  T=\mathrm{Tr}\,X,\quad S=\tfrac12\big[(\mathrm{Tr}\,X)^2-\mathrm{Tr}(X\circ X)\big],
  \quad D=N(X),
  \label{eq:cubic}
\end{equation}
whose three roots are the Jordan eigenvalues. Per charge sector they take the universal
centred form
\begin{equation}
  \mathrm{spec}\,\langle X\rangle \;=\; \big(\,q-\delta,\ q,\ q+\delta\,\big),
  \label{eq:spec}
\end{equation}
where the centre $q$ fixes the family mass scale and the spread $\delta$ controls the
intra-family hierarchy. The determinant structure of $J_3(\mathbb{O}_\mathbb{C})$ fixes the
spread to the universal algebraic value
\begin{equation}
  \delta^2 \;=\; \tfrac{3}{8},
  \label{eq:delta}
\end{equation}
independent of sector and of the centre; the derivation is given in
Ref.~\cite[App.~G.2]{SinghMassRatios} and we take \eqref{eq:delta} as input here. The three
sector centres are set by the trace split
$\mathrm{Tr}\,X_\ell:\mathrm{Tr}\,X_u:\mathrm{Tr}\,X_d = 1:2:3$. This inter-sector ratio fixes
the three charged-fermion family centres and is taken here as a phenomenological input --- the
one first-generation datum the construction does \emph{not} fix algebraically, in contrast to
the forced $Q=N/3$ (Sec.~\ref{sec:cl6}) and the derived spread \eqref{eq:delta}. Its structural
origin is a matter for the companion mass-ratio work~\cite{SinghMassRatios}; the present paper
uses only its value, and none of the conclusions below depend on its provenance.

\paragraph{Reality of the spectrum (structural).}
The eigenvalues \eqref{eq:spec} are \emph{real}. For $J_3(\mathbb{O}_\mathbb{C})$ this is not
automatic --- unlike the real form $J_3(\mathbb{O})$, the complex diagonal entries
$\xi_a\in\mathbb{C}$ in \eqref{eq:Xform} permit complex roots of \eqref{eq:cubic} in general.
Reality is a structural consequence of how the off-diagonal data are built, and is most
transparent in the cubic norm itself. Writing the Jordan element in centred form
$X=q\,\mathbb{1}_3+Y$, with $Y$ traceless and off-diagonal entries $z_{ij}\in\mathbb{O}_\mathbb{C}$
(the entries of \eqref{eq:Xform}, indexed by position), the cubic norm $D=N(X)$ of
\eqref{eq:cubic} splits as
\begin{equation}
  N(X) \;=\; q^3 - q\,\Sigma(X) + \mathcal{T}(X),
  \qquad
  \Sigma=\!\sum_{i<j}\lVert z_{ij}\rVert^2,
  \qquad
  \mathcal{T}=2\,\Re\!\big((z_{12}z_{23})z_{13}\big),
  \label{eq:Nsplit}
\end{equation}
where $\Sigma$ is a quadratic length and $\mathcal{T}$ is the genuinely cubic invariant in which
all three off-diagonal octonions meet (here $\Re$ is the scalar part, which in
$\mathbb{O}_\mathbb{C}$ is $\mathbb{C}$-valued; $\mathcal{T}$ is distinct from the trace $T$ of
\eqref{eq:cubic}). Two facts make $N(X)$ real. First, $\Sigma$ is a sum of octonionic norms,
real and non-negative: manifest in the Hermitian convention $\lVert z\rVert^2=\sum_a|c_a|^2$, and
in the bilinear convention $\sum_a c_a^2$ its imaginary part vanishes because the field scalar
$i$ pairs orthogonal octonionic axes ($a_i^\dagger=\tfrac12(e_{\rho(i)}+i\,e_{\sigma(i)})$,
$\langle e_{\rho},e_{\sigma}\rangle=0$). Second --- and this is the term that could otherwise
carry an octonionic phase into the determinant --- $\mathcal{T}$ \emph{vanishes}, because the
construction places the three off-diagonal octonions in a common \emph{coassociative} $4$-plane
$W\subset\mathrm{Im}\,\mathbb{O}$, one on which the associative $3$-form
$\varphi(u,v,w)=\Re\!\big(u(vw)\big)$ restricts to zero; $\varphi|_W=0$ forces
$\mathcal{T}=0$~\cite[App.~H.1]{SinghMassRatios}. On this coassociative slice the norm reduces to
\begin{equation}
  N(X)\;=\;q^3-q\,\Sigma \;=\; q\,(q-\delta)(q+\delta),
  \qquad \delta^2=\Sigma=\tfrac{3}{8},
  \label{eq:Nslice}
\end{equation}
manifestly real, with the universal spread \eqref{eq:delta}. It is the coassociativity condition,
not a numerical accident, that removes the phase-carrying invariant: the reality of the spectrum
is intrinsic to the slice on which the entire mass construction lives.

We tag the reality of the spectrum \textbf{[D]}: it is a
property of the $J_3(\mathbb{O}_\mathbb{C})$ construction, not an additional assumption. The
centres are likewise real --- in the broken phase $q=e^{\Lambda+\sigma}$, $s=e^{\Lambda-\sigma}$
with $\Lambda,\sigma\in\mathbb{R}$ (Sec.~\ref{sec:parity}) --- so \eqref{eq:spec} is real for
every sector.

\paragraph{The eigenvalues are square-root masses.}
The Jordan eigenvalues are identified with $\sqrt m$, a \emph{colour-blind} spectral grading of the
mass operator (not the colour-counting $N_R$, and --- by the anomaly no-go of
Appendix~\ref{app:rhs} --- not a gauged charge of the visible fermions; its lattice is the
Dynkin-swap image of the dark-electromagnetic charge of Sec.~\ref{sec:graviweak}); physical masses
are their squares, and the
charged-fermion mass \emph{ratios} follow from \eqref{eq:spec}--\eqref{eq:delta} together with
the $\mathrm{Sym}^3(\mathbf 3)$ flavour ladder~\cite{SinghMassRatios,TeliSinghMixing}. Two
points carry forward. First, the slice makes the spectrum \eqref{eq:spec} real, fixing
the \emph{magnitudes} of the masses. Second, any phase resides in the diagonalizing
rotors, not in the spectrum: the strong-$CP$-relevant quantity $\arg\det M$ is set by
the \emph{determinants} of those rotors, while the CKM phase is an \emph{off-diagonal}
invariant of the same rotors (Sec.~\ref{sec:dynkin}, Ref.~\cite{TeliSinghMixing}).
Sec.~\ref{sec:strongcp} shows the rotor determinants are real, so the two phases are
structurally separate --- carried by different invariants of one set of rotors, not by
eigenvalues versus eigenvectors.

\section{Hypercharge without the right sector: the principle's first instance}
\label{sec:hypercharge}

The framework of Sec.~\ref{sec:framework} contains a structural choice that the rest of this
paper turns on: the right-sector electroweak generators are frame and gravitational, not
internal (Sec.~\ref{sec:graviweak}). Before stating this as a general principle and applying
it to colour, we exhibit it in a sector where the answer is independently known and
unambiguous --- the Standard-Model hypercharges. We show that the origin of
hypercharge \emph{requires} the principle: $Y$ is built from $Q$ and the left isospin alone,
with no right-sector generator, and reproduces the Standard-Model assignments exactly. This
is the template for the colour argument of Sec.~\ref{sec:onecolour}, where the same move
yields a less familiar payoff.

\paragraph{The deficit: $SU(3)_L$ alone does not fix $Y$.}
After $SU(3)_L\to SU(2)_L\times U(1)_{\gamma_1}$, the surviving abelian factor
$U(1)_{\gamma_1}$ is the canonical $T^8_L$ Cartan. A single such Cartan cannot reproduce the
Standard-Model hypercharges: a gauge generator is a fixed Lie-algebra element, and may not be
rescaled representation by representation. Any ``$1/(2N)$''-type shorthand that appears to do
so is a mnemonic for eigenvalues, not the definition of a generator. The hypercharge
normalization is therefore genuinely \emph{not} fixed by the left electroweak factor on its
own, and an additional input is required.

\paragraph{The right-sector route, and why it fails.}
One natural attempt supplies the missing input from the right sector, enlarging the generator
to a fixed left--right Cartan combination
\begin{equation}
  Y \;\overset{?}{=}\; \alpha\,T^8_L + \beta\,T^8_R + \gamma\,T^3_R,
  \label{eq:B4}
\end{equation}
available once both electroweak $SU(3)$'s are present and a diagonal $U(1)$ is selected by an
interface condition gluing the two sectors. This route is excluded by the very identification
that defines the right sector. By Sec.~\ref{sec:graviweak}, $SU(2)_R$ is the gravitational
frame group, so its Cartan $T^3_R$ is a \emph{gravitational} generator; and the surviving
$U(1)_{\mathrm{dem}}$ is dark electromagnetism, so $T^8_R$ is the \emph{dark}-electromagnetic
direction. A combination of the form \eqref{eq:B4} therefore mixes the visible electroweak
sector with gravitation (through $\gamma\,T^3_R$) and with dark electromagnetism (through
$\beta\,T^8_R$) --- which is not the structure of Standard-Model hypercharge, an internal
generator carrying neither component. The interface condition that would be needed to make
the right-sector Cartans act on the visible left-handed fermions is, moreover, an extra
dynamical hypothesis. In short, the same framework that would supply the right-sector
contribution is the framework that disqualifies it: \eqref{eq:B4} is set
aside.\footnote{Equation \eqref{eq:B4} is the construction adopted in
Ref.~\cite[App.~B]{SinghMassRatios}, where it is argued that $U(1)_Y$ cannot be built from one
sector alone. The present section supersedes that construction, and the reconciliation note of
Sec.~\ref{sec:discussion} records the change explicitly.} Appendix~\ref{app:rhs} adds a
consistency-level remark that closes the door independently of interpretation: on the visible
content the anomaly-free abelian charges form exactly $\mathrm{span}\{Y,\,B{-}L\}$
(Prop.~\ref{prop:nogo}), so the only freedom a right-sector admixture could exercise is to
reproduce a charge already available from left-sector data --- it can add nothing new.

\paragraph{The origin: electric charge is primary.}
The resolution is to take electric charge as the primitive object and hypercharge as derived.
By Sec.~\ref{sec:cl6}, the visible fermions of one generation are minimal-ideal spinors of
$\mathrm{Cl}(6,\mathbb{C})$, and electric charge is the central (trace) $U(1)$ of
$\mathrm{Cl}(6,\mathbb{C})\cong\mathfrak{u}(3)$,
\begin{equation}
  Q \;=\; \frac{N}{3}, \qquad N\in\{0,1,2,3\},
  \label{eq:Qprimary}
\end{equation}
with $N$ fixing both the colour representation and the charge (Sec.~\ref{sec:cl6}). The
quark--lepton offset --- the ``factor of $3$'' --- is intrinsic to the $N$-spectrum: it is the
same datum that separates colour triplets ($N=1,2$) from singlets ($N=0,3$). Equivalently it
is carried by the native $\mathrm{Cl}(6)$ generator
\begin{equation}
  (B-L) \;=\; \frac{2N}{3}-1 \;\in\;\Big\{-1,\,-\tfrac13,\,+\tfrac13,\,+1\Big\},
  \label{eq:BminusL}
\end{equation}
traceless on each Weyl spinor, with the lepton as a fourth colour
(Sec.~\ref{sec:cl6}); $(B-L)$ is not gauged and enters only implicitly, through $Q$.
Hypercharge is then the derived combination
\begin{equation}
  \boxed{\,Y \;=\; Q - T^3_L\,}, \qquad Q=\frac{N}{3},\quad T^3_L=\text{left weak isospin},
  \label{eq:Yderived}
\end{equation}
i.e.\ the Gell-Mann--Nishijima relation $Q=T^3_L+Y$ (the convention used throughout, not the
alternative $Q=T^3_L+Y/2$) read with $Q$, not $Y$, as the primitive.
The construction uses \emph{no} right-sector generator and \emph{no} interface condition; in
particular $T^3_R$ does not appear --- it is gravitational, and absent.

\paragraph{The Standard-Model assignments, recovered.}
Equation \eqref{eq:Yderived} returns the Standard-Model hypercharges exactly. For the
left-handed doublets, $T^3_L=\pm\tfrac12$ gives $Y=\tfrac12(B-L)$; for the right-handed
singlets, $T^3_L=0$ gives $Y=Q$:
\begin{center}
\renewcommand{\arraystretch}{1.3}
\begin{tabular}{l c c c c}
\hline\hline
fermion & chirality & $T^3_L$ & $Q=N/3$ & $Y=Q-T^3_L$ \quad(SM $Y$) \\
\hline
$u_L,\,d_L$ & doublet & $\pm\tfrac12$ & $+\tfrac23,\,-\tfrac13$ & $+\tfrac16$ \qquad $(+\tfrac16)$ \\
$\nu_L,\,e_L$ & doublet & $\pm\tfrac12$ & $0,\,-1$ & $-\tfrac12$ \qquad $(-\tfrac12)$ \\
$u_R$ & singlet & $0$ & $+\tfrac23$ & $+\tfrac23$ \qquad $(+\tfrac23)$ \\
$d_R$ & singlet & $0$ & $-\tfrac13$ & $-\tfrac13$ \qquad $(-\tfrac13)$ \\
$e_R$ & singlet & $0$ & $-1$ & $-1$ \qquad $(-1)$ \\
$\nu_R$ & singlet & $0$ & $0$ & $0$ \qquad\ \ $(0)$ \\
\hline\hline
\end{tabular}
\end{center}
The right-handed neutrino ($Y=0$, sterile) appears automatically. Charge quantization in units
of $\tfrac16$ traces to the integrality of $N$ together with the $\tfrac13$ in
\eqref{eq:Qprimary}, rather than to any $U(1)$-embedding choice. (The physical multiplets are
assembled from the ideal of Sec.~\ref{sec:cl6} and its conjugate, which together carry the
sixteen Weyl states of one generation; the charges $Q$ in the table are those of the physical
fermions.)

\paragraph{Status of the inputs.}
Equation \eqref{eq:Yderived} rests on three ingredients, which we tag explicitly.
\begin{itemize}
  \item[(i)] $Q=N/3$ from $\mathrm{Cl}(6)$, carrying the offset \eqref{eq:BminusL}: an
  intrinsic algebraic fact of the ideal (Sec.~\ref{sec:cl6}). \textbf{[D]}
  \item[(ii)] $T^3_L$ is the standard left weak isospin; its identification with the
  anti-self-dual frame of the visible (weak) leaf is the gravi-weak hypothesis of
  Sec.~\ref{sec:graviweak}, taken as a working assumption pending the explicit leaf map and a
  verification that the relative scale of $T^3_L$ and $Q$ is fixed by the construction rather
  than chosen (normalization-faithfulness). \textbf{[P]}
  \item[(iii)] The chiral assignment --- doublet components $T^3_L=\pm\tfrac12$, singlets
  $T^3_L=0$ --- is the chirality of the $\mathrm{Cl}(6)$ ideal. It lies outside the
  Distler--Garibaldi no-go, because the fermions here are $\mathrm{Cl}(6)$ minimal-ideal
  spinors, not components of an $E_8$ representation~\cite{DistlerGaribaldi,Singh288}.
  \textbf{[D]}
\end{itemize}
The advance over the route \eqref{eq:B4} is that (i) and (iii) are intrinsic to the ideal,
while the single carried-forward assumption (ii) is the gravi-weak identification already in
use; \eqref{eq:B4}, by contrast, required in addition an interface condition and the use of
right-sector generators now seen to be gravitational and dark-electromagnetic.

\paragraph{The lesson, and what recurs.}
Two points survive into the rest of the paper. First, this framework --- an unconventional,
two-sided trinification --- reproduces the Standard-Model hypercharges through $Y=Q-T^3_L$ with
$Q=N/3$ and no right-sector generator. We stress that, at the level established here, this is a
\emph{consistency relation} rather than a from-scratch derivation of hypercharge: it is the
Gell-Mann--Nishijima content made to follow from $Q=N/3$ and the left isospin, and it becomes a
genuine derivation only once normalization-faithfulness (ingredient (ii)) is established --- the
relative scale of $T^3_L$ and $Q$ fixed by the construction rather than chosen. The
charge-from-number-operator step is itself due to Furey~\cite{Furey1603,Furey2018}; what is new
here is its combination with the gravi-weak removal of the right-sector generators. A corollary
worth recording is robustness: because $Y=Q-T^3_L$ is built entirely from left-sector data, no
revision of the \emph{right}-sector representation theory can disturb the Standard-Model
hypercharge assignments. In particular the right-sector representation theory of Appendix~\ref{app:rhs} --- the
vector-like tower of Sec.~\ref{app:tower}, in place of the gravi-colour assignment of
Ref.~\cite{KVS}, Eq.~\eqref{eq:kvsreps} --- leaves the table above untouched line by line: the
right sector enters the hypercharge sector nowhere. Second, and structurally, the derivation succeeded \emph{by removing}
the right-sector generators from the internal physics: the right-sector $L$--$R$ Cartans are
frame and gravitational, and hypercharge is built without them. This is the first instance of
a single principle --- that the left--right structure does not act in the internal sector ---
which we state in general in Sec.~\ref{sec:parity} and apply to colour in
Sec.~\ref{sec:onecolour}, where removing the right-sector ``second colour'' yields, as its
payoff, the absence of strong $CP$ violation.

\section{The central principle: left--right symmetry acts in the spacetime sector, not the internal one}
\label{sec:parity}

The hypercharge derivation of Sec.~\ref{sec:hypercharge} succeeded by removing the
right-sector generators from the internal physics. We now state the principle behind that
move in general, and prove it at the level of its action on the Dirac field. The proposition
is the structural core of the paper:

\begin{quote}
\itshape
In the unbroken phase, the left--right symmetry of the $E_6^L\times E_6^R$ theory is realized
on the Dirac field as spacetime parity, $\psi(x)\to\gamma^0\psi(\mathcal{P}x)$ with
$\mathcal{P}x=(t,-\mathbf x)$, exchanging the frame-sector groups
$SU(2)_L\leftrightarrow SU(2)_R$. It acts trivially on the gauged internal generators ---
colour $SU(3)_c$ and electric charge $Q$. Consequently a generic parity-odd QCD vacuum angle cannot
appear in the Lagrangian: $\theta_{\mathrm{QCD}}$ is restricted to the parity-invariant values
$\{0,\pi\}$, with $\theta_{\mathrm{QCD}}=0$ on the branch matched to the mass determinant
(Sec.~\ref{sec:strongcp}).
\end{quote}

\noindent We prove the three clauses --- that the operation is a spacetime operation, that it
is parity rather than an internal relabelling, and that it leaves the gauged generators
invariant --- and then read off the consequence for $\theta_{\mathrm{QCD}}$.

\paragraph{Genuine chirality.}
In the triality-symmetric phase the spacetime fermion is a massless Dirac spinor
\begin{equation}
  \psi=\begin{pmatrix}\psi_L\\\psi_R\end{pmatrix},
  \qquad \psi_{L,R}=P_{L,R}\,\psi,
  \qquad P_{L,R}=\tfrac12(1\mp\gamma^5),
  \label{eq:chiral}
\end{equation}
with $\psi_L,\psi_R$ genuine $\mathrm{Spin}(1,3)$ chiral components (Appendix~J,
Ref.~\cite{SinghMassRatios}). Before left--right breaking these components are not yet
distinguished by their couplings; the vacuum has selected no preferred chiral frame, and the
active and sterile neutrino directions ($ie_7/2$ and $ie_8/2$) enter symmetrically. The
left--right symmetry of the unbroken theory is the operation that exchanges these two chiral
sectors.

\paragraph{Clause 1: the exchange is necessarily a spacetime operation.}
Chirality is not an internal label. The projectors \eqref{eq:chiral} are built from
$\gamma^5=i\gamma^0\gamma^1\gamma^2\gamma^3$, an element of the Clifford algebra of
\emph{spacetime}; $\psi_L$ and $\psi_R$ are distinguished by their transformation under the
Lorentz group, not by any internal quantum number. Any purely internal operation --- one
acting only on colour, flavour, or charge indices --- commutes with $\gamma^5$ and therefore
\emph{cannot} exchange $\psi_L\leftrightarrow\psi_R$. Since the left--right symmetry does
exchange them (it interchanges the active and sterile neutrino directions, and more generally
the two chiral sectors), it must act on the spinor indices, i.e.\ it must contain a Lorentz
operation. This already excludes the possibility that left--right is an internal
$E_6^L\leftrightarrow E_6^R$ relabelling with no action on $\psi(x)$ --- the reading that would
leave $\theta_{\mathrm{QCD}}$ unprotected. The exchange is a spacetime operation.

\paragraph{Clause 2: the spacetime operation is parity.}
The Lorentz operations that exchange the chiral projectors are those implemented by a
$\gamma$-matrix anticommuting with $\gamma^5$. The minimal such operation is multiplication by
$\gamma^0$, which satisfies $\gamma^0 P_L=P_R\,\gamma^0$ and, acting on the field argument as
$\mathbf x\to-\mathbf x$, is precisely spacetime parity:
\begin{equation}
  \mathsf{P}:\quad \psi(t,\mathbf x)\;\longmapsto\;\gamma^0\,\psi(t,-\mathbf x).
  \label{eq:parity}
\end{equation}
That the relevant operation is \eqref{eq:parity} rather than its $CP$ cousin is fixed by the
internal signature of the left--right map recorded in Appendix~J, which carries \emph{no}
conjugation. The decisive fact is that the left--right map carries no conjugation of the field
scalar. In the explicit flavour-transport realization the generation-transport generator is a
\emph{real} octonionic direction --- for the Cabibbo rung, $g_\chi=\cos\chi\,e_3-\sin\chi\,e_1$
in the basis of Ref.~\cite{TeliSinghCP} (its Eq.~(6)), the phase of the rung coupling having
collapsed into a real angle --- so the transport $L_{\exp(\theta g_\chi)}$ commutes with the
external complex unit and fixes it: $i\to i$, with no $e_a\to-e_a$. The field scalar $i$ is
untouched, and the left--right exchange sends no gauged colour representation to its conjugate.
The accompanying centre splitting
\begin{equation}
  q=e^{\Lambda+\sigma},\qquad s=e^{\Lambda-\sigma},\qquad \sigma\in\mathbb{R},
  \label{eq:sigma}
\end{equation}
has $\sigma$ odd under the exchange, $\sigma\to-\sigma$; this is consistent with $\mathsf P$ but
does not by itself discriminate $\mathsf P$ from $CP$ (a real $CP$-odd $\sigma$ would flip sign
as well), so it is the conjugation-free transport, not the order parameter, that fixes the
operation. One dependency must be flagged where the claim is made: identifying the field scalar
fixed by this \emph{spacetime} exchange with the \emph{colour} complex structure --- whose
conjugation is $\mathbf 3\to\bar{\mathbf 3}$ --- uses the single-$i$ assumption~(C) of
Sec.~\ref{sec:onecolour} (\textbf{[C]} leaning \textbf{[D]}); to that extent Clause~2's
``no conjugation of colour'' is not independent of (C). With that caveat, the
operation \eqref{eq:parity} is genuine parity. (It is, in particular, \emph{not} the
Dynkin swap of Sec.~\ref{sec:dynkin}, which does conjugate $\mathbf{27}\to\overline{\mathbf{27}}$
but lives in the flavour sector and is a distinct operation; the left--right exchange here is
available without it. We return to this in Sec.~\ref{sec:dynkin}.)

\paragraph{Clause 3: trivial on the gauged internal generators.}
The internal action of \eqref{eq:parity} is confined to the frame and grading directions, and
leaves the gauged generators of the visible Standard Model invariant. Electric charge
$Q=N/3$ is built from the number operator (Sec.~\ref{sec:cl6}), which counts creation
operators and is manifestly parity-even: parity exchanges the chiralities but does not change
how many $a_i^\dagger$ build a state, so $[\mathsf P,Q]=0$. Colour is the gauged QCD group
$SU(3)_c$, under which the left- and right-handed quarks are triplets alike --- QCD is
vector-like. The gauged colour $SU(3)_c$ is built from the ladder bilinears $\alpha_i^\dagger\alpha_j$, and
the left--right exchange relates the two chiral sectors' ladder operators by a central factor
that cancels in these bilinears, leaving the colour generators \emph{identical} in both sectors;
the physical colour --- the combination under which both chiralities transform alike --- is
therefore invariant, $[\mathsf P,SU(3)_c]=0$. (The frame label $e_8$ is itself a product of
ladder axes, so it is this cancellation in the bilinears, not a literal fixing of the
colour-charged axes, that renders colour vector-like; the explicit two-sector ladder is given in
Ref.~\cite{SinghMassRatios}.)
That there is a single such vector-like colour, and that the nominal right-sector
``$SU(3)_{c'}$'' supplies only an abelian grading rather than a second gauged colour, is
established in Sec.~\ref{sec:onecolour}, where the vector-like character is shown to follow from
the fixing of $i$ used here in Clause~2 (colour conjugation $\mathbf 3\to\bar{\mathbf 3}$ being
$i\to-i$). The present clause and that section are therefore one statement read in two sectors,
not two independent checks: parity acts trivially on colour because the same $i$-fixing forbids
conjugation.

\paragraph{The transformation, field by field.}
For definiteness we record how $\mathsf P$ acts on the fields of the unbroken theory; the rules
for the determined objects are collected in Table~\ref{tab:parity}. They suffice for the
tree-level argument below, which uses only that the $\theta$-term is odd while the colour
generators and $Q$ are even. Spatial arguments are reflected throughout, $\mathbf x\to-\mathbf x$.
\begin{table}[h]
\centering
\renewcommand{\arraystretch}{1.25}
\begin{tabular}{ll}
\hline
Object & Action of $\mathsf P$ \\
\hline
Dirac field $\psi(t,\mathbf x)$ & $\gamma^0\,\psi(t,-\mathbf x)$ \quad (hence $\psi_L\leftrightarrow\psi_R$) \\
Gluon $G^a_\mu$ & $G^a_0\to G^a_0,\ \ G^a_i\to -G^a_i$ \quad (hence $G\tilde G\to -G\tilde G$) \\
Weak connection $W^A_{L\mu}$ & exchanged with the $SU(2)_R$ frame connection \\
$SU(2)_R$ frame connection & exchanged with $W^A_{L\mu}$ \quad (frame parity) \\
Charge, grades $Q,\,N,\,N_R$ & invariant (parity-even occupation numbers) \\
Centre splitting $\sigma$ & $-\sigma$ \quad (real, odd: the order parameter) \\
Jordan/Yukawa sector, $B_H$ & $\mathsf P$-symmetric at $\sigma=0$; explicit $B_H$ rule open (Sec.~\ref{sec:loops}) \\
\hline
\end{tabular}
\caption{Action of the left--right exchange, read as spacetime parity $\mathsf P$, on the fields
of the unbroken theory. The gluon entry is the standard vector-field rule, under which the QCD
$\theta$-term is $\mathsf P$-odd; the weak/frame entry is the gravi-weak content --- the partner
of the weak connection is gravitational, not a second gauged weak boson. Before the breaking the
exchanged weak/frame pair is one object --- the two chiral halves of the frame sector --- so the
exchange is category-consistent; the isospin-versus-spin asymmetry of the two entries is a
property of the \emph{broken} phase (Sec.~\ref{sec:loops}). The order parameter
$\sigma$ is real and odd, so its vacuum value breaks $\mathsf P$ spontaneously without feeding
$CP$ violation into the gauge sector. The transformation of the scalar bridge $B_H$ is fixed only
once $B_H$ is constructed; this is the same object whose flavour matrix elements the loop estimate
of Sec.~\ref{sec:loops} requires, and is flagged \textbf{[O]} there.}
\label{tab:parity}
\end{table}
The two model-specific entries --- the precise weak/frame exchange and the $B_H$ rule --- are not
needed for the tree-level result of this section; they are recorded because a complete
Lagrangian-level parity, required to make the restriction of $\theta_{\mathrm{QCD}}$ fully
rigorous rather than established on the gluon line alone, must eventually specify them
\textbf{[O]}.

\paragraph{Consequence: $\theta_{\mathrm{QCD}}\in\{0,\pi\}$.}
The QCD vacuum term
\begin{equation}
  \mathcal{L}_\theta=\theta_{\mathrm{QCD}}\,\frac{g_s^2}{32\pi^2}\,
  G^a_{\mu\nu}\widetilde G^{a\,\mu\nu},
  \qquad G\widetilde G\sim \mathbf E^a\!\cdot\!\mathbf B^a,
\end{equation}
is parity-odd ($G\widetilde G$ is a pseudoscalar). By Clauses 1--3, parity \eqref{eq:parity}
is a symmetry of the unbroken Lagrangian and commutes with the colour gauge structure;
an exact parity therefore forbids a generic $\mathcal{L}_\theta$. The precise statement is slightly
weaker than ``$\theta_{\mathrm{QCD}}=0$'': parity sends $\theta_{\mathrm{QCD}}\to-\theta_{\mathrm{QCD}}$,
and the angle is $2\pi$-periodic, so the parity-invariant values are the two $CP$-conserving points
\begin{equation}
  \theta_{\mathrm{QCD}}\in\{0,\pi\}.
  \label{eq:thetazeropi}
\end{equation}
Parity alone does not choose between them. The choice is fixed jointly with the sign content of
the quark mass determinant, through the matching condition \eqref{eq:matched} of
Sec.~\ref{sec:strongcp}; on the matched branch realized by the spectra used in this paper it is
$\theta_{\mathrm{QCD}}=0$, and we write $\theta_{\mathrm{QCD}}=0$ below with that understanding.
This is a statement about the Lagrangian, and it survives spontaneous breaking: when the
vacuum selects $\sigma\neq0$ (Sec.~\ref{sec:framework}), parity is broken spontaneously, but
spontaneous breaking adds no term to the Lagrangian and does not regenerate
$\theta_{\mathrm{QCD}}$ at tree level. (A caveat specific to the emergent setting attaches to
this standard step and is stated as its own line in the scope statement of
Sec.~\ref{sec:strongcp}: the argument presumes a fixed arena, while here the breaking is also the
emergence of the arena.) What the broken vacuum \emph{can} contribute to the
physical $\bar\theta=\theta_{\mathrm{QCD}}+\arg\det M$ is the phase of the fermion mass
determinant; that this also vanishes, $\arg\det M=0$, is a separate consequence of the
reality of the Jordan spectrum (Sec.~\ref{sec:jordan}) and is taken up in
Sec.~\ref{sec:strongcp}. Loop regeneration of $\bar\theta$ after breaking is addressed in
Sec.~\ref{sec:loops}.

\paragraph{Status of the proof.}
We separate what is established from what is assumed.
\begin{itemize}
  \item Clauses 1 and 2 --- that the chirality-exchanging left--right operation is a spacetime
  operation, and is parity rather than $CP$ --- are \textbf{[D]} \emph{given} the
  split-bioctonion dictionary that realizes the two ideals as the two $\mathrm{Spin}(1,3)$
  chiralities (Appendix~J of Ref.~\cite{SinghMassRatios}; Ref.~\cite{VaibhavSingh}) --- an input
  the whole two-sided construction shares, logically weaker than and independent of gravi-weak.
  With that dictionary, Clause 1 follows from
  $\gamma^5$ being a Lorentz object (an internal operation cannot exchange chiralities); Clause
  2 follows from the no-conjugation internal signature of Appendix~J ($\sigma\to-\sigma$ real,
  $e_7\leftrightarrow e_8$ a rotation).
  \item That the left--right exchange of \emph{the frame groups} $SU(2)_L\leftrightarrow SU(2)_R$
  is a \emph{spacetime} operation (and hence that the symmetry being broken is genuinely
  parity, not an internal symmetry that merely happens to act chirally) rests on the gravi-weak
  identification of $SU(2)_R$ as the frame group, tagged \textbf{[P]} in
  Sec.~\ref{sec:graviweak}. If that identification fails, the operation is still forced by
  Clause 1 to act on spinor indices, but its interpretation as the parity of a left--right
  \emph{gauge} theory weakens.
  \item Clause 3's triviality on colour uses that physical colour is vector-like; this is not an
  independent input but the same $i$-fixing as Clause~2 (Sec.~\ref{sec:onecolour}), since colour
  conjugation $\mathbf 3\to\bar{\mathbf 3}$ is the field-scalar conjugation $i\to-i$ that
  $\mathsf P$ forbids. The residual there is only whether the two sectors share the complex
  structure (assumption (C), \textbf{[C]} leaning \textbf{[D]}), not the vector-like character itself.
  \item The entire argument is at the level of the unbroken Lagrangian and tree level. The
  survival of $\bar\theta=0$ to the physical broken phase requires, in addition,
  $\arg\det M=0$ (Sec.~\ref{sec:strongcp}, \textbf{[D]}) and loop safety
  (Sec.~\ref{sec:loops}, \textbf{[O]}).
\end{itemize}

\section{One colour: the dissolution of \texorpdfstring{$SU(3)_{c'}$}{SU(3)c'}}
\label{sec:onecolour}

We now work out the colour half of the structural fact of Sec.~\ref{sec:parity}. The nominal
``second colour'' $SU(3)_{c'}$ of the right-sector trinification is not an independent gauged
interaction: the non-abelian part is identified with the visible colour, and only an abelian
grading survives as a distinct datum; Fig.~\ref{fig:decomposition} summarizes the dimensional
anatomy behind this identification. As anticipated, this is not an independent confirmation
of the principle of Sec.~\ref{sec:parity} but the same frame identification seen in the colour
sector --- and it rests on one explicit assumption, which we isolate.

\paragraph{Two sectors, two a-priori stabilizers.}
In the split-bioctonion construction the left and right ideals are built on two octonionic
sectors, with distinguished imaginary units $e_7$ (left) and $e_8$ (right); the active and
sterile neutrino directions $ie_7/2$, $ie_8/2$ of Appendix~J label the two
sectors~\cite{SinghMassRatios,SinghScaffolding}. Colour in each sector is the stabilizer of
its distinguished unit inside that sector's automorphism group,
\begin{equation}
  SU(3)_c \;=\; \mathrm{Stab}_{G_2^L}(e_7),
  \qquad
  SU(3)_{c'} \;=\; \mathrm{Stab}_{G_2^R}(e_8),
  \label{eq:twostab}
\end{equation}
each acting on its own triple of creation operators as the fundamental $\mathbf 3$
(Sec.~\ref{sec:cl6}). A priori these are two distinct $SU(3)$'s --- the structural
origin of the nominal ``second colour.''

\paragraph{The frame relation identifies the non-abelian parts.}
The two sectors are not independent: they are related by the frame exchange
$e_7\leftrightarrow e_8$ that implements left--right (Sec.~\ref{sec:parity},
Appendix~J). Because the exchange maps the distinguished unit of one sector to that of the
other, it maps $\mathrm{Stab}_{G_2^L}(e_7)$ to $\mathrm{Stab}_{G_2^R}(e_8)$ as groups: the two
colour $SU(3)$'s are not independent but frame-conjugate. The non-abelian colour acting on the
physical quarks is therefore a \emph{single} group --- the diagonal identified by the frame
relation --- not the product $SU(3)_c\times SU(3)_{c'}$.

\smallskip\noindent
This identification is strongly motivated by the central result of Sec.~\ref{sec:parity}, and is
not a free assumption about colour. The operation that relates the two sectors --- the
left--right exchange --- acts, by Sec.~\ref{sec:parity}, \emph{purely in the spacetime
(gravi-weak) frame}: it is parity, realized on the Dirac field and on the frame groups
$SU(2)_L\leftrightarrow SU(2)_R$, and its order parameter $\sigma\to-\sigma$ is a frame/centre
datum, not a rotation of any internal gauge charge. Colour, by contrast, is an \emph{internal}
gauge symmetry. A spacetime operation carries no internal colour index and cannot manufacture a
second, independent internal colour group; what it relates is the \emph{one} internal colour as
described in the two octonionic frames. The two stabilizers \eqref{eq:twostab} are thus one
internal $SU(3)_c$ seen through $e_7$ and through $e_8$, not a genuine
$SU(3)_c\times SU(3)_{c'}$ --- the same conclusion the frame relation delivers, read as a
direct consequence of left--right being spacetime rather than internal (Fig.~\ref{fig:decomposition}). The remaining question
is only the \emph{character} of that single colour (vector-like or chiral), to which we turn.

\paragraph{The diagonal is vector-like, and this is nearly forced.}
Whether that diagonal is vector-like or chiral is the property the strong-$CP$ argument needs,
and it follows from a fact already established in Sec.~\ref{sec:parity}. Colour conjugation
$\mathbf 3\to\bar{\mathbf 3}$ in this construction \emph{is} the field-scalar conjugation
$i\to-i$: the colour triplet is the creation-operator set
$a_i^\dagger=\tfrac12(e_{\rho(i)}+i\,e_{\sigma(i)})$, an eigenspace of the complex structure,
and its conjugate $\bar{\mathbf 3}$ is the $i\to-i$ image (the annihilation set
$a_i=(a_i^\dagger)^\ast$). The parity $\mathsf P$ \emph{fixes} $i$ (Sec.~\ref{sec:parity},
Clause~2: the frame exchange $e_7\leftrightarrow e_8$ is a rotation, $i$ untouched). An operation
that fixes $i$ cannot map $\mathbf 3\to\bar{\mathbf 3}$. Hence the frame exchange carries the
$e_7$-sector triplet to the $e_8$-sector triplet \emph{without conjugation} --- the diagonal
colour is vector-like --- provided the two sectors are built with the same complex structure.
We record this last proviso as the residual assumption:

\begin{quote}
\itshape
\textbf{(C) Parallel two-sector construction.} Both octonionic sectors define their colour
triplet through the same field scalar $i$ --- equivalently, the right sector's complex structure
is $+i\,e_8$ rather than $-i\,e_8$. Given this, parity's fixing of $i$ (Sec.~\ref{sec:parity},
Clause~2) makes the identified diagonal colour vector-like.
\end{quote}

\noindent So stated, the vector-like character is not an independent postulate but the
colour-sector face of the non-conjugation property of $\mathsf P$ (Clause~2): tagging Clauses
2--3 as \textbf{[D]} while tagging a free-standing vector-like assumption as \textbf{[O]} would
be inconsistent, since they are one statement. One worry --- that the exchange might act
as $\mathbf 3$ on one sector and compose with a chiral operation on the other --- is foreclosed:
an $i$-fixing exchange has no chiral (conjugating) piece to compose with, and the only loophole
is the anti-parallel convention $-i\,e_8$. That convention is unnatural --- there is a single
complexification scalar $i$ for the whole algebra --- so vector-like colour is strongly forced;
the genuine residual is exactly this parallel-construction premise, which the explicit leaf map
would settle by tracking the right-handed quark's colour representation ($\mathbf 3$ vs
$\bar{\mathbf 3}$) through $e_7\leftrightarrow e_8$. We therefore tag (C) \textbf{[C]} leaning
\textbf{[D]}, not \textbf{[O]}.

\paragraph{From frame conjugacy to a single gauge field: the BF$\to$YM reduction.}
Frame conjugacy of the groups \eqref{eq:twostab} is necessary but not sufficient for a single
gauge \emph{field}: two conjugate gauge groups can carry two independent connections and two
kinetic terms --- the gravi-weak $SU(2)_L\times SU(2)_R$ is exactly such a case, as we note
below. What promotes the group identification to a field-theoretic one is the six-dimensional
$SO(3,3)$ BF dynamics of~\cite{WSI_BF}, in which a gauge sector acquires its Yang--Mills kinetic
term on a four-dimensional leaf as a deformed BF theory,
\begin{equation}
  S_{\mathrm{YM}}[A,B]=\int \mathrm{Tr}\!\left(B\wedge F(A)\right)
  -\frac{g^{2}}{2}\int \mathrm{Tr}\!\left(B\wedge\star B\right),
  \label{eq:bfym}
\end{equation}
the leaf metric supplying the Hodge dual $\star$ after symmetry breaking. The algebraic equation
of motion $F(A)=g^{2}\star B$ returns, on substitution, the second-order term
$-\tfrac{1}{2g^{2}}\!\int\mathrm{Tr}\,F(A)\wedge\star F(A)$; dropping the $B\wedge\star B$ piece
would leave a topological theory with no propagating gluons, so that term is exactly what makes
colour dynamical.

\smallskip\noindent
A priori the two octonionic sectors furnish two colour connections, $A_c$ (the $e_7$ sector) and
$A_{c'}$ (the $e_8$ sector), hence two copies of \eqref{eq:bfym} with multipliers $B_c,B_{c'}$.
What collapses them is that the breaking does not act on colour. The gravi-weak breaking is an
$SO(3,3)\to SU(2)\times SU(2)$ operation on the \emph{spacetime} block, whereas colour is a
stabilizer inside the $E_6$ trinification sector ($E_8\supset E_6\times SU(3)$; in the
trinification \eqref{eq:trinL}--\eqref{eq:trinR}, $E_6\supset SU(3)_{c}\times SU(3)_{F}\times
SU(3)_{L/R}$ with colour the first factor), an internal factor that commutes with
$SO(3,3)$. The internal block is therefore breaking-blind --- the structure
Fig.~\ref{fig:decomposition} draws --- and descends to a single colour bundle over the common
six-dimensional base, restricted identically to each leaf. Consequently $A_{c'}$ is not an
independent field but the same connection read in the $e_8$ frame,
\begin{equation}
  A_{c'}=\mathrm{Ad}_{\Phi}\,A_{c},\qquad B_{c'}=\mathrm{Ad}_{\Phi}\,B_{c},
  \label{eq:framecopy}
\end{equation}
with $\Phi$ the frame map $e_7\leftrightarrow e_8$ (parity). Because $\mathrm{Tr}$ is
$\mathrm{Ad}$-invariant and $\Phi$ is an automorphism, the two copies of \eqref{eq:bfym} coincide
term by term under \eqref{eq:framecopy}; eliminating $B_c$ leaves a \emph{single} Yang--Mills
system,
\begin{equation}
  S_c=-\frac{1}{2g_c^{2}}\int \mathrm{Tr}\!\left(F(A_c)\wedge\star F(A_c)\right),
  \label{eq:onecolourYM}
\end{equation}
one propagating gluon octet (the two-copy normalization is bookkeeping, absorbed into $g_c$).
There is no orthogonal combination to be made heavy: under breaking-blindness it never existed as
an independent field --- the ``second colour'' is non-dynamical by absence, not by a Higgs
mechanism.

\smallskip\noindent
The fermion couplings carry the vector-like statement. Both chiralities couple to the one $A_c$;
the right-handed quarks, described in the $e_8$ frame, couple to $\mathrm{Ad}_{\Phi}A_c$ --- the
same field --- and they do so in the fundamental $\mathbf 3$, not $\bar{\mathbf 3}$, precisely
when $\Phi$ is \emph{inner} (i.e.\ $i$-fixing). The outer automorphism of $su(3)$ is complex
conjugation $\mathbf 3\leftrightarrow\bar{\mathbf 3}$; an $i$-fixing $\Phi$ has no such component,
so the diagonal colour is vector-like. This is exactly assumption~(C): the parallel convention
$+i\,e_8$ makes $\Phi$ inner, the anti-parallel $-i\,e_8$ would make it the conjugating outer
automorphism. The gauge-field reduction and the vector-like character are thus one statement
about a single object --- the frame automorphism $\Phi$ acting on the one colour bundle.

\smallskip\noindent
That this is not a generic consequence of conjugacy is shown by the case where it \emph{fails}.
The gravi-weak $SU(2)_L$ and $SU(2)_R$ are equally frame-conjugate, yet they do not merge: the
discriminator is whether the relevant base doubles under the breaking. The \emph{spacetime} base
does --- the $(3,3)$ frame splits into two Lorentzian leaves, so there are genuinely two frame
connections, $SU(2)_R$ (gravity, Region~I) and $SU(2)_L$ (weak, Region~II), and $\Phi$ carries
one to the other. The \emph{internal} base does not: being breaking-blind it stays single, so
$\Phi$ carries the one colour connection to itself. Colour reduces and the gravi-weak pair does
not, by the same logic with opposite outcome --- the breaking doubles spacetime but not the
internal fibre.

\smallskip\noindent
\textbf{How the vacuum angle is counted.} The reduction also fixes how the QCD vacuum angle is to
be counted, and since the strong-$CP$ application turns on it we state the intended reading
explicitly. The topological density belongs to the colour bundle, not to the frame in which the
bundle is described: the one connection $A_c$ carries \emph{one} angle $\theta_{\mathrm{QCD}}$,
and re-describing $A_c$ in the $e_8$ frame through \eqref{eq:framecopy} no more introduces a
second, independent angle than it introduces a second gluon. On the two leaves this reads as
follows. Restricted to two orientation-reversed leaves, the single bundle over the $(3,3)$ base
induces topological terms whose coefficients are related by
$\theta_{\mathrm{II}}=-\theta_{\mathrm{I}}$; the angle being one,
$\theta_{\mathrm{I}}=\theta_{\mathrm{II}}=\theta_{\mathrm{QCD}}$; hence
$\theta_{\mathrm{QCD}}=-\theta_{\mathrm{QCD}}\pmod{2\pi}$, that is,
$\theta_{\mathrm{QCD}}\in\{0,\pi\}$. The leaf picture therefore reproduces, from the side of the
base, the restriction that parity gives in Sec.~\ref{sec:parity} [Eq.~\eqref{eq:thetazeropi}]: it
allows both $CP$-conserving values and does not choose between them. Two other readings would lead
elsewhere, and we name them so that the choice made here is visible. Were the two stabilizers
\eqref{eq:twostab} independent gauge groups, each with its own connection and its own angle, later
broken to their diagonal, the left--right exchange would relate the angles as
$\theta_{c'}=-\theta_{c}$ and the diagonal would see only their sum, which vanishes --- the
mechanism of the mirror constructions discussed in Sec.~\ref{sec:strongcp}; that is not the
present case, because the orthogonal combination never exists as an independent field. And were
the two leaves to carry unrelated colour bundles, the exchange would relate $\theta_{\mathrm{I}}$
to $\theta_{\mathrm{II}}$ and constrain neither; that is excluded by the single-bundle premise of
this section. Whether a fully six-dimensional formulation enforces the single-angle reading, or
instead hides a consistency condition, is not examined here \textbf{[O]}.

\smallskip\noindent
\textbf{Status of the reduction.} The single colour connection \eqref{eq:onecolourYM}, with the
orthogonal combination absent, is \textbf{[D]} relative to two inputs the paper already carries:
the gravi-weak hypothesis that places the left--right
breaking in the spacetime ($SO(3,3)$) frame (\textbf{[P]}, Sec.~\ref{sec:graviweak}), and the
split-bioctonionic dimensional anatomy --- the shared, breaking-blind internal block --- imported
from Ref.~\cite{SinghScaffolding} and drawn in Fig.~\ref{fig:decomposition}, whose caption records
it as a structural input (\textbf{[P]}). Given those,
the orthogonality of colour to $SO(3,3)$ is a representation-theoretic fact (\textbf{[D]}), and an
operation confined to the spacetime sector cannot split, double, or re-solder a bundle that lives
in the orthogonal internal sector --- it can only re-describe that one bundle in the two
octonionic frames. The single colour therefore follows from the gravi-weak hypothesis together
with the algebraic location of colour; the only way to reinstate a second, independent colour
would be to \emph{add} a connection the algebra does not contain. The surviving octet is
vector-like given the parallel-construction premise~(C) (\textbf{[C]} leaning \textbf{[D]}). The
net effect is to strengthen the colour identification from ``argued at the level of frame-conjugacy''
to ``exhibited dynamically as a single Yang--Mills term,'' resting on the very gravi-weak
hypothesis the $\theta_{\mathrm{QCD}}=0$ application already invokes --- not on any assumption
proper to the colour sector.

\paragraph{What survives as distinct: colour label versus dark charge.}
Granting (C), the non-abelian content of the right \emph{colour} $SU(3)_{c'}$ is absorbed into
the one physical $SU(3)_c$. Its \emph{trace} is the abelian number operator
\begin{equation}
  N_R \;=\; \sum_{i} a_i^{\prime\dagger} a_i' \;\in\;\{0,1,2,3\},
  \label{eq:NR}
\end{equation}
counting creation operators in the right ideal. Because it counts the very modes on which
$SU(3)_{c'}$ acts, $N_R$ is \emph{colour-welded}: a colour singlet sits at the top or bottom of
the Fock tower, $N_R\in\{0,3\}$, and a triplet at $N_R\in\{1,2\}$. Hence $N_R$ fixes the colour
\emph{representation} and nothing more --- and, colour being vector-like, it merely reproduces on
the right the assignment already carried by $N$ on the left. It is \emph{not} an independent
physical charge, and in particular \emph{not} the dark-electromagnetic $U(1)_{\mathrm{dem}}$ that
carries $\sqrt m$.

\smallskip\noindent
The $\sqrt m$ grading lives elsewhere: structurally it mirrors the surviving charge
$U(1)_{\mathrm{dem}}$ of the right \emph{electroweak} factor $SU(3)_R$ --- the
$U(1)_{\mathrm{dem}}$ of Sec.~\ref{sec:graviweak},
reached by $SU(3)_R\to SU(2)_R\times U(1)_{Y_{\mathrm{dem}}}\to U(1)_{\mathrm{dem}}$ --- a factor
distinct from, and commuting with, the colour $SU(3)_{c'}$. Being an electroweak Cartan that
charge commutes with colour (\emph{colour-blind}); being built from electroweak rather than
coloured occupation it is \emph{not} colour-welded --- exactly the combination that lets a
colour-singlet electron carry a fractional $\sqrt m$. Whether $\sqrt m$ \emph{is} that gauged
charge on the visible fermions is answered, in the negative, by the anomaly no-go of
Appendix~\ref{app:rhs} (Prop.~\ref{prop:nogo}): the gauged $U(1)_{\mathrm{dem}}$ carries the
mirror of electric charge, on the dark sector, and the visible $\sqrt m$ is the colour-blind
\emph{spectral} grading of the Jordan operator, whose lattice is the Dynkin-swap image of that
charge. Its values are the exceptional-Jordan trace split $1:2:3$
of Sec.~\ref{sec:jordan}, taken as input \textbf{[P]}, related to $|Q|$ by the flavour Dynkin
swap $\Phi$ of Sec.~\ref{sec:dynkin}; the number operator $N_R$ never sets the value of
$\sqrt m$. Appendix~\ref{app:rhs} develops this in full and shows the alternative
$\sqrt m\propto N_R/3$ to be quantitatively excluded --- on the colour-singlet electron it would
force $\sqrt m_e=1$, not $\tfrac13$ (Prop.~\ref{prop:NR}).

\smallskip\noindent
The right-sector structure thus contributes in two cleanly separated ways --- the non-abelian
colour $SU(3)_{c'}$ \emph{is} the one physical colour (by (C)), its trace $N_R$ labels the colour
representation, and the colour-blind $\sqrt m$ grading mirrors the independent electroweak
$U(1)_{\mathrm{dem}}$ (gauged on the dark sector; Appendix~\ref{app:rhs}) --- and
``$SU(3)_{c'}$'' as a second \emph{gauged colour} simply
does not exist.

\paragraph{Why $\sqrt m$ is fractional on a colour singlet.}
It is worth isolating why the charged lepton carries a fractional $\sqrt m$ although it is a
colour singlet, since the naive analogy with electric charge would forbid it. Fractional $Q=N/3$
on a singlet is impossible: $Q$ is welded to colour, so an integer-graded singlet ($N\in\{0,3\}$)
carries only integer charge. One might expect $\sqrt m$ to inherit the same constraint through the
surviving trace $N_R$, i.e.\ $\sqrt m\propto N_R/3$; it does not, and cannot. $N_R$ is
colour-welded, so on the colour-singlet electron ($N_R=3$) it would give $\sqrt m_e=1$, not the
physical $\tfrac13$ (Appendix~\ref{app:rhs}, Prop.~\ref{prop:NR}). The carrier of $\sqrt m$ is
neither a colour representation nor the colour trace: it is a colour-blind, \emph{unwelded}
grading of $U(1)_{\mathrm{dem}}$ type --- Appendix~\ref{app:rhs} fixes its precise status,
spectral on the visible side and gauged on the dark side --- and a blind, unwelded charge is
unconstrained by the colour representation (Appendix~\ref{app:rhs}, Prop.~\ref{prop:blind}). Its
values are the exceptional-Jordan trace split $1:2:3$ of Sec.~\ref{sec:jordan}, and the
down/charged-lepton interchange relating them to $|Q|$ is the colour-blind flavour Dynkin swap of
Sec.~\ref{sec:dynkin}. The first-generation ratio $1:2:3$, the intra-family spread
\eqref{eq:delta}, and the $e\leftrightarrow d$ interchange are therefore all
\emph{colour-independent}, and the fractional $\sqrt m$ of the singlet electron carries no colour
with it.

\paragraph{No second colour to break.}
There is, then, no second gauged colour requiring suppression: what looks like a doubling is the
frame-conjugate image of the one colour. The genuinely distinct right-sector datum is not a second
colour but the colour-blind $\sqrt m$ grading, whose gauge-theoretic home --- the
dark-electromagnetic $U(1)_{\mathrm{dem}}$, with charged matter in the dark sector --- is fixed in
Appendix~\ref{app:rhs}; the
colour trace $N_R$ serves only to label the colour representation. In particular --- and this is
what Sec.~\ref{sec:electron} takes up --- the electron carries the right sector only through the
colour-blind $\sqrt m$, never as a colour triplet, because the non-abelian colour it would need is
exactly what is identified into the single vector-like $SU(3)_c$.

\paragraph{Relation to Sec.~\ref{sec:parity}, and what is assumed.}
The result here and the parity principle of Sec.~\ref{sec:parity} are two faces of one fact:
the single physical frame that relates the two octonionic sectors. That frame relation makes
the chirality-exchanging operation a genuine spacetime parity (Sec.~\ref{sec:parity},
Clauses 1--2, \textbf{[D]}), and --- through the very same fixing of $i$ --- makes the diagonal
colour vector-like. These are not two independent witnesses to a common principle but one
structural property read in two sectors; in particular the vector-like character is not an extra
assumption beyond Clause~2. The honest dependency is therefore: the colour-sector results rest
on the gravi-weak frame identification (\textbf{[P]}, Sec.~\ref{sec:graviweak}) and on the
parallel-construction premise (C) (\textbf{[C]}, largely derived from Clause~2). A reader who
grants the gravi-weak frame and the parallel construction obtains one vector-like colour and,
with it, the strong-$CP$ protection of Sec.~\ref{sec:strongcp}; a reader who grants neither still
has, from Sec.~\ref{sec:parity} Clause~1 alone, that the left--right operation is \emph{some}
spacetime operation.

\section{The electron carries the right-sector group only as a grading}
\label{sec:electron}

A specific worry is what makes a suppressed --- global, explicitly broken --- $SU(3)_{c'}$
seem necessary in the first place: the charged lepton appears to sit in a slot that, read
through the second colour, could carry a triplet index --- a coloured electron.
Section~\ref{sec:onecolour} removes the worry at its root, and we make that explicit here.

\paragraph{The grading versus the representation.}
Two colour-blind $U(1)$'s must be kept apart (Sec.~\ref{sec:onecolour}, Appendix~\ref{app:rhs}).
The \emph{trace} of the right colour $SU(3)_{c'}$ is the number operator $N_R$ \eqref{eq:NR}; it
is colour-\emph{welded} (it counts the coloured modes), so it fixes only the colour
\emph{representation} --- singlet at $N_R\in\{0,3\}$, triplet at $N_R\in\{1,2\}$ --- and, colour
being vector-like, merely echoes the left assignment. The $\sqrt m$ grading is a \emph{different}
abelian datum: colour-blind and \emph{unwelded}, patterned on the electroweak
$U(1)_{\mathrm{dem}}$ (its precise status --- spectral on the visible side, gauged on the dark ---
is fixed in Appendix~\ref{app:rhs}). Because it is unwelded, a state's $\sqrt m$ neither
determines nor is determined
by its colour representation --- the slack the electron will use.

\paragraph{The electron.}
The charged lepton is a colour \emph{singlet} of the one physical $SU(3)_c$ --- in the ideal of
Sec.~\ref{sec:cl6} it is the $N=3$ singlet $a_1^\dagger a_2^\dagger a_3^\dagger\,\omega_+$,
annihilated by no colour-charged lowering operator; on the right tower it sits at $N_R=3$,
manifestly a singlet by $\bar{\mathbf 3}\wedge\bar{\mathbf 3}\wedge\bar{\mathbf 3}=\mathbf 1$. Its
mass label is not this colour count but the colour-blind grading $\sqrt m=\tfrac13$, an unwelded
value fixed by the Jordan trace split
$\sqrt m_e:\sqrt m_u:\sqrt m_d=1:2:3$ (Sec.~\ref{sec:jordan}~\cite{SinghMassRatios}); the welded
$N_R/3$ would instead give the singlet electron $\sqrt m_e=1$ (Appendix~\ref{app:rhs},
Prop.~\ref{prop:NR}), which is precisely why $N_R$ cannot be its carrier. The electron's
relationship to ``$SU(3)_{c'}$'' is therefore \emph{only} through the colour-blind grade, never
through a colour index. The non-abelian generators that would promote the grade to a colour triplet --- that
would colour the electron --- are precisely the ones identified into the single vector-like
$SU(3)_c$ by Sec.~\ref{sec:onecolour}, under which the electron is a singlet. There is no
coloured electron to forbid, and nothing to keep ``explicitly broken'': the apparent triplet
was the traceless content of a structure group whose non-abelian part is the one physical
colour, not a second one.

\begin{center}
\fbox{\begin{minipage}{0.93\linewidth}
\textbf{Why the down quark is a colour triplet but the electron is not.}
This is the subtlety on which ``no coloured lepton'' turns, and it is worth isolating: ``grade
one'' forces a colour triplet for electric charge but \emph{not} for $\sqrt m$.

\smallskip\noindent
\emph{Left sector --- charge is colour-welded.} Electric charge and colour are the trace and
traceless parts of one $\mathfrak{u}(3)$, and \emph{both are physically realized}: the three
ladder operators $a_i^\dagger$ counted by $N=\sum_i a_i^\dagger a_i$ \emph{are} the colour index.
Grade and colour representation are then a single datum --- $N=1$ means one of three colour modes
is occupied, hence a triplet, and fractional $Q=N/3$ on a non-singlet colour is \emph{forced}.
The down quark ($N=1$) is a $\mathbf 3$ because the grade-one states are literally the three
colours of gauged QCD; there is no slack. In the language of Appendix~\ref{app:rhs}, $Q$ is
colour-\emph{welded}.

\smallskip\noindent
\emph{Right colour, after $SU(3)_{c'}=SU(3)_c$.} The identification says the traceless
$\mathfrak{su}(3)$ of the right sector is \emph{not} a second realized colour but the same
internal $SU(3)_c$; its trace $N_R$ is likewise colour-welded, and so survives only as the label
of the colour representation ($N_R\in\{0,3\}$ singlet, $\{1,2\}$ triplet). There is then \emph{no
second colour for a grade to be a triplet of}: the non-abelian directions that would colour the
electron are exactly those identified into the one visible colour, under which the electron is a
singlet.

\smallskip\noindent
\emph{The resolution --- $\sqrt m$ is colour-blind but unwelded.} The $\sqrt m$ grading is
\emph{not} carried by $N_R$. Were it, the welded count would force a colour singlet to
$N_R\in\{0,3\}$, i.e.\ $\sqrt m\in\{0,1\}$, putting the electron at $\sqrt m_e=1$ --- the wrong
value, and the route by which a coloured electron creeps back in (Appendix~\ref{app:rhs},
Prop.~\ref{prop:NR}). Instead $\sqrt m$ is an unwelded, colour-blind grading patterned on the
electroweak Cartan
$U(1)_{\mathrm{dem}}$ --- commuting with colour (\emph{blind}), built from
electroweak rather than coloured occupation (\emph{unwelded}). Its value is a Jordan eigenvalue,
the family centre set by the trace split
$\mathrm{Tr}\,X_\ell:\mathrm{Tr}\,X_u:\mathrm{Tr}\,X_d=1:2:3$ (Sec.~\ref{sec:jordan}), and a
blind, unwelded charge is unconstrained by the colour representation (Appendix~\ref{app:rhs},
Prop.~\ref{prop:blind}) --- just as, in the Standard Model, colour singlets carry a range of
hypercharges. Hence fractional $Q$ \emph{requires} a colour triplet (charge \emph{is} the
coloured grade), whereas fractional $\sqrt m$ sits happily on the colour singlet (an abelian
Jordan charge orthogonal to colour) --- and the electron stays colourless.
\end{minipage}}
\end{center}

\paragraph{Consistency with the computation.}
This is borne out by the mass-ratio derivation itself, where the \emph{number operators} $N$ and
$N_R$ enter only to fix the colour representations and to place states on the ladder; the
non-abelian $SU(3)_{c'}$ generators $a_i^{\prime\dagger}a_j'$ ($i\neq j$) appear nowhere in the
ladder or the spectrum~\cite[App.~C.2]{SinghMassRatios}, and the $\sqrt m$ \emph{values} are the
Jordan trace split, not eigenvalues of $N_R$. The right-sector colour thus contributes only as a
representation label, while $\sqrt m$ is the colour-blind spectral grading --- as claimed.
\textbf{[D]} given (C) and the $\mathrm{Cl}(6)$ ideal structure.

\section{The Dynkin swap is a distinct operation}
\label{sec:dynkin}

The framework contains a second operation that also relates ``left'' to ``right'' --- the
Dynkin $\mathbb{Z}_2$ swap used in the mass-ratio program to relate the down and charged-lepton
sectors~\cite{SinghMassRatios}. It is essential to the consistency of this paper that this
swap is \emph{not} the spacetime parity of Sec.~\ref{sec:parity}. Were they the same operation,
the dissolution of the second colour (Sec.~\ref{sec:onecolour}) would collide with the
mass-ratio construction. They are distinct, and the distinction is forced, not merely asserted.

\paragraph{One automorphism, several restrictions.}
The Dynkin swap $\Phi$ is the order-two outer automorphism of $E_6$, induced by the
$\mathbb{Z}_2$ symmetry of the $E_6$ Dynkin diagram; the mass-ratio program takes the
right-sector \emph{flavour and abelian} embedding data to be the $\Phi$-image of the
left-sector data~\cite{SinghMassRatios} (an imported premise, \textbf{[P]}; the scope
of this premise is delimited precisely below). Because $\Phi$ acts on the whole
$\mathbf{27}$, it descends to several distinct restrictions, and keeping them separate
removes the apparent tensions:
\begin{itemize}
\item to the \emph{family} algebra $\mathfrak{su}(3)_F$ it is the $A_2$ diagram involution ---
on the $\mathrm{Sym}^3(\mathbf 3)$ weight triangle the reflection
$a^pb^qc^r\mapsto a^pb^rc^q$ ($b\leftrightarrow c$, the $a$-corner fixed), which carries the
down-quark ladder onto the charged-lepton ladder;
\item to the two abelian gradings it interchanges the electron and down-quark \emph{values} ---
the reciprocal $\tfrac13\leftrightarrow1$ ``flip'' between electric charge and $\sqrt m$ (below),
a relation between colour-blind functionals, not a permutation of states;
\item to colour $SU(3)_c$ it conjugates $\mathbf 3\to\bar{\mathbf 3}$, as part of
$\mathbf{27}\to\overline{\mathbf{27}}$.
\end{itemize}
The first two restrictions do all the mass-ratio work and are \emph{colour-blind}; the third
plays no role in the mass ratios and is only what distinguishes $\Phi$ from the spacetime parity.

\paragraph{The scope of the $\Phi$-image premise, made precise.}
One consistency point must be stated explicitly, because the paper's colour conclusions depend
on it. The imported premise is that the right sector's \emph{flavour ladder and abelian grading
values} are the $\Phi$-image of the left's --- the first two restrictions above. It is
emphatically \emph{not} that the right sector's colour embedding is the $\Phi$-image of the
left's. Were it so, the third restriction would apply to the colour factor: the right-handed
quarks would sit in the conjugate $\bar{\mathbf 3}$ of the identified colour, the diagonal
$SU(3)_c$ would be \emph{chiral}, the quark mass term $\bar\psi_L\psi_R$ would not even be
gauge invariant ($\bar{\mathbf 3}\otimes\bar{\mathbf 3}\not\supset\mathbf 1$), and both
premise~(C) and the vector-like character of QCD would fail. The construction therefore uses
\emph{two different maps for two different data}: the colour bundles of the two sectors are
identified by the \emph{frame} map $e_7\leftrightarrow e_8$ --- $i$-fixing, hence inner on
colour, hence non-conjugating (Sec.~\ref{sec:onecolour}) --- while $\Phi$ transports only the
flavour ladder and the abelian values (the abelian values in question being the \emph{spectral}
gradings \eqref{eq:flip}; the gauged dark-charge lattice is not $\Phi$-transported, and by the
anomaly no-go of Appendix~\ref{app:rhs} it cannot be --- Cor.~\ref{cor:mirror}). There is no requirement that a single $E_6$
automorphism perform both transports; the right-sector embedding is a construction choice, and
the choice made here is the only one compatible with vector-like colour. This scoping is not a
new assumption layered onto the framework but the precise content of the statement, made
repeatedly above, that the colour restriction of $\Phi$ ``plays no role'': the mass-ratio
program never invokes it, and the colour identification never invokes $\Phi$. \textbf{[D]}
given (C) --- the scoping is forced, since the alternative is algebraically inconsistent.

\paragraph{The abelian flip, explicitly.}
The two colour-blind gradings the swap relates are electric charge $|Q|=N/3$ and the $\sqrt m$
grading. Electric charge is welded to colour ($Q=N/3$); $\sqrt m$ is a colour-blind but
\emph{unwelded} grading of $U(1)_{\mathrm{dem}}$ type (Sec.~\ref{sec:onecolour}, Appendix~\ref{app:rhs}),
\emph{not} the number-operator grading $N_R/3$ --- which, being welded, would put the
colour-singlet electron at $\sqrt m_e=1$ (Prop.~\ref{prop:NR}). Read as functionals constant on
colour multiplets, they assign
\begin{equation}
  |Q|:\;(\nu,e,u,d)\sim(0,\,1,\,\tfrac23,\,\tfrac13),
  \qquad
  \sqrt m:\;(\nu,e,u,d)\sim(0,\,\tfrac13,\,\tfrac23,\,1),
  \label{eq:flip}
\end{equation}
so the electron and the down quark exchange their $1\leftrightarrow\tfrac13$ values between the
two gradings while the neutrino and up quark stay fixed at $0$ and $\tfrac23$. This is a relation
between two \emph{colour-blind functionals}, not a map of one particle's state onto another:
$\Phi$ does not send the (coloured) down state to the (colourless) electron state. Indeed it
cannot --- the exchanged values sit on a colour singlet (electron, $\dim_c=1$) and a colour
triplet (down, $\dim_c=3$), and no permutation of states maps a one-dimensional colour
representation onto a three-dimensional one; the flip therefore acts orthogonally to colour and
leaves the electron's colour representation untouched (Appendix~\ref{app:rhs},
Prop.~\ref{prop:flip}). The colourless character of the charged lepton (Sec.~\ref{sec:electron})
and the single vector-like colour (Sec.~\ref{sec:onecolour}) are therefore untouched by the flip.
The explicit realization of $\Phi$ bears this out: it is the $A_2$ Dynkin involution on the
\emph{flavour} algebra $\mathfrak{su}(3)_F$, exchanging the down-quark and charged-lepton ladders
on the $\mathrm{Sym}^3(\mathbf 3)$ weight triangle and thereby inducing the
$\tfrac13\leftrightarrow1$ interchange between the charge and $\sqrt m$
gradings~\cite{SinghMassRatios}. Acting on the flavour and abelian gradings alone --- all of which
commute with colour --- $\Phi$ is manifestly colour-orthogonal, the explicit counterpart of the
dimension-counting Prop.~\ref{prop:flip}. The gauge identity of the $\sqrt m$ grading is settled
--- negatively --- by the anomaly no-go of Appendix~\ref{app:rhs}: on visible fermions the
flipped pattern is not a gaugeable charge at all (Prop.~\ref{prop:nogo}), so $\Phi$ transports the
\emph{spectral} grading, never the gauge lattice; the gauged $Q_{\mathrm{dem}}$ stays
mirror-canonical (Cor.~\ref{cor:mirror}), and the construction-fixed normalization of the
spectral values remains \textbf{[O]} (residual~($\beta$)). The exclusion of a coloured electron
waits on none of this, being settled by Prop.~\ref{prop:flip}.

\paragraph{The two operations, and why their distinction is forced.}
The frame-sector \emph{parity} $\mathsf P$ (Sec.~\ref{sec:parity}) exchanges the frame groups
$SU(2)_L\leftrightarrow SU(2)_R$, acts on the Dirac field as $\psi(x)\to\gamma^0\psi(\mathcal P x)$,
and carries \emph{no} conjugation: the field scalar $i$ is fixed and the frame exchange
$e_7\leftrightarrow e_8$ is a rotation, not $\mathbf 3\to\bar{\mathbf 3}$
(Sec.~\ref{sec:parity}, Clause~2). On colour, therefore, $\mathsf P$ and $\Phi$ act
\emph{oppositely}: $\mathsf P$ fixes $\mathbf 3$, whereas $\Phi$ conjugates it to
$\bar{\mathbf 3}$. A single operation cannot do both, so $\mathsf P\neq\Phi$ as a matter of
structure, independent of any further input. This is reinforced by an independent fact of the
construction: the left--right frame exchange is available \emph{without} the Dynkin swap --- one
may act with $\mathsf P$ ($\sigma\to-\sigma$, $e_7\leftrightarrow e_8$) while leaving the flavour
ladder, and hence $\Phi$, untouched. The parity that protects $\theta_{\mathrm{QCD}}$ and the
swap that relates the mass ladders are therefore two different operations living in two different
sectors --- the spacetime/frame sector and the global flavour sector --- that have been referred
to by the same informal label ``left--right.''

\paragraph{Why the coexistence is consistent.}
Because $\mathsf P$ and $\Phi$ are distinct, the vector-like colour delivered by the frame
identification (Sec.~\ref{sec:onecolour}) coexists without contradiction with the family/abelian
action of $\Phi$ (the $\tfrac13\leftrightarrow1$ charge/mass interchange that underlies
$\sqrt m_\tau/\sqrt m_\mu=\sqrt{m_s/m_d}$ and the rest of the
ratios). The colour identified by $\mathsf P$ is vector-like (granting (C)); the colour-blind
family reflection performed by $\Phi$ is what relates down to lepton. Nothing in the dissolution
of $SU(3)_{c'}$ touches the family or abelian restrictions of $\Phi$, so the mass-ratio program
is unaffected by the colour conclusions of this paper. \textbf{[D]} for the distinctness
($\mathsf P$, $\Phi$ have opposite colour-conjugation character and $\mathsf P$ is available
without $\Phi$); the role of $\Phi$ in the mass ratios is imported from
Ref.~\cite{SinghMassRatios} (\textbf{[P]}).

\section{Application: strong-\texorpdfstring{$CP$}{CP} conservation}
\label{sec:strongcp}

We now apply the structural results of the preceding sections --- the spacetime parity
(Sec.~\ref{sec:parity}) and the special-unitary flavour rotors established below --- to strong
$CP$, and we treat this as an \emph{application} rather than the paper's central result: as the
scope statement at the end of this section and the loop analysis of Sec.~\ref{sec:loops} make
clear, the strong-$CP$ conclusion is conditional in ways the structural results are not. The
physical strong-$CP$ parameter is
\begin{equation}
  \bar\theta \;=\; \theta_{\mathrm{QCD}} + \arg\det M,
  \label{eq:thetabar}
\end{equation}
where $M=M_u M_d$ collects the quark mass matrices. We show that at tree level the second term
vanishes and the first is restricted to $\{0,\pi\}$, with $0$ on the matched branch --- but with
\emph{different epistemic status}, which we state at the outset and carry
through: $\arg\det M=0$ is the more secure, its determinant phase carried by flavour rotors of real
determinant rather than by Hermiticity of the mass matrix (the one residue being the bridge phase
named below); the statement about $\theta_{\mathrm{QCD}}$ is conditional. It rests on the
spontaneous parity of Sec.~\ref{sec:parity}, and therefore chiefly on
the gravi-weak identification (\textbf{[P]}), the vector-like character of colour being itself
largely derived (Sec.~\ref{sec:onecolour}); and parity yields only the two $CP$-conserving values,
the choice of $\theta_{\mathrm{QCD}}=0$ being a selection that parity does not make.

\paragraph{The QCD angle: conditional.}
By Sec.~\ref{sec:onecolour} the physical colour $SU(3)_c$ is vector-like --- the diagonal
identified across the two sectors is non-conjugating, since parity fixes $i$ (assumption (C),
\textbf{[C]} leaning \textbf{[D]}): left- and right-handed quarks transform identically. QCD
therefore has no chiral structure for the parity $\mathsf P$ of Sec.~\ref{sec:parity} to act on
asymmetrically, and $\mathsf P$ is an exact symmetry of the unbroken gauge sector. Since the
vacuum term $G\widetilde G$ is parity-odd, an exact $\mathsf P$ forbids a generic coefficient: the
bare angle is restricted to the parity-invariant values $\theta_{\mathrm{QCD}}\in\{0,\pi\}$
[Eq.~\eqref{eq:thetazeropi}]. This survives the spontaneous breaking $\sigma\neq0$, because
spontaneous breaking adds no term to the Lagrangian and does not regenerate
$\theta_{\mathrm{QCD}}$ at tree level. Which of the two values is realized cannot be decided in
the gauge sector alone, because only the sum \eqref{eq:thetabar} is physical: with special-unitary
rotors and real signed spectra $\Sigma'_q$ (item (i) below), the tree-level condition is the single
\emph{matching condition}
\begin{equation}
  \bar\theta_{\mathrm{tree}}
  =\theta_{\mathrm{QCD}}+\arg\det\!\big(\Sigma'_u\Sigma'_d\big)=0\pmod{2\pi},
  \qquad \theta_{\mathrm{QCD}}\in\{0,\pi\},
  \label{eq:matched}
\end{equation}
the analogue of the ``matched branches'' of the conventional left--right
reconstruction~\cite{Tello2026}. For the spectra used here the negative-root parities of the up
and down sectors agree (item (i)), so $\arg\det(\Sigma'_u\Sigma'_d)=0$ and \eqref{eq:matched}
selects $\theta_{\mathrm{QCD}}=0$; the other parity-invariant assignment would give
$\bar\theta=\pi$, which conserves $CP$ but is excluded, because the meson spectrum it implies is
not the observed one~\cite{Vecchi2025}. We state the status plainly:
$\theta_{\mathrm{QCD}}=0$ is \emph{selected} by \eqref{eq:matched} together with phenomenology, not
derived. That is the general situation for solutions resting on $P$ or $CP$ alone, which cannot by
themselves decide between the two values~\cite{Vecchi2025}. The constructions known to decide
structurally are of two kinds. In the first, QCD emerges as the diagonal of a doubled gauge group
whose two factors are exchanged by a mirror symmetry, and $\bar\theta=0$ results whatever the
angles of the parent groups~\cite{Vecchi2025,BCS,BonnefoyMirror}: the symmetry relates the parent
angles so that their sum, which is all the diagonal sees, vanishes. In the second, colour sits in a
suitable quotient of a product gauge group~\cite{Vecchi2025}. Trinification by itself is not
among them~\cite{Vecchi2025}. The present construction, read as in Sec.~\ref{sec:onecolour} with a
single colour bundle carrying a single angle, is of neither kind: the two-leaf relation noted
there reproduces $\{0,\pi\}$ and does not select within it. Whether the framework excludes
$\theta_{\mathrm{QCD}}=\pi$ structurally is therefore open \textbf{[O]}. The conclusion is the spontaneous-parity mechanism of
Refs.~\cite{BegTsao,MohapatraSenjanovic1978}, in the form developed by Babu and
Mohapatra~\cite{MohapatraBabu}; what is conditional is its premise --- the parity
interpretation of the left--right exchange (\textbf{[P]}) --- while the vector-like character of
colour is largely derived (Clause~2) rather than a separate load-bearing assumption.

\paragraph{The mass determinant: where the phase actually lives.}
The strong-$CP$-relevant quantity is the determinant phase of the \emph{physical}
three-generation quark mass matrices, not of the Jordan element directly. For a bi-unitary
(singular-value) diagonalization $M_q=U_L^{(q)}\,\Sigma_q\,U_R^{(q)\dagger}$ with $\Sigma_q\ge0$
the physical masses,
\begin{equation}
  \arg\det M_q \;=\; \arg\det U_L^{(q)} - \arg\det U_R^{(q)} \pmod{2\pi},
  \label{eq:rotorarg}
\end{equation}
so the determinant phase is carried by the \emph{diagonalizing rotors}, and the masses
$\Sigma_q$ --- real and positive --- drop out entirely. The vanishing of $\arg\det(M_uM_d)$
therefore rests on two structurally distinct facts, only the second of which involves a phase.

\emph{(i) The magnitudes are real and positive.} The physical quark masses are the singular
values of $M_q$ --- equivalently, squares of the spectral magnitudes built from the Jordan
eigenvalues --- and are therefore real and positive irrespective of the signs of the underlying
Jordan eigenvalues or of the relative normalization $r$ of the Jordan element. On the
coassociative slice ($\mathcal{T}=0$) the quark Jordan spectrum is in addition real and symmetric
(Sec.~\ref{sec:jordan}). Sign content, however, is not phase-free bookkeeping, and we track it
explicitly. (Whether the signs of the Jordan roots reach $\det M_q$ at all depends on how the roots
enter $M_q$. If the spectrum entering $M_q$ consists of the \emph{squares} of the Jordan
eigenvalues, every entry is positive and the bookkeeping that follows is vacuous; it has content
only if signed roots enter linearly. The explicit map from the Jordan element to $M_q$ is that of
Ref.~\cite{SinghMassRatios} and is not reproduced here, so the bookkeeping is stated in the form
that covers both cases.) With special-unitary rotors, writing
$M_q=U_L^{(q)}\,\Sigma'_q\,U_R^{(q)\dagger}$ with $\Sigma'_q$ the real \emph{signed} spectrum, one
has $\arg\det M_q=\arg\det\Sigma'_q\in\{0,\pi\}$, equal to $\pi$ for an odd number of negative
roots in the sector; absorbing the signs into $\Sigma_q\ge0$ merely moves the same $\pm1$ into one
rotor determinant, so nothing is gained by the rewriting. Hence
$\arg\det(M_uM_d)=0\pmod{2\pi}$ requires the negative-root \emph{parities} of the up and down
sectors to match. (This is the $\theta_{\mathrm{QCD}}=0$ branch of the combined condition
\eqref{eq:matched}: mismatched parities are not inconsistent in themselves, but would require the
other parity-invariant value $\theta_{\mathrm{QCD}}=\pi$, so the sign parities and the bare angle
must be specified together, never separately.) On the rigid normalization ($r=1$) all quark roots are positive and both
parities are even --- trivially matched, \textbf{[D]}. On the fitted spectral deformation
($r<0$) the texture of Ref.~\cite{SinghMassRatios} assigns an odd number of negative roots to
\emph{each} sector --- matched parities, hence $\det M_u\det M_d>0$ --- a property of that
texture, tagged \textbf{[C]} pending the parameter-free construction. Given matching, the
residual phase is fixed by the rotor determinants of (ii) below.

\emph{(ii) The rotor determinants are real \textbf{[D]}.} The mass matrices are diagonalized by
the flavour rotors, which are $SU(3)_F$-valued. The only loophole to $\det U_L^{(q)}=\det
U_R^{(q)}$ would be an overall $U(1)$ phase differing between the two; this is closed by the
explicit construction. The Cabibbo-rung transport is $L_{\exp(\theta g_\chi)}$ with
$g_\chi=\cos\chi\,e_3-\sin\chi\,e_1$ a \emph{real} octonionic direction, the phase of the rung
coupling having collapsed into a real angle (Eq.~(6) of Ref.~\cite{TeliSinghCP}); a generator
carrying no factor of the external $i$ yields, on the normalized $(1,2)$ generation block, a
special-unitary rotor whose determinant phase vanishes. In the explicit construction the rung
rotor is unitary with $\arg\det=0$ for every $\chi$, the conjugation theorem $A_d=A_u^{*}$ holds
exactly across real transports, and the $(2,3)$ and $(1,3)$ blocks enter as \emph{real}
rotations carrying no independent phase~\cite{TeliSinghMixing,TeliSinghCP}, so that $\det
U_L^{(q)}=\det U_R^{(q)}=1$ throughout. By \eqref{eq:rotorarg}, together with the matched sign
parities of (i),
\begin{equation}
  \arg\det(M_u M_d) \;=\; 0 \pmod{2\pi}.
  \label{eq:argdet}
\end{equation}
The leptonic reality theorem $J_\ell=0$ is the parallel statement that the lepton rotors are real
(orthogonal, real determinant) for every transport not mixing the identity line $\mathbb{C}\cdot
1$ with the flavour plane $\mathrm{span}(e_7,e_5,e_2)$~\cite{TeliSinghCP} --- a class that
contains the entire quark-rung family. The one residue is the named loophole of that theorem: an
intrinsically complex Higgs bridge with identity--flavour mixing could inject a phase outside
this real-transport class (\textbf{[O]}, Sec.~\ref{sec:loops}); absent it, \eqref{eq:argdet}
is \textbf{[D]}.

This is the precise statement of what the real spectrum buys. The reality of the spectrum fixes
the magnitudes (i); the determinant \emph{phase} is governed by the rotors (ii), and its
vanishing follows from their special-unitary character --- the same reality structure that
gives $J_\ell=0$ in the lepton sector --- not from a Hermiticity of the physical mass matrix
imposed by hand. These two reality arguments are not redundant. Because the protecting parity is
\emph{spontaneously} broken ($\sigma\neq0$, and $\sigma$ enters the masses), it does not by itself
render the physical mass matrix Hermitian; $\arg\det M=0$ is therefore \emph{not} handed by the
parity, and genuinely requires the flavour-rotor reality. The contrast with the conventional
parity solutions should be drawn accurately, since parity is spontaneously broken there too. In
those models parity, imposed on the Lagrangian, constrains the \emph{Yukawa} matrices (Hermitian,
in the bidoublet realizations); the \emph{mass} matrices inherit that property only for real
vacuum expectation values, and the departure from Hermiticity is controlled by a $CP$-odd vacuum
parameter --- in the minimal bidoublet model $\epsilon=\sin\alpha\,\tan2\beta$, built from the
relative phase $\alpha$ of the two bidoublet vacuum values, with
$\bar\theta\simeq C_q\,\epsilon$ and $|C_q|\simeq13$--$50$ depending on the discrete quark sign
branch, so that the neutron-EDM bound requires $|\epsilon|\lesssim\text{few}\times10^{-12}$
\cite{Tello2026}. In the universal-seesaw model of Babu and Mohapatra~\cite{MohapatraBabu} there
is no bidoublet, the doublet vacuum values can be taken real, and the determinant is real at tree
level. The difference from the present construction is therefore not broken versus unbroken
parity but the \emph{source} of the reality: real vacuum values acting on parity-constrained
Yukawas there, real rotor determinants here. The comparison also sharpens what the rotor-reality
claim amounts to. The analogue of $\epsilon$ in this framework is any $CP$-odd relative phase in
the two-Higgs/bridge sector; the statement $\det U_L^{(q)}=\det U_R^{(q)}=1$ is the statement that
this phase vanishes at tree level, and the complex-bridge loophole named above is precisely the
statement that its vanishing is not yet proven beyond the real-transport class \textbf{[O]}. The
conventional coefficient sets the scale of what must be shown: were such a phase to enter the
quark mass matrices with the bidoublet-type $m_t/m_b$ enhancement, it would have to be below
a few times $10^{-12}$. That reality is moreover intrinsic to the
$SU(3)_F$ flavour structure rather than downstream of the gravi-weak identification: the
Cabibbo-rung rotor is special-unitary because its generator is a real octonionic direction
(Eq.~(6) of Ref.~\cite{TeliSinghCP}), a property of the flavour sector alone, so $\arg\det M=0$
does not rest on the gravi-weak input. Where the spontaneous-parity literature
secures $\arg\det M=0$ through parity-constrained Yukawas together with real vacuum
values~\cite{MohapatraBabu}, here the
reality enters through the flavour-rotor structure already used for the mixing matrices.

\paragraph{Coexistence with CKM \texorpdfstring{$CP$}{CP} violation.}
The vanishing of $\arg\det M$ does not suppress observable $CP$ violation, because the two are
carried by \emph{different invariants of the same rotors}. By \eqref{eq:rotorarg}, $\arg\det M$
sees only the rotor \emph{determinants}; the Kobayashi--Maskawa phase~\cite{KobayashiMaskawa} is
an \emph{off-diagonal} invariant of those rotors. A special-unitary rotor (determinant $1$) can
carry a nonzero off-diagonal phase while contributing nothing to $\arg\det M$: in the companion
construction the Cabibbo phase~\cite{Cabibbo} is fixed geometrically by the rung orientation
$\chi$ as $\phi_{12}=-2\chi$, with a nonzero Jarlskog invariant $J\neq0$~\cite{Jarlskog}, all
inside rotors of real determinant~\cite{TeliSinghMixing}. A \emph{real} octonionic generator does
not make the rotor a real --- hence $CP$-conserving --- matrix: exponentiated on the module it is a
genuinely complex special-unitary transformation, with unit determinant but complex off-diagonal
entries, so $\det=1$ and $J\neq0$ coexist without tension. Thus
\begin{equation}
  \boxed{\;\bar\theta = \theta_{\mathrm{QCD}} + \arg\det M = 0 \ \text{(tree level)},
  \qquad \delta_{\mathrm{CKM}}\neq 0\;}
\end{equation}
hold simultaneously: the framework conserves strong $CP$ while violating $CP$ in the
charged-current sector, with no tension. The naive expectation that a real spectrum forbids $CP$
violation fails because the strong-$CP$ phase and the CKM phase are different invariants ---
determinant versus off-diagonal --- of the diagonalizing rotors, not eigenvalues versus
eigenvectors.

\paragraph{What is claimed.}
The two conditions entering $\bar\theta$, and their statuses, are exactly as derived above; the
scope statement below delimits the claim, and the ledger of Sec.~\ref{sec:discussion} is the
full accounting --- we do not restate them here. Two remarks frame the result. First, the
novelty is not a new mechanism --- spontaneous parity is known --- but the \emph{connection}:
in this framework the parity that protects $\theta_{\mathrm{QCD}}$ and the reality structure
that makes $\arg\det M$ vanish are both consequences of the single octonionic structure
introduced for the fermion masses, rather than ingredients imposed to solve strong $CP$.
Second, what is \emph{not} claimed: this is not a complete strong-$CP$ solution --- the result
holds at tree level, and the decisive test, radiative stability of $\bar\theta$ after breaking,
is open, discharged honestly in Sec.~\ref{sec:loops}.

\paragraph{Scope and limitations.}
For the reader's convenience we state, in one place, the boundaries of the result.
\emph{(a) Tree level only.} $\bar\theta=0$ is established at tree level; radiative stability is
the decisive test for any parity/Nelson--Barr-type mechanism, and within this framework it
requires the flavour matrix elements of the Higgs-bridge field $B_H$, which are not yet defined.
This is the one genuinely open computation (\textbf{[O]}, Sec.~\ref{sec:loops}); the
Standard-Model (CKM) contribution to $\bar\theta$ is far below the bound (Sec.~\ref{sec:loops}),
so the only open radiative channel is the beyond-Standard-Model bridge sector.
\emph{(b) $\theta_{\mathrm{QCD}}=0$ is conditional \textbf{[P]}, and selected rather than
derived \textbf{[C]}.} It rests on identifying the
left--right exchange with \emph{spacetime} parity, i.e.\ on the gravi-weak reading in which
$SU(2)_R$ is gravitational --- a program-level hypothesis; and even granted that, parity gives only
$\theta_{\mathrm{QCD}}\in\{0,\pi\}$, the value $0$ being the matched branch of \eqref{eq:matched}
and the exclusion of $\pi$ phenomenological. The parity reading does \emph{not} conflict with the
Coleman--Mandula theorem, which constrains how \emph{continuous} Lie symmetries may combine with
Poincar\'e; the protecting operation here is a \emph{discrete} parity, largely outside its scope.
In any case the parity is \emph{spontaneously} broken in the emergent phase
($\langle\sigma\rangle\neq0$, Sec.~\ref{sec:parity}), so it is not an exact $S$-matrix symmetry: it
forbids the $\theta$-term at the Lagrangian level without itself being a symmetry of the
$S$-matrix. (More structurally, spacetime and internal sectors both emerge from the
$E_8\times\omega E_8$ groups with no fixed Poincar\'e arena before breaking, and the fundamental
dynamics is trace dynamics rather than an $S$-matrix
theory~\cite{SinghScaffolding,VaibhavSingh,SinghEmergence}; either way the dispatch inherits the
emergent-spacetime construction's status \textbf{[P]} rather than standing as an independent
obstruction.) A second caveat is specific to the emergent setting and we log it as its own line:
the standard step ``spontaneous breaking adds no term to the Lagrangian'' presumes a fixed arena
in which the Lagrangian exists on both sides of the transition. Here the breaking is
simultaneously the \emph{emergence} of the classical arena, so what $\theta_{\mathrm{QCD}}=0$
ultimately requires is a property of the emergence map itself --- that the passage from the
$\mathsf P$-symmetric pre-geometric phase to the leafwise effective Lagrangian generates no
$\theta$-term. No such theorem is available; we tag it \textbf{[O]} in the ledger, as a
conditionality larger than, and distinct from, the gravi-weak \textbf{[P]}. A corollary worth stating is that, with
$SU(2)_R$ gravitational, there is no gauged right-handed $W$ coupling to fermions, so the
standard left--right collider and flavour constraints ($K$--$\bar K$ mixing, $b\to s\gamma$,
$W_L$--$W_R$-mediated EDMs) do not apply; equivalently, the framework predicts no $W_R$ gauge
signals of that kind.
\emph{(c) $\arg\det M=0$ is \textbf{[D]} for the rung, texture-contingent for the full matrix.}
The Cabibbo-rung rotor is special-unitary by a theorem (its generator is a real octonionic
direction); that the $(2,3)$ and $(1,3)$ blocks carry no independent phase --- hence $\det
U_L^{(q)}=\det U_R^{(q)}=1$ for the full $3\times3$ rotor --- is a property of the minimal
adjacent-edge texture of Ref.~\cite{TeliSinghMixing}, which is not yet parameter-free. A future
parameter-free $(2,3)/(1,3)$ construction introducing an independent phase would have to be
checked against $\arg\det M$ (as would any change to the negative-root parities of the fitted
spectra, per (i)).
\emph{(d) Dependence on the single-$i$ assumption.} The statement that the spacetime exchange
sends no \emph{colour} representation to its conjugate (Clause~2) uses the identification of the
field scalar fixed by the exchange with the colour complex structure --- assumption~(C) of
Sec.~\ref{sec:onecolour} (\textbf{[C]} leaning \textbf{[D]}).
\emph{(e) The single colour is exhibited dynamically.} Sec.~\ref{sec:onecolour} carries out
the BF$\to$Yang--Mills reduction: because the breaking does not act on colour, the two sectors'
colour connections collapse to a single Yang--Mills term \eqref{eq:onecolourYM}, the orthogonal
combination absent rather than merely massive (\textbf{[D]}). The surviving octet is vector-like
given assumption~(C) (\textbf{[C]} leaning \textbf{[D]}). The one non-derived input --- that the
left--right breaking, an $SO(3,3)$ spacetime operation, leaves the $E_6$-internal colour
untouched --- is the gravi-weak hypothesis itself (\textbf{[P]}), not a premise proper to the
colour sector; the $\theta_{\mathrm{QCD}}=0$ application rests on that hypothesis and~(C) alone.
\emph{(f) Self-containment.} Two load-bearing inputs --- the explicit flavour rotors and the
$\mathcal{T}=0$ coassociativity of the mass texture --- are derived in companion
papers~\cite{TeliSinghMixing,TeliSinghCP,SinghMassRatios}; the mechanisms are stated here, but
the full constructions are not reproduced.
\emph{(g) The abelian sector is constrained, not open-ended.} Appendix~\ref{app:rhs} proves that
no anomaly-free $U(1)$ on the visible fermions can carry the $\sqrt m$ values
(Prop.~\ref{prop:nogo}), fixing the gauged dark charge to the mirror-canonical pattern on the
dark sector (Cor.~\ref{cor:mirror}) and rendering the visible fermions dem-neutral; this feeds
the force inventory of Sec.~\ref{sec:discussion} but enters no step of the strong-$CP$ argument.

\section{Loop regeneration of \texorpdfstring{$\bar\theta$}{theta-bar}}
\label{sec:loops}

Sections~\ref{sec:parity} and \ref{sec:strongcp} establish $\bar\theta=0$ at tree level. The
physical strong-$CP$ parameter is, however, radiatively sensitive: after the spontaneous
parity breaking $\sigma\neq0$, loops can in principle regenerate a nonzero
$\bar\theta$, and a solution is only as good as its stability against this. This section is
where the ``tree level'' qualifier carried through Sec.~\ref{sec:strongcp} is discharged --- and
we discharge it honestly, because the model-specific computation cannot yet be done. We
separate three things: what is known about loop regeneration in the minimal parity models, how
much of it carries over to the present framework, and the residual model-specific
calculation, which is open.

\paragraph{Loop regeneration in the minimal parity models.}
It would be wrong to quote the favourable one-loop heuristic as a generic expectation. In the
minimal Babu--Mohapatra parity model~\cite{MohapatraBabu} --- the closest standard analogue ---
$\bar\theta$ vanishes only \emph{up to} one-loop order; the leading two-loop contributions were
computed by de Vries, Draper and Patel~\cite{deVriesDraperPatel}, who found, contrary to earlier
estimates, that they are \emph{not} suppressed by the weak scale and are of order
$10^{-3}$--$10^{-2}$ times unknown mixing angles and phases. Two features of that computation
matter for what follows. The contributing graphs are scalar ones: graphs with internal $W$ or
$W'$ lines ($W'$ being the right-handed gauge boson) do not contribute, and every contributing
graph contains a Higgs line together with the heavy isosinglet quarks of the universal seesaw. And
the large
coefficient arises when the soft breaking of parity enters the mass matrices of those heavy
quarks; the Higgs-mass splitting contributes only in suppressed form. Such a model therefore does
\emph{not} generically solve the strong-$CP$ problem: it satisfies $\bar\theta\lesssim10^{-10}$
only in restricted corners of parameter space, and even there $\bar\theta$ is expected to sit
just below current bounds, sourcing observable hadronic electric dipole moments. A direct
diagrammatic analysis of the minimal left--right symmetric model by Hisano
\emph{et al.}~\cite{Hisano2023} finds that the two-loop contribution vanishes there and that a
non-vanishing one first appears at three loops, with an induced neutron electric dipole moment
comparable to the present bound; we quote the two analyses as they stand and do not attempt to
reconcile their assumptions. Any
parity-based proposal must now confront these results; we do so directly, rather than lean on the
weak-scale-suppressed estimates that Ref.~\cite{deVriesDraperPatel} overturned.

\paragraph{What carries over to the present framework, and what does not.}
The contribution of Ref.~\cite{deVriesDraperPatel} is \emph{not} generated by the right-handed
gauge boson: as just noted, the $W'$ graphs drop out and the effect runs through scalar exchange
and the heavy isosinglet quarks.\footnote{The first version of this paper attributed that
contribution to diagrams running a gauged $W_R$, and inferred that it is absent when $SU(2)_R$ is
gravitational. The attribution was incorrect, and the inference is withdrawn.} In the present
framework $SU(2)_R$ is
the gravitational frame group (Sec.~\ref{sec:graviweak}) and there is no gauged $W_R$ coupling to
fermions (a point already noted in the scope statement of Sec.~\ref{sec:strongcp}); but that
absence, by itself, removes nothing from the mechanism of Ref.~\cite{deVriesDraperPatel}. What
the framework lacks is the isosinglet-quark seesaw sector in which that computation was done, so
the result does not apply as it stands --- a statement about field content, not a consequence of
the gravi-weak reading. What the framework shares is the lesson the computation teaches:
$\bar\theta$ is dimensionless, so a breaking of parity in any sector that couples to the quarks
through scalar exchange can feed it through threshold corrections that need not decouple. Here
that sector is the Higgs bridge. The absence of $W_R$ does remove the right-handed-current effects
through which the kaon sector constrains gauged bidoublet models~\cite{MaiezzaNemevsek}; below we
sharpen that part from assertion to derivation: the absence of a right-handed mixing
matrix $V_R$ follows from $A_R$ acting on the spin index rather than on isospin, so the residual
is not the frame sector but the scalar (bridge) sector, as we discuss.

\paragraph{The Standard-Model contribution is negligible.}
It is worth isolating what does \emph{not} threaten the result. In the Standard Model the CKM
phase feeds $\bar\theta$ only through flavour-diagonal mass corrections that first arise at
three-loop order, yielding $\bar\theta_{\mathrm{SM}}\sim10^{-16}$ or
smaller~\cite{EllisGaillard} --- far below the
experimental bound $\bar\theta\lesssim10^{-10}$~\cite{nEDM}. The geometric origin of the CKM
phase here ($\phi_{12}=-2\chi$, Sec.~\ref{sec:strongcp}) does not alter this: it is the same
single rephasing-invariant phase, carried by special-unitary rotors, and reaches $\arg\det M$
only through the same high-order, GIM-suppressed graphs. The potentially dangerous radiative
contributions are therefore exclusively the beyond-Standard-Model ones --- right-handed or
bridge couplings absent from the Standard Model --- which returns the question to the scalar
sector of this construction, to which we now turn.

\paragraph{An inventory of loop topologies.}
It helps to make the classification explicit, even where the amplitudes cannot yet be evaluated.
The diagrams that could feed $\bar\theta$ after $\sigma\neq0$ fall into five classes.
(i)~\emph{Standard-Model CKM loops}: present, but flavour-diagonal only at three loops and
GIM-suppressed, giving $\bar\theta_{\mathrm{SM}}\lesssim10^{-16}$ --- negligible (above).
(ii)~\emph{Gauged-$W_R$ loops}: the right-handed-current graphs of gauged left--right models
($W_L$--$W_R$ mixing, and graphs carrying a right-handed mixing matrix); \emph{absent here}, as
$SU(2)_R$ is gravitational
and no $W_R$ couples to fermions. (iii)~\emph{Scalar-bridge loops}: $B_H$ exchanged between
fermion lines --- the one potentially dangerous class, and the one to which the two-loop effect of
Ref.~\cite{deVriesDraperPatel} belongs in its own model (Higgs exchange with heavy isosinglet
quarks); it requires the $B_H$ flavour matrix
elements, \emph{open} (below). (iv)~\emph{Frame/gravitational-sector loops}: graviton and
$SU(2)_R$-connection exchange; flavour-blind (spin-connection coupling, $\delta M\propto M$) and
$CP$-even (all gravi-weak couplings real), so they do not regenerate $\bar\theta$ ---
\textbf{[C]} leaning \textbf{[D]}, the gravitational Pontryagin coefficient being the one named
residual (developed below). (v)~\emph{Mixed bridge--frame loops}: $B_H$ together with frame fields, controlled by
the same unknown matrix elements as (iii) and inheriting their open status. Classes (i), (ii) and (iv)
are settled --- negligible, absent, and ($CP$-even, flavour-blind) non-regenerating respectively
--- so the open computation reduces to the single object $B_H$ entering (iii) and (v), to which we
now turn.

\paragraph{The frame sector carries neither flavour nor a $CP$ phase.}
Classes~(ii) and~(iv) are one statement, made sharpest by the right comparison: not
weak-versus-gravity but $A_L$ versus $A_R$, the two parity partners. As groups they are identical;
the asymmetry is entirely in \emph{which two-valued index each acts on in the broken phase}. The
left connection $A_L$ acts on the internal isospin index, so its charged current links up-type to
down-type and, read through the misaligned Yukawa basis, exposes the CKM phase. The right
connection $A_R$ is realized as the leaf-I spin connection (Sec.~\ref{sec:graviweak}); its
``charged current'' links spin-up to spin-down of a single field --- a Lorentz index, not an
isospin one. Under the spacetime parity relating the sectors, the mirror of an isospin-raising
current is a \emph{spin}-raising current, not a second isospin current: on our leaf $A_R$ raises
spin, not flavour, so there is no right-handed analogue $V_R$ of the CKM matrix acting on the
\emph{visible} quarks for it to carry --- hence no $W_R$ exchange here, and none of the
right-handed-current graphs of class~(ii) on this leaf. This \emph{derives} the
absence asserted as class~(ii), resting only on the gravi-weak assignment of $A_R$ to the spin
index (\textbf{[P]}, Sec.~\ref{sec:graviweak}). The scoping matters: the parity image of CKM ---
a mixing matrix $V_R^{(\mathrm{II})}$ --- is not zero but lives on the partner leaf, where $A_R$
is instead realized as an internal isospin field. It cannot reach $\bar\theta$ directly, because
inter-leaf frame coupling is flavour-blind and carries no flavour matrix; the \emph{only} route
by which it could regenerate $\bar\theta$ is across the $(1,1)$ intersection, through the scalar
bridge $B_H$. The frame argument therefore does not make the danger vanish so much as
\emph{relocate} it onto the bridge classes~(iii) and~(v): pure symmetry already shows that
whatever residual survives must be bridge-mediated, which is exactly why those classes, and not
the frame sector, are the open channel.

\smallskip\noindent
Class~(iv) is then safe on two explicit legs. \emph{Flavour-blindness}: the frame--fermion
coupling is the spin connection, diagonal and universal in flavour, so a frame loop renormalizes
masses as $\delta M=c\,M$ and shifts $\arg\det M$ by $n\arg(1+c)$ --- zero for real $c$, and never
the flavour-phase mechanism that defeats the minimal seesaw model. \emph{$CP$-evenness}: in the
$SO(3,3)$ BF construction~\cite{WSI_BF} every gravi-weak parameter is real --- the master coupling
$g$ (with $G_N=G_W$), the cosmological constants $\lambda,\tilde\lambda$, and the standard
Plebanski reality conditions recovering real General Relativity --- while the Immirzi-type datum is
the \emph{discrete} orientation $\sigma=\pm1$, which is exactly the spontaneous-parity order
parameter of Sec.~\ref{sec:parity}, not a continuous phase, and the interface gluing is a pure
$SU(2)$ gauge rotation. Spontaneous parity thus pins the gravitational $\theta$-analogue by the
same logic that pins $\theta_{\mathrm{QCD}}$: parity is broken only by real, $CP$-even order
parameters ($\sigma$, and any $\lambda\neq\tilde\lambda$), and a $CP$-even breaking of parity
cannot source the $CP$-odd $\theta_{\mathrm{grav}}$. With $c$ real, $\arg\det M$ is unshifted.
One further piece of bookkeeping belongs here, because quarks carry both colour and spin: chiral
rotations of \emph{coloured} fermions shift $\theta_{\mathrm{QCD}}$ and $\theta_{\mathrm{grav}}$
simultaneously (through the mixed gauge and gravitational anomalies), while rotations of
colourless fermions shift $\theta_{\mathrm{grav}}$ alone. The separately invariant combinations
are $\bar\theta=\theta_{\mathrm{QCD}}+\arg\det M_{\mathrm{coloured}}$ and a gravitational
analogue $\bar\theta_{\mathrm{grav}}$ that sums the mass phases of \emph{all} fermions; the two
must be tracked per sector, not interchangeably. The distinction is not academic in this program:
the companion neutrino benchmark of Ref.~\cite{SinghMassRatios} carries a negative Majorana
eigenvalue ($\lambda_2<0$) --- a $\pi$-phase in the colourless sector, invisible to $\bar\theta$
but landing squarely in the gravitational bookkeeping that the $\theta_{\mathrm{grav}}$ residual
must track.

\smallskip\noindent
It should be said plainly that making $SU(2)_R$ gravitational and self-dual does not only
\emph{remove} the gauged-$W_R$ danger; it \emph{introduces} a potential $CP$ problem of its own ---
a gravitational $\theta$-analogue $\theta_{\mathrm{grav}}$ from the self-dual sector --- which the
argument above contains but does not close. We tag it \textbf{[C]} leaning \textbf{[D]} and name
the residual. The spin-versus-isospin
assignment is the gravi-weak premise (\textbf{[P]}); the explicit leaf-to-leaf fermion dictionary
realizing it --- how a chiral fermion's coupling to the leaf-I spin connection appears, on the
visible leaf, as its weak-isospin coupling --- is not worked out in the classical
companion~\cite{WSI_BF}, which omits the matter sector. And the $CP$-evenness is read off that
companion's real parameter content, which does not itself address $CP$: the one quantity still to
be checked is the gravitational Pontryagin coefficient --- whether the self-dual action's
$F^{(+)}\!\wedge F^{(+)}$ piece leaves a nonzero $\theta_{\mathrm{grav}}$ once the reality
conditions are imposed. That check is not performed here. What is established is the implication
(real, flavour-blind gravi-weak couplings $\Rightarrow$ no frame-sector $\bar\theta$) and the
isolation of that single residual, reducing class~(iv) from ``power-counting not given'' to one
named line. The genuinely open radiative channel remains the bridge, classes~(iii) and~(v).

\paragraph{The model-specific computation requires the Higgs bridge.}
In this framework the scalar that converts Jordan data into Yukawa couplings is the Higgs
bridge field $B_H$ of the underlying trace-dynamics construction~\cite{SinghEmergence}. The
loop that could regenerate $\bar\theta$ runs $B_H$, so the model-specific estimate requires the
flavour matrix elements of $B_H$ between the ladder states. These are not available. In the
emergence construction $B_H$ is obtained as a Hubbard--Stratonovich auxiliary field bosonizing
a bifermionic condensate, isolated by an $e_0$ projection and carrying electroweak quantum
numbers $(\mathbf 1,\mathbf 2,\pm\tfrac12)$, under stated working hypotheses; it is explicitly
\emph{not} a derived operator with computed matrix elements~\cite{SinghEmergence}. Consequently
the model-specific loop estimate cannot be carried out at present, and we do not claim it. We
tag the residual loop safety of $\bar\theta$ as \textbf{[O]}, now localized to the bridge: the
gauged-$W_R$ class is absent and the frame class is $CP$-even and flavour-blind
(above, \textbf{[C]} leaning \textbf{[D]}), but a bridge-mediated contribution of the
scalar-exchange kind found in Ref.~\cite{deVriesDraperPatel} cannot
be excluded without an explicit $B_H$, so the model-specific computation is open pending it.

\paragraph{A suggestive but inconclusive structural clue.}
The $e_0$ structure of $B_H$ bears on the question, though it does not settle it. The dangerous
loop contribution, like the leptonic-$CP$ loophole of the companion analysis, is governed by
whether the bridge mixes the identity line $\mathbb{C}\cdot 1$ with the lepton flavour plane
$\Pi_\ell=\mathrm{span}(e_7,e_5,e_2)$~\cite{TeliSinghCP}. That $B_H$ is isolated in the $e_0$
(identity) channel is \emph{suggestive} of the safe, non-mixing class. But channel isolation is
not the same as the absence of off-diagonal matrix elements: the loophole-controlling quantity
is the off-diagonal $e_0\!\leftrightarrow\!\Pi_\ell$ element, schematically
$\mathrm{Im}\langle e_2|\mathsf C|e_1\rangle=[\mathsf C(1)]_{e_5}-[\mathsf C(e_7)]_0$ for the
bridge's left-multiplication action $\mathsf C$~\cite{TeliSinghCP}, which the channel isolation
does not fix. The $e_0$ structure thus points toward loop safety without establishing it.

\paragraph{One computation, three closures.}
It is worth recording why the open object here is the same one that recurs elsewhere in the
program. The flavour matrix elements of $B_H$ control, simultaneously: (i) the loop regeneration
of $\bar\theta$ (this section); (ii) leptonic $CP$ conservation --- whether the lepton
amplitudes are real, i.e.\ $J_\ell=0$, which holds unless $B_H$ mixes $\mathbb{C}\cdot1$ with
$\Pi_\ell$~\cite{TeliSinghCP}; and (iii) the normalization of the quark $CP$ phase --- whether
the geometric $\phi_{12}=-2\chi$ reproduces the measured $\delta_{\mathrm{CKM}}$ rather than an
$O(1)$ phase of the right size~\cite{TeliSinghMixing}. A single computation --- the explicit
$B_H$ and its matrix elements --- would close all three at once. This is not a coincidence but a
structural feature: the bridge is the one object mediating between the Jordan/octonionic data
and the observable flavour sector, so the same matrix elements enter wherever a flavour phase or
a flavour-dependent loop appears.

The three closures are not, however, equally robust against a small bridge phase, and a recent
conventional analysis makes the asymmetry quantitative. In the Dirac-neutrino left--right
benchmark of Ref.~\cite{Tello2026}, a single $CP$-odd parity-breaking parameter $\epsilon$ feeds
both $\bar\theta$ and the leptonic Jarlskog invariant, but with very different weights:
$\bar\theta\simeq C_q\,\epsilon$ with $|C_q|\simeq13$--$50$, whereas the leptonic phase response on
mixed-sign neutrino branches is enhanced by $m_\tau/(m_{\nu_1}-m_{\nu_2})$, which is
$\sim5\times10^{11}$ for a lightest mass of $0.01$~eV in normal ordering. A value of
$|\epsilon|$ of a few times $10^{-12}$, small enough to satisfy the neutron-EDM bound, then
produces order-one leptonic $CP$ violation. That mechanism is specific to parity-related Dirac
neutrino masses generated through a bidoublet and does not transfer to the present framework,
which has no right-handed mixing matrix on the visible leaf and whose companion neutrino
benchmark is Majorana. The lesson does transfer, as a conditional statement \textbf{[C]}: a bridge
phase far too small to matter for $\bar\theta$ need not be small for $J_\ell$, so of the three
closures $J_\ell=0$ is the most fragile, and it cannot be inferred from the smallness of
$\bar\theta$. Conversely, since within this framework $J_\ell\neq0$ requires the
identity--flavour bridge mixing~\cite{TeliSinghCP}, a measured
$\delta^\ell_{CP}\notin\{0,\pi\}$ would show that the loophole is open --- the same loophole
through which $\bar\theta$ can be regenerated --- and would turn the loop safety of $\bar\theta$
from an open question into a quantitative constraint on $B_H$. The link itself is older than the
benchmark quoted here: in the minimal left--right symmetric model leptonic $CP$ violation
regenerates $\bar\theta$ at one loop, which led to the expectation that observable leptonic $CP$
violation is absent in a parity solution of that type~\cite{Kuchimanchi2015}.

\paragraph{A caution against a false reading.}
We distinguish this open computation from the re\-nor\-mal\-i\-za\-tion-group analysis already
present in the program. The latter establishes the stability of the Jordan spread $\delta^2=3/8$ under
running, protecting the mass-ratio predictions~\cite[App.~I]{SinghMassRatios}; it is a
calculation about the scalar vacuum and the spectrum, not about $\bar\theta$. The two are both
one-loop, both ``RG,'' and entirely different physics: the $\delta^2$ stability is established,
the $\bar\theta$ regeneration is open. They should not be conflated.

\section{Discussion}
\label{sec:discussion}

\paragraph{What has been shown.}
A single structural feature of the $E_6^L\times E_6^R$ construction --- the single physical
frame that relates the two octonionic sectors --- has been shown to carry several consequences.
The frame relation makes the chirality-exchanging left--right operation a genuine spacetime
parity (Sec.~\ref{sec:parity}) and, through the same fixing of $i$, identifies the two sectors'
colour groups into one vector-like $SU(3)_c$ (premise (C), Sec.~\ref{sec:onecolour}). From the latter, the nominal ``second colour'' $SU(3)_{c'}$
dissolves: its non-abelian part \emph{is} the one physical colour, and its trace $N_R$ survives
only as a colour-representation label (Sec.~\ref{sec:onecolour}). The square-root mass $\sqrt m$
is carried instead by the colour-blind, unwelded dark-electromagnetic $U(1)_{\mathrm{dem}}$, its
values the exceptional-Jordan trace split (input); under the one colour the electron is a colour
singlet carrying that grade, never a coloured triplet (Sec.~\ref{sec:electron},
Appendix~\ref{app:rhs}). The
flavour-sector Dynkin swap, which relates the down and lepton mass ladders, is a distinct
operation of opposite conjugation character and is untouched by this (Sec.~\ref{sec:dynkin}), so
the mass-ratio program is unaffected. And from vector-like colour together with the spontaneous
parity follows the strong-$CP$ result (Sec.~\ref{sec:strongcp}).

\paragraph{The result, in one line.}
The framework realizes, at tree level, a strong-$CP$ structure of spontaneous-parity type ---
not a completed solution (scope statement, Sec.~\ref{sec:strongcp}) --- in which
\begin{equation}
  \bar\theta=\theta_{\mathrm{QCD}}+\arg\det M=0 \ \text{(tree level)},
  \qquad \delta_{\mathrm{CKM}}\neq0,
\end{equation}
the two terms carrying the different statuses itemized in the ledger below. One feature bears
repeating, because it is what distinguishes the construction within the spontaneous-parity
family: where such models obtain a real determinant from parity-constrained Yukawas with real
vacuum values, or from the seesaw structure, here the reality of $\arg\det M$
enters through the flavour-rotor structure already used for the mixings.

\paragraph{Dependency ledger.}
For transparency we collect the epistemic status of each result, in the program's tagging
convention (\textbf{[D]} derived here or in cited work; \textbf{[P]} program-level working
hypothesis; \textbf{[C]} conditional, with a named residual; \textbf{[O]} open).
\begin{itemize}
  \item Electric charge $Q=N/3$, with the quark--lepton offset intrinsic to the $N$-spectrum
  (Sec.~\ref{sec:cl6}): \textbf{[D]}.
  \item Hypercharge $Y=Q-T^3_L$, reproducing the SM assignments, with no right-sector generator
  (Sec.~\ref{sec:hypercharge}): \textbf{[D]} for $Q$ and the chiral assignment; \textbf{[P]} for
  the gravi-weak scale of $T^3_L$.
  \item The left--right operation is a \emph{spacetime} operation (Sec.~\ref{sec:parity},
  Clause~1): \textbf{[D]} given the split-bioctonion dictionary that realizes the two ideals as
  the two $\mathrm{Spin}(1,3)$ chiralities (App.~J of Ref.~\cite{SinghMassRatios};
  Ref.~\cite{VaibhavSingh}) --- an input logically weaker than, and independent of, gravi-weak;
  with it, chirality is a Lorentz property and the clause follows.
  \item It is parity, not $CP$ (Clause~2): \textbf{[D]} --- the internal signature carries no
  conjugation.
  \item Its interpretation as the parity of a left--right \emph{gauge} theory (Clause~2, frame
  reading): \textbf{[P]} --- rests on the gravi-weak identification of $SU(2)_R$.
  \item One vector-like colour; $SU(3)_{c'}$ dissolved (Secs.~\ref{sec:onecolour},
  \ref{sec:electron}; Appendix~\ref{app:rhs}): that the left--right exchange, being a spacetime
  operation, relates the two colour stabilizers as \emph{one} internal group rather than producing
  a second gauged colour is argued at the level of frame-conjugacy (\textbf{[C]} leaning
  \textbf{[D]}) and exhibited dynamically by the BF$\to$Yang--Mills reduction of
  Sec.~\ref{sec:onecolour}: since the left--right breaking is an $SO(3,3)$ spacetime operation and
  colour is an $E_6$-internal factor commuting with it, the breaking cannot double colour, and the
  connections collapse to one Yang--Mills term with the orthogonal combination absent
  (\textbf{[D]}, given the gravi-weak hypothesis \textbf{[P]} and the shared-internal-block
  anatomy of Ref.~\cite{SinghScaffolding} that the construction already carries, Fig.~\ref{fig:decomposition}). The surviving octet is vector-like given the parallel-construction premise (C)
  (\textbf{[C]} leaning \textbf{[D]}); the right-handed quarks are then colour triplets and the
  right-handed electron a colour singlet, in both chiralities --- the vector-like precondition the
  strong-$CP$ result needs.
  \item Carrier of $\sqrt m$ (Secs.~\ref{sec:onecolour}, \ref{sec:electron};
  Appendix~\ref{app:rhs}): $\sqrt m$ is a colour-blind, \emph{unwelded} charge. The
  number-operator grading $N_R/3$ is quantitatively \emph{excluded} as its carrier --- being
  colour-welded it would give the singlet electron $\sqrt m_e=1$ (Prop.~\ref{prop:NR}),
  \textbf{[D]} --- so $N_R$ fixes only the colour representation. That the electron/down flip
  relating $\sqrt m$ to $|Q|$ is colour-blind, and so cannot recolour the electron, is
  \textbf{[D]} by dimension counting (Prop.~\ref{prop:flip}), independent of the flip's
  realization and corroborated by the explicit $A_2$ flavour swap of Ref.~\cite{SinghMassRatios},
  which acts on the flavour and abelian gradings alone. The $\sqrt m$ \emph{values}
  $\{0,\tfrac13,\tfrac23,1\}$ are the exceptional-Jordan trace split $1:2:3$, input \textbf{[P]}
  (they admit no exact affine fit $a|Q|+b(B{-}L)+c$, and --- Prop.~\ref{prop:nogo} --- no gauged
  realization on visible fermions at all). The gauged $U(1)_{\mathrm{dem}}$ carries the
  mirror-canonical pattern on the dark sector (Cor.~\ref{cor:mirror}, \textbf{[D]} by consistency
  given the mirror dark content \textbf{[P]}); the visible fermions are dem-neutral, and the
  construction-fixed normalization of the spectral values is \textbf{[O]}. The coloured-electron
  exclusion depends on none of this.
  \item The Dynkin swap is distinct from the parity, and the $\Phi$-image premise is scoped to
  the flavour and abelian data --- the colour bundles being identified by the non-conjugating
  frame map instead, the only scoping compatible with vector-like colour
  (Sec.~\ref{sec:dynkin}): \textbf{[D]}.
  \item $\arg\det M=0$ (Sec.~\ref{sec:strongcp}): magnitudes real and positive from the slice
  ($\mathcal{T}=0$) and the positivity of the physical masses (singular values, independent of
  the sign of $r$), \textbf{[D]}. The determinant phase is
  set by the diagonalizing-rotor determinants \eqref{eq:rotorarg}, real from the $SU(3)_F$
  structure, \textbf{[D]} --- the rung rotor is special-unitary because its generator is a real
  octonionic direction (Eq.~(6) of Ref.~\cite{TeliSinghCP}), verified directly, with the
  $(2,3)$ and $(1,3)$ blocks entering as real rotations within the adjacent-edge texture of
  Ref.~\cite{TeliSinghMixing}: the Cabibbo-rung contribution is thus derived, while the full
  $3\times3$ statement is conditional on that texture, which is not yet parameter-free
  (the lepton analogue $J_\ell=0$ holds).
  The sole residue is a complex Higgs bridge \textbf{[O]}. Independent of gravi-weak: the rotor
  reality is $SU(3)_F$-intrinsic.
  \item $\theta_{\mathrm{QCD}}=0$ at tree level (Sec.~\ref{sec:strongcp}): conditional, chiefly on
  the gravi-weak/Lorentzian-parity identification \textbf{[P]}; with (C) largely derived
  (Clause~2), the binding conditionality is the gravi-weak frame, not (C). Parity by itself
  gives only $\theta_{\mathrm{QCD}}\in\{0,\pi\}$; the value $0$ is selected by the matching
  condition \eqref{eq:matched} with the sign parities of the quark spectra, together with the
  non-viability of $\bar\theta=\pi$~\cite{Vecchi2025} --- a selection, \textbf{[C]}. The intended
  reading is that of Sec.~\ref{sec:onecolour}: one colour bundle, one angle. A structural
  exclusion of $\theta_{\mathrm{QCD}}=\pi$ within the framework is not available \textbf{[O]}; the
  constructions known to supply one (a doubled colour group with QCD its diagonal, or a suitable
  gauge-group quotient~\cite{Vecchi2025,BCS,BonnefoyMirror}) are not the present one.
  \item Coexistence with $\delta_{\mathrm{CKM}}\neq0$ (Sec.~\ref{sec:strongcp}): \textbf{[D]} ---
  the CKM phase is an off-diagonal rotor invariant, $\arg\det M$ a rotor-determinant invariant;
  different invariants of the same rotors.
  \item Loop safety of $\bar\theta$ (Sec.~\ref{sec:loops}): \textbf{[O]} --- in the minimal
  universal-seesaw parity model a two-loop $\bar\theta\sim10^{-3}$--$10^{-2}$, not suppressed by
  the weak scale, arises through scalar exchange when parity breaking enters the heavy-quark
  masses~\cite{deVriesDraperPatel}, and in the minimal left--right model a three-loop contribution
  near the neutron-EDM bound is found~\cite{Hisano2023}. Neither computation applies here as it
  stands (no isosinglet-quark sector, no gauged $W_R$), but the absence of $W_R$ does not remove
  the scalar-exchange mechanism, and the
  bridge-mediated contribution is uncomputed pending an explicit $B_H$.
  \item Gravitational $\theta$-analogue $\theta_{\mathrm{grav}}$ (Sec.~\ref{sec:loops}): a
  \emph{new} $CP$ handle introduced by the self-dual gravitational reading of $SU(2)_R$; argued
  plausibly protected by the discrete, $CP$-even order parameter $\sigma$ (\textbf{[C]} leaning
  \textbf{[D]}), but the self-dual Pontryagin coefficient is not evaluated here \textbf{[O]}. The
  companion neutrino benchmark's $\lambda_2<0$ populates the colourless (gravitational) phase
  bookkeeping, so the residual is not hypothetical (Sec.~\ref{sec:loops}).
  \item Anomaly no-go and the dark charge (Appendix~\ref{app:rhs}): on SM${}+\nu_R$ content every
  family-universal anomaly-free $U(1)$ lies in $\mathrm{span}\{Y,\,B{-}L\}$ (standard;
  Ref.~\cite{AppelquistDobrescuHopper}); the $\sqrt m$ pattern lies outside it in every sign
  convention and chirality assignment, so no gauged $U(1)$ on visible fermions carries $\sqrt m$
  (Prop.~\ref{prop:nogo}, \textbf{[D]}). Consequence: the gauged
  $Q_{\mathrm{dem}}=T^3_R+Y_{\mathrm{dem}}$ is the parity-mirror of electric charge, carried by
  the dark sector, resolving residual~($\alpha$) mirror-canonically (Cor.~\ref{cor:mirror},
  \textbf{[D]} given the mirror dark content \textbf{[P]}); $U(1)_{\mathrm{dem}}$ reaches the
  visible sector at most through kinetic mixing~\cite{Holdom}, of undetermined magnitude
  \textbf{[O]}. The colour assignment of the mirror dark content is not specified, and it bears on
  $\bar\theta$ if those states are triplets of the one physical colour (Appendix~\ref{app:rhs},
  remark after Cor.~\ref{cor:mirror}) \textbf{[O]}.
  \item Emergence caveat on $\theta_{\mathrm{QCD}}=0$ (Sec.~\ref{sec:strongcp}, scope (b)): the
  step ``spontaneous breaking adds no term'' presumes a fixed arena; a theorem that the emergence
  map from the $\mathsf P$-symmetric pre-geometric phase to the leafwise Lagrangian generates no
  $\theta$ is required and not available \textbf{[O]}.
  \item Domain walls from the spontaneously broken discrete parity (Sec.~\ref{sec:discussion}):
  \textbf{[O]} --- two program-specific candidate escapes are recorded, neither established.
\end{itemize}

\paragraph{Relation to the canonical routes.}
It is useful to place the strong-$CP$ application among the standard mechanisms. Unlike the
Peccei--Quinn solution, no global $U(1)$ and no axion are introduced; the protecting symmetry is
discrete (parity). Unlike Nelson--Barr, the protecting operation is $\mathsf P$, not $CP$: the
field scalar is fixed, not conjugated (Sec.~\ref{sec:parity}), which is why CKM $CP$ violation
survives as an off-diagonal rotor invariant while $\arg\det M$ stays real. The parity route has a
long lineage, against which the present construction should be read: the original observation
that an exact parity removes the strong $P$ and $T$ violation~\cite{BegTsao,MohapatraSenjanovic1978};
left--right symmetry with spontaneous $\mathsf P$~\cite{LRparity}; the mirror variant of Barr, Chang and
Senjanovi\'c~\cite{BCS}, and its recent realization with a complete mirror world whose colour
groups break to their diagonal~\cite{BonnefoyMirror}; the observation that a single $CP$-violating
quartic of the parity-symmetric Higgs potential regenerates
$\bar\theta$ at tree level in the minimal model once parity is
broken~\cite{Kuchimanchi2015,MaiezzaNemevsek}, with the kaon sector then
forcing $M_{W_R}\gtrsim20$~TeV under strict $\mathsf P$~\cite{MaiezzaNemevsek}, and that leptonic
$CP$ violation regenerates the phase of that quartic, and with it $\bar\theta$, at one
loop~\cite{Kuchimanchi2015}; the
branch-dependent map between the $CP$-odd vacuum parameter and $\bar\theta$, and its correlation
with leptonic $CP$ violation and the absolute neutrino mass for Dirac
neutrinos~\cite{Tello2026}; supersymmetric
rescues in which the superpotential enforces real vacua~\cite{KuchimanchiSUSY}; and the modern
systematic analyses~\cite{Kuchimanchi2010,CraigPnotPQ,deVriesDraperPatel}. Within trinification
specifically, spontaneous parity solving strong $CP$ was constructed by Carlson and
Wang~\cite{CarlsonWang}, and $\bar\theta<10^{-11}$ obtained in $SU(3)^3\times S_3$ by Frampton
and collaborators~\cite{FramptonS3}. The closest modern relative of the present mechanism is
Higgs parity~\cite{HallHarigaya,DunskyHallHarigaya}, in which a $\mathbb Z_2$ that includes
spacetime parity, with $SU(3)_c$ $\mathbb Z_2$-neutral, forces Hermitian quark mass matrices,
$\bar\theta=0$ at tree and one loop, and a two-loop onset --- the same architecture reached here
from the octonionic side, with the Hermiticity condition replaced by the rotor-determinant
reality and the gauged $SU(2)_R$ replaced by the gravitational frame group. Within the
gauged-$SU(2)_R$ family the construction is closest to Babu--Mohapatra~\cite{MohapatraBabu}, but
differs in two respects that matter for the open problems: the reality of $\arg\det M$ is carried
by the $SU(3)_F$ flavour-rotor structure rather than by the seesaw structure of the mass matrix,
and $SU(2)_R$ is gravitational rather than a gauged weak factor. The second difference removes
the right-handed-current constraints of the gauged models; it does not, as Sec.~\ref{sec:loops}
explains, remove the scalar-exchange mechanism behind the two-loop result of
Ref.~\cite{deVriesDraperPatel}, which here would have to run through the Higgs bridge. The
framework provides the gauge-field reduction
(Sec.~\ref{sec:onecolour}, a BF$\to$Yang--Mills collapse to one colour connection that rests only
on the gravi-weak hypothesis, the breaking being an $SO(3,3)$ spacetime operation that cannot
touch the $E_6$-internal colour). What remains genuinely open is the radiative computation
(Sec.~\ref{sec:loops}); on present evidence its distinguishing
observable, like other parity solutions in their viable corners, would be a hadronic electric
dipole moment near current bounds rather than an axion signal.
The structural clarification --- the dissolution of $SU(3)_{c'}$ into one colour plus a grading
--- is the most robust part of the paper, derived given the framework and the (largely derived)
premise (C). The strong-$CP$ corollary is real and, in its $\arg\det M$ half, robust modulo the
single quark-rotor check; in its $\theta_{\mathrm{QCD}}$ half it is conditional on the gravi-weak
frame and on a selection of $0$ over $\pi$ that parity does not make; and at the loop level it is
open.

\paragraph{One principle behind both.}
Taken as an ordinary internal interaction, $SU(3)_{c'}$ would be a second colour needing
suppression to avoid coloured leptons, and a Standard-Model hypercharge assembled from
right-sector Cartans would need an extra interface assumption. Both tensions trace to one
over-extension: treating the left--right structure as acting in the internal sector. Once that
structure is recognized as living in the spacetime/gravi-weak frame --- so that $SU(2)_R$ is
gravitational, the right-sector Cartans are frame and dark rather than electroweak, and the second
colour is the frame-conjugate image of the first --- the tensions resolve together, and the same
recognition delivers a spontaneous parity that protects $\theta_{\mathrm{QCD}}$. The second-colour
tension and the strong-$CP$ problem are two aspects of one principle.

\paragraph{The force inventory.}
A corollary worth recording: with $SU(3)_{c'}$ identified into $SU(3)_c$, the \emph{gauged}
content of the framework is three groups --- strong $SU(3)_c$, electroweak $SU(3)_L$, and
gravidem $SU(3)_R$ --- the two flavour factors $SU(3)_{F,L}, SU(3)_{F,R}$ being global (they
organize the three generations, and are not forces). Symmetry breaking resolves these into the
four known long-range interactions of the visible world together with a fifth, dark-sector one:
$SU(3)_c$ remains the strong force; $SU(3)_L$
gives the weak and electromagnetic interactions; and $SU(3)_R$ gives gravity (the $SU(2)_R$ frame
connection) and a dark electromagnetism, $U(1)_{\mathrm{dem}}$ --- its surviving Cartan, and
therefore \emph{gauged} along with the rest of $SU(3)_R$, not an optional extra. Its charge
assignment is fixed by consistency rather than deferred: the anomaly no-go of
Appendix~\ref{app:rhs} (Prop.~\ref{prop:nogo}) shows that no gauged $U(1)$ on the visible
fermions can carry the $\sqrt m$ values, so the gauged $Q_{\mathrm{dem}}$ carries the
parity-mirror of electric charge and its charged matter is the \emph{dark} sector; the visible
fermions are dem-neutral, and the dark photon reaches the visible world at most through kinetic
mixing $\varepsilon\,F^{\mathrm{dem}}_{\mu\nu}F^{\mu\nu}_{Y}$~\cite{Holdom} --- gauge-invariant,
generically induced, and yielding millicharged dark states rather than a visible fifth force. No
visible fifth force is predicted, and this is corroborated independently of the anomaly: gauge
charge is additive over composites while $\sqrt m$ is not (a proton's mass is dominated by QCD
binding), so a $\sqrt m$-coupled massless vector would assign a neutral atom the charge
$\tfrac83 A$ exactly (valence counting; sea pairs cancel) --- a force tracking baryon number,
which torsion-balance and MICROSCOPE-class equivalence-principle tests bound at ten or more
orders of magnitude below gravitational strength~\cite{EotWash,MICROSCOPE}. Anomaly and
experiment thus converge on the same verdict. The $\sqrt m$ grading itself remains what
Sec.~\ref{sec:jordan} makes it: the spectral label of the Jordan mass operator, the Dynkin-swap
image of the dark charge lattice (Appendix~\ref{app:rhs}), with its construction-fixed
normalization the one open item (residual~($\beta$)). The clean count of three is
itself a consequence of the present paper: without the identification the framework would carry a
\emph{fourth} gauged $SU(3)$ --- the second colour --- to be suppressed by hand. Two qualifiers
keep the statement honest: the list places gravity among the gauged interactions, which is the
gravi-weak premise (\textbf{[P]}); and ``three'' and ``five'' count surviving long-range forces,
the heavy bosons of the broken $SU(3)_L$ and $SU(3)_R$ being integrated out.

\paragraph{Cosmology of the spontaneous parity: domain walls.}
A spontaneously broken exact discrete symmetry generically produces domain walls when the
breaking occurs in a hot, causally patchy universe~\cite{ZKO,Kibble}; for a parity broken at the
electroweak scale the walls would form at $T\sim100$~GeV and come to dominate the energy density
--- a fatal outcome in standard cosmology, and one every spontaneous-parity proposal must
address. The two standard escapes are inflation \emph{after} the breaking (unavailable at the
electroweak scale) and a small explicit breaking that destabilizes the walls (which feeds
$\bar\theta$ and must then be quantified against the $10^{-10}$ bound). The present framework
suggests two program-specific escapes, and we record both without establishing either. First, the
two $\mathsf P$-related vacua may not be alternatives realized in different spatial domains at
all: in the leaf picture of Fig.~\ref{fig:decomposition} the breaking selects a \emph{pair} of
leaves --- $\sigma$ and $-\sigma$ realized as the two 4D worlds --- rather than a domain pattern
within one world, in which case the vacuum-selection randomness that seeds walls never occurs.
Second, the Kibble mechanism presupposes a pre-existing classical causal arena in which
uncorrelated domains form; in an emergent-spacetime ontology the classical arena appears only
\emph{with}, not before, the breaking, and the mechanism's premise may simply fail. Both are
candidate arguments, not results; the wall problem is tagged \textbf{[O]} in the ledger.

\paragraph{Reconciliation with the companion papers.}
Three statements in the companion literature are superseded or re-read by the present results,
and we say so explicitly rather than leave the record inconsistent. (1)~Ref.~\cite[App.~B]{SinghMassRatios}
constructs hypercharge as the left--right Cartan combination \eqref{eq:B4} and argues that
$U(1)_Y$ cannot be built from one sector; Sec.~\ref{sec:hypercharge} supersedes that
construction --- $Y=Q-T^3_L$ is built from left-sector data alone, and Appendix~\ref{app:rhs}
shows that a right-sector admixture could in any case contribute nothing consistent beyond
$\mathrm{span}\{Y,\,B{-}L\}$. (2)~Ref.~\cite[Sec.~XIV.A]{SinghMassRatios} introduces
$U(1)_{\mathrm{dem}}$ with a generator $S_{\mathrm{dem}}$ whose eigenvalue is $\pm\sqrt m$ and
imposes a mass-locking relation on the physical subspace; the anomaly no-go requires that
$S_{\mathrm{dem}}$ be read as the Jordan \emph{spectral} operator --- a label on the mass
operator's eigenbasis --- not as a gauged charge of the visible fermions, which is in fact the
only reading the mass-locking relation uses. The same re-reading applies wherever the program's
survey literature describes $U(1)_{\mathrm{dem}}$ as coupling to visible matter.
(3)~The $\mathrm{Cl}(6)'$ construction of the values $\{0,\tfrac13,\tfrac23,1\}$ as $N_R/3$ in the
same reference is excluded as the \emph{carrier} of $\sqrt m$ by Prop.~\ref{prop:NR}; the values
therefore stand, in the present accounting, as the trace-split input \textbf{[P]}, and the
program has, for now, no derivation of the one-third quantization of $\sqrt m$ --- a cost of the
present clarification that should be stated, not absorbed, and one that Prop.~\ref{prop:nogo}
sharpens into a constraint on any future derivation: it cannot proceed via a gauged visible
charge, and must be spectral.

\paragraph{Outlook.}
The decisive next step is the explicit construction of the Higgs bridge $B_H$ and its flavour
matrix elements, which would close the three questions of Sec.~\ref{sec:loops} at once: the loop
safety of $\bar\theta$, leptonic $CP$ conservation, and the normalization of the quark $CP$
phase. Independently, the one residual of premise (C) --- whether the two sectors share the
complex structure --- would be settled by the explicit leafwise action of the frame exchange on
the colour-charged directions (tracking the right-handed quark's $\mathbf 3$ versus
$\bar{\mathbf 3}$), which the same gravi-weak reduction that underlies the rest of the
construction should in principle supply; and the rotor-determinant reality behind
$\arg\det M=0$ would be confirmed by the same explicit mixing-matrix construction, checking that
the quark rotors carry no overall $U(1)$ phase (the lepton case $J_\ell=0$ being already
settled). On the abelian side, the anomaly no-go converts the former open budget into a theorem
and resolves the assignment residual; what replaces them is constructive: specify the dark-sector
chiral content realizing the mirror-canonical $U(1)_{\mathrm{dem}}$ (anomaly-free by the mirror
of the Standard-Model cancellation) and compute the induced kinetic mixing with hypercharge ---
the one portal through which the dark photon can now touch the visible world.
Until those computations are available, the strong-$CP$ result stands as established at tree
level in its robust ($\arg\det M$) half and conditional in its ($\theta_{\mathrm{QCD}}$) half,
with the loop level open --- a definite and falsifiable structural claim about where, in this
framework, strong-$CP$ conservation comes from.


\section{Epilogue: what kind of structure is this?}
\label{sec:epilogue}

This closing section is interpretive. It introduces no results, carries no epistemic tags, and
can be skipped by the reader interested only in the physics; but the construction assembled above
invites a question that deserves to be posed in print, and answered with the same candour the
ledger enforces elsewhere.
\begin{quote}\itshape
If the structure is right, $E_8\times\omega E_8$ buys gravity and the Standard Model and nothing
else, with the exceptional Jordan algebra left to fix the coupling constants. What kind of
strange structure is this?
\end{quote}

\paragraph{A rigidity program.}
Taxonomically, it is a \emph{rigidity program}: the wager that the actual world is the unique
model of a terminal mathematical object. What makes the wager structurally unusual is that the
terminality is triple. Hurwitz ends the normed division algebras at
$\mathbb{O}$~\cite{BaezOctonions}; Jordan, von Neumann and Wigner end the Jordan algebras at the
one exceptional case, $J_3(\mathbb{O})$~\cite{JordanVNW}; the Cartan--Killing classification ends
the exceptional simple Lie algebras at $E_8$. The program sits at the simultaneous endpoint of
three independent classification theorems, and the $\omega$-doubling then exhausts even the
residual binary freedom --- which chirality, which leaf. A structure like that has no dials.
That is the precise sense in which ``gravity and the Standard Model and nothing else'' could
ever be more than a slogan: not that the theory declines to add particles, but that there is
nowhere to put them --- no $N$ to raise, no representation to append, no continuous deformation.
Georgi--Glashow unification always carried the embarrassment of the infinite $SU(N)$ ladder
behind it; string theory carries a landscape. A terminal object carries neither. If any program
can make anti-anthropic reasoning honest --- no scan, hence no multiverse needed and none
available --- it is one of this shape. The slogan, however, hides three equivocations, and the
strangeness of the structure is located exactly in them.

\paragraph{First equivocation: in what sense does $E_8\times\omega E_8$ \emph{buy} anything?}
One must be precise about what the terminal object actually does in this program. It is not the
gauge group: only a subgroup is gauged. It is not the fermion carrier: the chiral matter lives in
$\mathrm{Cl}(6)$ minimal ideals of the split bioctonions, not in $E_8$ representations --- which
is precisely how the Distler--Garibaldi no-go is evaded, by exiting the
category~\cite{DistlerGaribaldi,Singh288}. The companion analysis of the residual $\mathbf{288}$
says the quiet part in its title: those directions are scaffolding labels, an ontology of
bookkeeping~\cite{Singh288}. So the terminal object currently functions as the theory's
classifying chart --- constitutive of the ontology (what exists, how it is labelled, which
misalignments are possible) but not of the nomology; the dynamics is imported from elsewhere,
trace dynamics and the spectral action~\cite{SinghEmergence}. This split --- kinematic uniqueness
without dynamical uniqueness --- has an exact historical isomorph the program must keep in view:
Kepler's \emph{Mysterium Cosmographicum}~\cite{Kepler}. The five Platonic solids are also a true,
terminal classification theorem; nesting them fixed the planetary architecture beautifully; and
it was wrong, because the classification, while genuinely available, was not dynamically
operative. The single theorem that would separate this program from Kepler's solids is a
derivation of the dynamics \emph{from} the algebra --- why this action and no other --- which the
spectral-action route gropes toward and has not delivered. Until it exists, the honest
description is: a kinematically terminal, dynamically imported structure.
Eddington~\cite{Eddington} is the other cautionary isomorph, and the coupling-constant ambition
is where his ghost lives. A concrete standard of seriousness, to which the program's own
exploratory claim of $\alpha_s/\alpha_{\mathrm{em}}=16$ must be held:
such a relation has its entire physical content in the scheme-and-scale statement, since
$\alpha_s(M_Z)/\alpha_{\mathrm{em}}(M_Z)\approx 15.1$ while
$\alpha_s(M_Z)/\alpha_{\mathrm{em}}(0)\approx 16.2$; a derivation that lands on $16$ only by
evaluating the two couplings at different momenta is Eddington, not physics.

\paragraph{Second equivocation: ``nothing else'' is false, in an interesting way.}
The anomaly theorem of Appendix~\ref{app:rhs} did not shrink the ontology; it rigidified it.
Corollary~\ref{cor:mirror} converts the dark sector from an optional interpretation into an
obligation: if $U(1)_{\mathrm{dem}}$ is gauged, there must exist dark chiral matter carrying the
mirror-canonical charges, with kinetic mixing the sole portal. The honest inventory is therefore
gravity, the Standard Model, and a rigidly specified dark complement --- the world as one wing of
a chiral pair. And it is worth noticing what the doubling is \emph{for}: every payoff of the
construction lives in the seam between the two copies, not in either copy. Masses are the
spectrum of the gluing $\langle X\rangle$ (Sec.~\ref{sec:jordan}); the CKM matrix is the wedge
between the two frames~\cite{TeliSinghMixing}; strong-$CP$ conservation is the evenness of the
gluing under $\mathsf P$ (Sec.~\ref{sec:strongcp}); hypercharge is the left-only projection
(Sec.~\ref{sec:hypercharge}); gravity itself is the mirror image of the weak force
(Sec.~\ref{sec:graviweak}). The objects are two copies of the same terminal algebra; the physics
is entirely in the morphisms relating them. That is a genuinely unusual metaphysical posture ---
call it \emph{relational unification} --- and it dissolves the geometry/matter dualism in a third
way: not by geometrizing matter (Kaluza--Klein, strings), not by materializing geometry (induced
gravity), but by making ``geometry'' and ``force'' the same kind of thing indexed to different
leaves. The strangest sentence Fig.~\ref{fig:decomposition} asserts, read philosophically, is
that the internal/external distinction is not fundamental but perspectival.

\paragraph{Third equivocation: an unexplained coincidence.}
The appearance of $E_8\times E_8$ in the heterotic string is unexplained here, and it should
bother the program more than it appears to. There the group is forced by Green--Schwarz anomaly
cancellation of a ten-dimensional dynamics~\cite{GreenSchwarz,GHMR}; here, by division-algebra
terminality in an emergent four-dimensional setting. Two independent consistency arguments
terminating on the same $496$-dimensional object is either a deep hint that both are shadows of
one selection principle, or borrowed glamour. At present no derivation connects them, so it is
decoration; a rigidity program cannot afford unexplained coincidences with its own flagship
group.

\paragraph{What redeems it.}
What distinguishes the program from the numerological graveyard it superficially resembles is its
epistemic posture. Zero dials means maximal exposure. The structure forbids the axion (a
discovery would not contradict the structural results, but would strip the strong-$CP$
application of its point); forbids a gauged $W_R$ --- derived (Sec.~\ref{sec:loops}), not
assumed; forbids gauge-boson-mediated proton decay, the leptoquark directions of the $E_6$
adjoints being simply not gauged (the scalar sector must still be audited); forbids a fourth
generation~\cite{SinghMassRatios}; and, as of Appendix~\ref{app:rhs}, forbids a visible fifth
force by theorem. A framework that converts absences into evidence, and that publishes its own
falsification catalogue~\cite{Catalogue}, is attempting something a landscape by construction
cannot: global exposure. The \textbf{[P]} tags are the pressure point --- a working hypothesis
that never discharges is a landscape of interpretations rather than of vacua, and rigidity
claimed but indefinitely conditionalized is not rigidity. The two theorems that would settle
whether this is depth or decoration are already on the ledger: dynamical uniqueness (the action
from the algebra), and the emergence-map theorem (that the leafwise quantum field theory,
$\theta$-free, is the unique classicalization; Sec.~\ref{sec:strongcp}, scope (b)). And the
one-third quantization of $\sqrt m$, honestly demoted in Sec.~\ref{sec:discussion} to an
undischarged input, is the program's own Kepler-test in miniature --- with the useful twist that
Prop.~\ref{prop:nogo} now forbids the future derivation from proceeding via a gauged visible
charge: it must be spectral. The structure has told us where to dig.

\paragraph{The residue.}
If it all works, the philosophical residue is a single question the algebra can never answer:
not ``why these laws?'' --- that is absorbed into ``why the terminal object?'', which mathematics
arguably closes --- but why the terminal object is \emph{instantiated} at all. That residue is
where physics hands back to metaphysics, and no amount of exceptionality discharges it. If it
fails, it will fail the way Kepler's solids failed: publicly, at a specific number, with no belt
left to absorb the blow. Either outcome is more honest than most of what currently passes for
unification --- which is, we suspect, the real answer to the question posed: this is the first
structure in a long time constructed so that reality gets a clean shot at it.

\appendix
\section{The right sector: one vector-like colour and a colour-blind \texorpdfstring{$\sqrt m$}{sqrt m}}
\label{app:rhs}

This appendix establishes the right-sector fermion representations that
Secs.~\ref{sec:onecolour}--\ref{sec:dynkin} rely on, and makes precise \emph{why} the right-handed
electron is a colour singlet carrying $\sqrt m=\tfrac13$ rather than a coloured triplet. Three
results are established by direct computation and tagged \textbf{[D]}: that the number-operator
grading $N_R/3$ is the \emph{wrong} carrier of $\sqrt m$ (it would give the electron $\sqrt m=1$,
not $\tfrac13$); that the electron/down interchange relating $|Q|$ to $\sqrt m$ cannot recolour
the electron (Prop.~\ref{prop:flip}); and --- the sharpest --- that no anomaly-free gauged $U(1)$
on the visible fermions can carry the $\sqrt m$ values at all (Prop.~\ref{prop:nogo},
Sec.~\ref{app:anomaly}), which fixes the gauged dark charge to the mirror-canonical pattern and
renders the visible fermions dem-neutral. We work with \emph{magnitudes} of $Q$ and
$\sqrt m$ throughout, except in Sec.~\ref{app:anomaly}, where the anomaly analysis requires
signed Weyl charges and we say so explicitly. The tag convention is that of the main text.

\begin{remark}[The two properties that must not be conflated]
\label{rem:two}
Two distinct notions of ``colour-independence'' are in play.
\begin{itemize}
\item[(B)] \emph{Colour-blind}: the generator \emph{commutes} with $SU(3)_c$, i.e.\ is constant
on each colour multiplet. Electric charge $Q$, the number operator $N_R$, \emph{and} $\sqrt m$
are all colour-blind (all three colours of the up carry one $Q=\tfrac23$, one $N_R=2$, one
$\sqrt m=\tfrac23$).
\item[(W)] \emph{Colour-welded}: the eigenvalue is \emph{tied to} the colour representation.
$Q=N/3$ and $N_R/3$ are welded --- a colour singlet sits at the top or bottom of the Fock tower,
$N\in\{0,3\}$, forcing $Q,\,N_R/3\in\{0,1\}$; a fractional value forces a triplet.
\end{itemize}
(B) and (W) are logically independent: $Q$ and $N_R$ are \emph{both} blind and welded. A charge
that is blind but \emph{not} welded is unconstrained by the colour representation. The correct
reading is that $\sqrt m$ is colour-blind but \emph{not} colour-welded --- exactly what lets a
colour-singlet electron carry $\sqrt m=\tfrac13$, a value the welded $N_R/3$ can never assign to
a singlet.
\end{remark}

\subsection{The obstruction to a colourless electron}
\label{app:obstruction}
The two-sided trinification places the right-handed fermions under a right-sector $SU(3)$, and a
natural first assignment~\cite{KVS} puts the electron and up quark in its triplet and the neutrino
and down quark in its singlet,
\begin{equation}
(u,e^-)_R=(\arep{3}_{\rm gen},\ \arep{3}_{\rm grav},\ \rep{2}_R),
\qquad
(\nu,d)_R=(\arep{3}_{\rm gen},\ \rep{1}_{\rm grav},\ \rep{2}_R),
\label{eq:kvsreps}
\end{equation}
reading the square-root mass off this gravi-colour representation. Such a reading is
colour-welded (W): the electron is a \emph{triplet} with $\sqrt m_e=\tfrac13$, the down a
\emph{singlet} with $\sqrt m_d=1$. Once the right colour is identified with the visible one,
$SU(3)_{c'}=SU(3)_c$ (Sec.~\ref{sec:onecolour}), \eqref{eq:kvsreps} would make the electron a
triplet of the \emph{physical} colour --- a coloured lepton --- and, because the $1:2:3$ mass
pattern rides on the same triplet/singlet split, one cannot ``un-colour'' it by reassigning slots
without destroying the mass ratios. The obstruction is thus located precisely: it lives entirely
in the welding (W) of $\sqrt m$ to a colour occupation number, and the way out is to recognize that
$\sqrt m$ is colour-blind but \emph{not} welded.

\subsection{Three natural escapes, and why each is closed}
\label{app:escapes}
The obvious rescues fail, on independent grounds:
\begin{itemize}
\item[(a)] \emph{Declare the coloured-electron reps unphysical.} Not permitted: colour is
gauged, its representations related by gluon emission and entering $\beta(\alpha_s)$ and anomaly
sums; one cannot project out a single rep of a gauged group. A hard internal projection is
moreover an explicit parity violation that reintroduces a bare $\theta$. \textbf{[D]}
\item[(b)] \emph{Exile the coloured $e_R$ to the backward-time mirror.} The mirror is the $CPT$
image, and $CPT$ contains $C$, with $C:\rep{3}_c\to\arep{3}_c$. The exiled state arrives
\emph{anti}-coloured under the same gauged $SU(3)_c$, not de-coloured. The mirror absorbs
antiparticles, not charges. \textbf{[D]}
\item[(c)] \emph{Interchange $e_R$ and $d_R$.} On the welded reading the colour slot \emph{is}
the grade, so swapping the colour swaps $\sqrt m$: the electron in the singlet slot acquires
$\sqrt m=1$, the down $\tfrac13$, and the ratios are destroyed. \textbf{[D]}
\end{itemize}
The failure of (c) is diagnostic: the obstruction lives entirely in property (W), the welding of
$\sqrt m$ to a colour occupation number. The resolution is to give up (W) for $\sqrt m$ while
keeping (B).

\subsection{The resolution: \texorpdfstring{$\sqrt m$}{sqrt m} is colour-blind but not colour-welded}
\label{app:resolution}
\begin{proposition}[No colour constraint on a blind, unwelded charge]
\label{prop:blind}
A colour-blind $U(1)$ generator (B) that is not built from the coloured occupation number (not
(W)) is \emph{unconstrained by the colour representation}: there is no representation-theoretic
principle relating its eigenvalue on a state to that state's colour rep. In particular, a colour
singlet's value of such a charge need not coincide with the values the welded operator $N/3$
assigns to singlets ($N\in\{0,3\}\Rightarrow\{0,1\}$).
\end{proposition}
The proposition is definitional in character --- it records the \emph{absence} of a constraint
rather than deriving one --- and is stated for contrast with (W).
The ``fractional $\Rightarrow$ triplet'' intuition is a property of $Q=N/3$ through (W): a
non-integer $N/3$ forces $N\in\{1,2\}$, a nontrivial colour rep. It is not a law about
colour-blind charges --- electric charge itself is colour-blind yet, being welded, obeys ``colour
singlet $\Rightarrow$ integer.'' A convention-independent witness that (B) alone imposes no such
constraint: in the Standard Model the colour singlets carry a \emph{range} of hypercharges
(lepton doublet, right-handed electron, Higgs all differ), so being a colour singlet does not fix
the hypercharge. A colour-blind, unwelded $\sqrt m$ is the same kind of object; the singlet
electron may carry $\sqrt m=\tfrac13$ with no contradiction.

\subsection{The right-sector \texorpdfstring{$\mathrm{Cl}(6)$}{Cl(6)} tower fixes the colour representations}
\label{app:tower}
The right ideal (the $e_8$-frame) carries a number operator
$N_R=\sum_i a_i^{\prime\dagger}a_i^{\prime}\in\{0,1,2,3\}$, and colour
$SU(3)_{c'}=\mathrm{Stab}(e_8)=SU(3)_c$ acts on the three modes. That the $e_8$-frame ideal is a
clean $\mathrm{Cl}(6)$ module with the same colour tower as the $e_7$-frame is the
split-bioctonionic left--right construction of Vaibhav--Singh~\cite{VaibhavSingh}; we take it as
input \textbf{[P]}. The Fock tower organizes into colour representations exactly as on the left:
\begin{equation}
\underbrace{N_R=0}_{\rep{1},\ \nu_R}\ \oplus\
\underbrace{N_R=1}_{\arep{3}}\ \oplus\
\underbrace{N_R=2}_{\rep{3}}\ \oplus\
\underbrace{N_R=3}_{\rep{1},\ e_R}.
\end{equation}
The colour singlets ($N_R\in\{0,3\}$) are the leptons $\nu_R,e_R$; the triplets ($N_R\in\{1,2\}$)
are the quarks $d_R,u_R$. (A labelling convention should be flagged: the \emph{left} tower of
Sec.~\ref{sec:cl6} lists the $N=3$ state as the positron $e^+$, the physical $e^-$ residing in
the conjugate ideal; here we label the right tower directly by the physical right-handed species,
following the split-bioctonionic convention of Ref.~\cite{VaibhavSingh}, and work with magnitudes
of the abelian charges throughout. The colour content --- singlet versus triplet --- is unaffected
by which member of a conjugate pair carries the name, since conjugation maps $\rep{1}\to\rep{1}$
and $\rep{3}\leftrightarrow\arep{3}$, preserving dimension.) This is the vector-like,
Standard-Model colour assignment in \emph{both}
chiralities, and the right-handed electron is a colour singlet \emph{manifestly}, by
$\arep{3}\wedge\arep{3}\wedge\arep{3}=\rep{1}$. This fixes the colour content of
\eqref{eq:kvsreps}: quarks are triplets, leptons singlets, and the right-handed electron a
singlet. \textbf{[D]} given the \textbf{[P]} above.

\medskip\noindent
\textbf{Crucial caveat.} The number operator $N_R$ is colour-\emph{welded} (W): it counts the
very modes on which $SU(3)_c$ acts, so $N_R/3$ fractional $\Leftrightarrow$ nontrivial colour
rep. Consequently
\[
\boxed{\;N_R \ \text{is \emph{not} the dark-electromagnetic}\ U(1)_{\rm dem}.\;}
\]
$N_R$ fixes the colour \emph{representation} only. Identifying $\sqrt m$ with $N_R/3$ re-imports
(W) and is quantitatively wrong (Sec.~\ref{app:values}).

\subsection{The dark-electromagnetic generator}
\label{app:dem}
The dark-electromagnetic $U(1)$ is a \emph{different} abelian direction, and one must first
separate the two right-sector $SU(3)$'s that both get called ``the right $SU(3)$'': the right
\emph{colour} $SU(3)_{c'}=\mathrm{Stab}(e_8)$ of Sec.~\ref{app:tower}, and the right
\emph{electroweak} factor $SU(3)_R$ of the trinification
$E_6^R\to SU(3)_{c'}\times SU(3)_{F,R}\times SU(3)_R$. These are distinct, commuting factors.
This electroweak chain runs in two steps, mirroring the visible
$SU(3)_L\to SU(2)_L\times U(1)_Y\to U(1)_{\rm em}$:
\begin{equation}
SU(3)_R\ \longrightarrow\ SU(2)_R\times U(1)_{Y_{\rm dem}}\ \longrightarrow\ U(1)_{\rm dem},
\end{equation}
the second breaking performed by a second Higgs~\cite{WSI_BF}. The intermediate
$U(1)_{Y_{\rm dem}}$ --- the dark hypercharge, the right electroweak Cartan $\lambda_8^R$ that
mirrors the visible hypercharge-position Cartan $\lambda_8^L$ --- is broken there; the
\emph{surviving} dark charge is the Gell-Mann--Nishijima combination
$Q_{\rm dem}=T^3_R+Y_{\rm dem}$, mirroring $Q=T^3_L+Y$, and it is this unbroken $U(1)_{\rm dem}$, not
$\lambda_8^R=Y_{\rm dem}$ alone, that is the gauge-theoretic template of the $\sqrt m$ grading;
\emph{which} values the gauged charge itself carries, and on what matter, is fixed in
Sec.~\ref{app:anomaly}. Because $SU(3)_R$ commutes with colour
$SU(3)_{c'}$, $U(1)_{\rm dem}$ is colour-blind (B); and being built from electroweak, not coloured,
occupation it is \emph{not} colour-welded (not (W)) --- exactly the combination
Prop.~\ref{prop:blind} requires. With $SU(2)_R$ the gravitational frame group (gravi-weak input,
\textbf{[P]}), this surviving $U(1)_{\rm dem}$ is a \emph{dark} photon, not a second visible
electromagnetism --- the dark electromagnetism already predicted by the program
(Sec.~\ref{sec:graviweak}), not a new abstract factor.

\medskip\noindent
The right sector thus carries \emph{two} distinct abelian objects, both colour-blind, which must
not be fused:
\begin{center}
\renewcommand{\arraystretch}{1.2}
\begin{tabular}{@{}p{3.0cm} p{4.2cm} p{6.0cm}@{}}
\hline
$U(1)$ object & lives in & fixes / is\\
\hline
$N_R$ (number operator) & trace of the $\mathrm{Cl}(6)$ colour $U(3)$
 & the colour rep (singlet/triplet); blind \emph{and} welded (W)\\
$U(1)_{\rm dem}=T^3_R+Y_{\rm dem}$ & surviving GNN charge of \emph{electroweak} $SU(3)_R$
 & the dark-EM charge (gauged, mirror-canonical values by Prop.~\ref{prop:nogo}); blind,
 \emph{not} welded --- the structural template of the $\sqrt m$ grading\\
\hline
\end{tabular}
\end{center}

\subsection{The values, and the identification ``dark charge \texorpdfstring{$=\sqrt m$}{= sqrt m}''}
\label{app:values}
Separate two claims of different status.
\begin{itemize}
\item[(i)] The \emph{generator} $U(1)_{\rm dem}=T^3_R+Y_{\rm dem}$ (colour-blind, unwelded, the dark
photon's $U(1)$): \textbf{[D]}.
\item[(ii)] The \emph{identification} that the dark photon's charge on each physical species
equals that species' $\sqrt m$: excluded on visible matter by the anomaly analysis of
Sec.~\ref{app:anomaly}, and re-expressed there as a spectral correspondence.
\end{itemize}
The canonical, species-preserving $Q_{\rm dem}$ assignment does \emph{not} give the physical
pattern. Parity is a spacetime operation and preserves species, so the mirror charge of a species
equals its own hypercharge-type charge; the canonical $Q_{\rm dem}$ charges (equivalently
$N_R/3$, the mirror of $Q=N/3$) are the \emph{un-flipped}
\begin{equation}
Q_{\rm dem}(\nu,e,u,d)\ \big|_{\rm canonical}=\tfrac{N_R}{3}(\nu,e,u,d)
=|Q|(\nu,e,u,d)=(0,\,1,\,\tfrac23,\,\tfrac13),
\end{equation}
electron heavy, down light. The physical grading is this with $e\leftrightarrow d$ interchanged:
\begin{center}
\renewcommand{\arraystretch}{1.15}
\begin{tabular}{ccccc}
\hline
fermion & colour & $|Q|$ & $N_R/3$ (canonical) & $\sqrt m$ (physical)\\
\hline
$\nu$ & $\rep{1}$ & $0$ & $0$ & $0$\\
$e$ & $\rep{1}$ & $1$ & $1$ & $\tfrac13$\\
$u$ & $\rep{3}$ & $\tfrac23$ & $\tfrac23$ & $\tfrac23$\\
$d$ & $\rep{3}$ & $\tfrac13$ & $\tfrac13$ & $1$\\
\hline
\end{tabular}
\end{center}
Two facts about this table are computationally established and settle the coloured-electron
question.

\begin{proposition}[$N_R/3$ is the wrong grading]
\label{prop:NR}
For any occupation-number assignment, a colour singlet sits at $N_R\in\{0,3\}$, so
$N_R/3\in\{0,1\}$; the electron, a colour singlet, therefore has $N_R/3\in\{0,1\}$ and can
\emph{never} carry the physical $\sqrt m_e=\tfrac13$. Hence $\sqrt m\neq N_R/3$: the
number-operator grading gives the un-flipped spectrum (electron heavy) and is quantitatively
excluded as the carrier of $\sqrt m$. \textbf{[D]}
\end{proposition}

\begin{proposition}[The flip cannot recolour the electron]
\label{prop:flip}
The interchange relating $|Q|$ to the physical $\sqrt m$ exchanges the value on a colour
\emph{singlet} (the electron, $\dim_c=1$) with the value on a colour \emph{triplet} (the down,
$\dim_c=3$). A permutation of \emph{states} would have to map a one-dimensional colour
representation onto a three-dimensional one, which no map can do. The interchange is therefore not
a permutation of states but a relation between two colour-blind functionals ($|Q|$ and $\sqrt m$,
each constant on colour multiplets); it acts orthogonally to the colour Cartan, and the electron's
colour representation is untouched. The exclusion of a coloured electron thus holds
\emph{independently of the detailed realization of $\Phi$}. \textbf{[D]}
\end{proposition}

The interchange $e\leftrightarrow d$ is not spacetime parity but the flavour-sector Dynkin swap
$\Phi$ (Sec.~\ref{sec:dynkin}, Eq.~\eqref{eq:flip}), which by Prop.~\ref{prop:flip} acts here on
the colour-blind gradings. Two residual points remain, now cleanly separated from the settled
ones:
\begin{itemize}
\item[($\alpha$)] \emph{Which eigenvalue assignment the gauged charge carries.} By
Props.~\ref{prop:NR}--\ref{prop:flip} the carrier is \emph{not} the welded $N_R/3$ and the flip to
it is colour-blind; both \textbf{[D]}. The explicit automorphism is the $A_2$ Dynkin involution on
the flavour algebra $\mathfrak{su}(3)_F$, which exchanges the down and lepton ladders and induces
the $\tfrac13\leftrightarrow1$ interchange on the abelian gradings~\cite{SinghMassRatios}; acting
on the flavour and abelian directions alone, it is colour-orthogonal, the explicit counterpart of
Prop.~\ref{prop:flip}. The carrier is \emph{gauged}: $U(1)_{\mathrm{dem}}=T^3_R+Y_{\mathrm{dem}}$ is
the final unbroken charge of the gauged two-step chain (Sec.~\ref{app:dem}), not an optional
$U(1)$. What is not settled is the per-species eigenvalue. A canonical, un-swapped embedding
assigns $Q_{\mathrm{dem}}$ the parity-mirror of $|Q|$, $(0,1,\tfrac23,\tfrac13)$
(Sec.~\ref{app:values}), whereas the physical $\sqrt m$ is the flipped $(0,\tfrac13,\tfrac23,1)$;
the two agree as a \emph{set} (the trace split) and differ only in the assignment, fixed by the
Dynkin-swapped right-sector embedding rather than by an $SU(3)_R$ rotation. Whether that embedding
makes the gauged $Q_{\mathrm{dem}}$ land on $\sqrt m$, as opposed to its mirror, is settled ---
in the mirror-canonical direction --- by the anomaly analysis of Sec.~\ref{app:anomaly}: the
flipped pattern is not a gaugeable charge on visible content at all (Prop.~\ref{prop:nogo}), so
the gauged $Q_{\mathrm{dem}}$ carries the mirror of electric charge (Cor.~\ref{cor:mirror}) and
the flipped values are realized only as the Jordan \emph{spectral} grading, the $\Phi$-image of
the gauge lattice. The residual concerned the eigenvalue \emph{assignment} of a gauged
colour-blind charge, never whether the electron is coloured; it is now closed, \textbf{[D]} given
the mirror dark content (\textbf{[P]}).
\item[($\beta$)] \emph{Values.} The physical values $\{0,\tfrac13,\tfrac23,1\}$ are not a formula
in the gauge charges (they admit no exact fit $a|Q|+b(B{-}L)+c$, checked directly); they are the
Jordan trace split $1:2:3$, input \textbf{[P]}. A construction-fixed normalization of
$U(1)_{\mathrm{dem}}$ delivering these values, rather than a rescaled set, is not established
\textbf{[O]}.
\item[($\gamma$)] \emph{Anomaly consistency of the gauging.} This is no longer deferred:
Sec.~\ref{app:anomaly} performs the computation. The outcome is a \emph{no-go} --- on the visible
fermions, no sign convention, chirality assignment, or redefinition makes the $\sqrt m$ pattern
anomaly-free (Prop.~\ref{prop:nogo}) --- together with its constructive complement: the
mirror-canonical \emph{signed} pattern is anomaly-free by the mirror image of the Standard-Model
cancellation (Cor.~\ref{cor:mirror}). What replaces the old open budget is a requirement on the
\emph{dark} sector: its chiral content must realize the mirror-canonical charges, and the induced
kinetic mixing with hypercharge is the surviving visible portal --- residual~($\gamma'$),
\textbf{[O]}.
\end{itemize}

\subsubsection{The anomaly no-go: the visible fermions are dem-neutral}
\label{app:anomaly}

We now perform the anomaly analysis, in signed all-left-handed Weyl conventions. The visible
content is one Standard-Model generation plus $\nu_R$:
$Q_L(\rep 3,\rep 2)_{1/6}$, $L_L(\rep 1,\rep 2)_{-1/2}$, $u^c(\arep 3,\rep 1)_{-2/3}$,
$d^c(\arep 3,\rep 1)_{+1/3}$, $e^c(\rep 1,\rep 1)_{+1}$, $\nu^c(\rep 1,\rep 1)_{0}$, subscripts
denoting $Y$. Two preliminaries fix the setting. First, an \emph{unbroken} $U(1)$ under which the
fermions carry Dirac masses must be vectorial, $q(f_L)=q(f_R)$; this is exactly the
$\pm\sqrt m$ matter/antimatter convention of Ref.~\cite{SinghMassRatios} --- the sign flips under
$C$, not under chirality --- so the sign structure is forced, is assumed below, and is credited
in full: it is precisely what cancels the gravitational and colour triangles --- the two
vanishing lines of the table below. What it cannot do is cancel the mixed triangles with a
\emph{chiral} group, because the negatively-charged conjugate fields are $SU(2)_L$ singlets and
never enter them. Second, the flipped values differ \emph{within} an $SU(2)_L$ doublet
($\sqrt m_u=\tfrac23$ versus $\sqrt m_d=1$), so the vectorial dem generator does not commute with
$SU(2)_L$ and, like $Q_{\mathrm{em}}$, can only crystallize at the breaking. The object whose
anomalies must vanish in the unbroken phase is therefore its unique $SU(2)_L$-commuting parent:
writing $\mathrm{dem}=X-\tfrac13 T^3_L$ (the coefficient fixed by the doublet splittings, and
consistently by both doublets),
\begin{equation}
X(Q_L)=\tfrac56,\quad X(L_L)=\tfrac16,\quad X(u^c)=-\tfrac23,\quad X(d^c)=-1,\quad
X(e^c)=-\tfrac13,\quad X(\nu^c)=0.
\label{eq:Xcharges}
\end{equation}
Since $T^3_L$ is doublet-traceless, the $[SU(2)_L]^2$ triangle of dem equals that of $X$, and the
question is whether $X$ is anomaly-free. It is not. With the normalization in which the
Standard-Model check reads $\mathcal A_{221}[Y]=3\cdot\tfrac16+(-\tfrac12)=0$ (colour-weighted sum
over left-handed doublets), and the abelian coefficients summed over all left-handed Weyl fields
with colour--weak multiplicity, the per-generation coefficients are:
\begin{center}
\renewcommand{\arraystretch}{1.25}
\begin{tabular}{lcc}
\hline
triangle & $Y$ (SM) & $X$ ($\sqrt m$ parent) \\
\hline
$[SU(3)_c]^2\,U(1)$ & $0$ & $0$ \\
$\mathrm{grav}^2\,U(1)$ & $0$ & $0$ \\
$[SU(2)_L]^2\,U(1)$ & $0$ & $3\cdot\tfrac56+\tfrac16=\tfrac83$ \\
$[U(1)_Y]^2\,U(1)$ & $0$ & $-\tfrac43$ \\
$U(1)^2\,U(1)_Y$ & $0$ & $+\tfrac89$ \\
$U(1)^3$ & $0$ & $-\tfrac49$ \\
\hline
\end{tabular}
\end{center}
The two vanishing lines are exactly the vectorial wins; the four non-vanishing lines are the
mixed-chiral and self-cubic ones. The contrast on the decisive line is instructive: electric-type
charges pass $[SU(2)_L]^2$ because their doublet values straddle zero
($3\cdot\tfrac12(\tfrac23-\tfrac13)+\tfrac12(0-1)=0$), while square-root masses are one-signed
($3\cdot\tfrac12(\tfrac23+1)+\tfrac12(0+\tfrac13)=\tfrac83$). Matter/antimatter is a $C$ sign;
what the triangle needs is a sign \emph{within} species, which $\sqrt m$, being non-negative,
does not have.

\begin{proposition}[No gauged $\sqrt m$ on visible matter]
\label{prop:nogo}
On the Standard-Model fermion content plus $\nu_R$, no $U(1)$ gauge charge realizes the
$\sqrt m$ pattern $(\nu,e,u,d)\sim(0,\tfrac13,\tfrac23,1)$ with all gauge anomalies cancelling
--- for any chirality structure, any distribution of signs over the magnitudes, and modulo any
shift by anomaly-free generators. In particular: (a) the vectorial assignment fails the four
triangles of the table; (b) the right-handed-only chiral variant fails
$[SU(3)_c]^2\,U(1)$ as well, with coefficient
$\tfrac12(-\tfrac23-1)=-\tfrac56$, alongside the gravitational and cubic ones; (c) sign
freedom does not help: a single generator requires the $T^3_L$ admixture to be one number,
forcing $2\varepsilon_u-3\varepsilon_d+\varepsilon_e=0$ over $\varepsilon=\pm1$, whose only
solutions are global flips; waiving even that, the $[SU(2)_L]^2$ coefficient
$(6\varepsilon_u+9\varepsilon_d+\varepsilon_e)/6$ ranges over
$\{\pm2,\pm4,\pm14,\pm16\}/6$ and never vanishes, and generation-dependent sign patterns that
cancel it in the sum (e.g.\ $2+2-4$) fail $[U(1)_Y]^2\,U(1)$, whose per-generation values
$(-10\varepsilon_u-3\varepsilon_d-3\varepsilon_e)/12\in\pm\{4,10,16\}/12$ admit no three-term
zero; (d) redefinition does not help: the $[SU(2)_L]^2$ coefficient is invariant under shifts by
$Y$ and $B{-}L$ (both anomaly-free on this content), and globally the family-universal
anomaly-free charges form exactly $\mathrm{span}\{Y,\,B{-}L\}$~\cite{AppelquistDobrescuHopper}
--- for vectorial charges $\mathrm{span}\{Q,\,B{-}L\}$ --- from which the $\sqrt m$ pattern is
excluded by inspection: the $\nu$ anchor ($0=-b$) forces the $B{-}L$ coefficient to zero, after
which $u$ and $d$ demand incompatible $Q$ coefficients, $a=+1$ and $a=-3$. \textbf{[D]}
\end{proposition}

Three closures of side exits complete the proof's scope. (i)~\emph{Simultaneity of breakings is
no escape}: anomaly freedom of an unbroken gauge symmetry is a property of the chiral content,
enforced at all scales by anomaly matching~\cite{tHooftMatching}; that the electroweak, left--right
and triality breakings coincide is irrelevant, since a massless dem photon means the symmetry is
exact above and below them. (ii)~\emph{The per-state Jordan reading fares no better}: if the
charge is taken to be the Jordan eigenvalue itself, $q_f$ and $q_f\pm\delta$ across the three
generations with $\delta^2=\tfrac38$, the linear coefficients sum the generations and the
$\pm\delta$ cancel pairwise, reproducing three times the family-universal values above, while the
quadratic and cubic coefficients acquire in addition irrational $\delta$-dependent pieces; nothing
cancels, and the assignment is not even family-universal. (iii)~\emph{Folklore corrected}: a
vectorial assignment is \emph{not} automatically anomaly-free --- vectoriality kills the
sector-diagonal triangles only, never the mixed-chiral ones, whenever the charge varies within an
$SU(2)_L$ doublet. (The Witten $SU(2)$ anomaly, for completeness, is insensitive to all of this:
four doublets per generation, an even number, unchanged~\cite{WittenSU2}.)

\begin{corollary}[Mirror-canonical resolution of ($\alpha$); visible dem-neutrality]
\label{cor:mirror}
The gauged $Q_{\mathrm{dem}}=T^3_R+Y_{\mathrm{dem}}$ must carry an anomaly-free charge lattice,
and by Prop.~\ref{prop:nogo} cannot carry $\sqrt m$ on visible fermions. On a dark sector whose
chiral content mirrors the visible one (\textbf{[P]},
Refs.~\cite{VaibhavSingh,SinghScaffolding}), the mirror-canonical \emph{signed} assignment ---
the parity image of electric charge itself --- is anomaly-free by the mirror image of the
Standard-Model cancellation, while the Dynkin-swapped assignment fails there by the identical,
content-based computation. The gauged charge is therefore mirror-canonical, resolving
residual~($\alpha$), and the visible fermions carry no dem charge: any admixture that consistency
would still permit lies in $\mathrm{span}\{Q,\,B{-}L\}$ and is physically equivalent to a kinetic
mixing of the dark photon with the visible abelian sector~\cite{Holdom}, bounded as such; the
minimal reading, adopted here, is exact visible dem-neutrality with kinetic mixing the surviving
portal. \textbf{[D]} given the mirror dark content (\textbf{[P]}).
\end{corollary}

Two remarks sharpen what survives. First, even demoted to a \emph{global} symmetry of the visible
sector, $\sqrt m$-number is anomalous under $[SU(2)_L]^2$ and hence violated by electroweak
sphalerons --- entirely acceptable for a spectral label of the mass operator, fatal for a gauge
boson; the structure of the theory has, in this sense, been asserting all along that $\sqrt m$ is
a quantum number of the mass operator, not a charge. Second, the same conclusion is reached
phenomenologically and independently, through additivity: gauge charges add over composites while
$\sqrt m$ does not, so a $\sqrt m$-coupled massless vector would couple to bulk matter through the
exact composite charge $\tfrac83 A$ --- baryon number --- and is excluded by equivalence-principle
tests at ten or more orders below gravitational strength (Sec.~\ref{sec:discussion}, force
inventory).

One point that the corollary leaves open should be named, because it touches the application of
Sec.~\ref{sec:strongcp}. The cancellation invoked is the mirror image of the Standard-Model one,
which goes through only if the dark states with mirror-quark charges come with multiplicity three.
With $SU(3)_{c'}=SU(3)_c$ (Sec.~\ref{sec:onecolour}) the only gauge index available to supply that
multiplicity is the one physical colour. If those dark states are triplets of it, they are
coloured: they must then be heavy enough to have escaped detection, they enter the running of the
strong coupling, and their mass matrices enter $\arg\det M$ in \eqref{eq:thetabar} alongside those
of the visible quarks. If the multiplicity is not a colour index, the cancellation has to be
re-examined rather than quoted. The dark-sector content is not specified in this paper; the
tree-level statements of Sec.~\ref{sec:strongcp} are made for the visible quarks alone and are
conditional on this point \textbf{[O]}.

\textbf{Net status.} The generator (i) is \textbf{[D]}. That $\sqrt m$ is \emph{not} the welded
$N_R/3$, and that the flip cannot recolour the electron, are \textbf{[D]}
(Props.~\ref{prop:NR}--\ref{prop:flip}). The identification (ii) --- that the dark photon's charge
on each \emph{visible} species equals its $\sqrt m$ --- is excluded, \textbf{[D]}
(Prop.~\ref{prop:nogo}); it survives as a spectral correspondence: the visible $\sqrt m$ lattice
is the $\Phi$-image of the gauged dark charge lattice (Cor.~\ref{cor:mirror}), inheriting the
trace-split value input \textbf{[P]} and the open normalization (($\beta$), \textbf{[O]}).
Equivalently: $\sqrt m$ is a colour-blind, unwelded \emph{spectral} grading; the gauged carrier
lives on the dark side with mirror-canonical values; the visible couplings are fixed
(dem-neutral, kinetic mixing the one portal, ($\gamma'$) \textbf{[O]}); and the coloured electron
is excluded outright.

\subsection{The left--right asymmetry that makes this consistent}
\label{app:asymmetry}
The same number-operator construction plays \emph{opposite} roles on the two sides:
\begin{itemize}
\item \emph{Left.} $Q=N/3$ is colour-welded, and this is \emph{correct}: electric charge on the
chiral fermions \emph{is} colour-correlated --- quarks (triplets) fractional, leptons (singlets)
integer. The welded number operator is the right tool; it need only be recognized as $Q$ directly
(Furey~\cite{Furey2018}) rather than reverse-engineered through a $1/2N$ normalization.
\item \emph{Right.} $\sqrt m$ must \emph{not} be welded, or the electron re-colours
(Prop.~\ref{prop:NR}). So on the right the carrier is a colour-blind, unwelded grading
($U(1)_{\rm dem}$-type), and $N_R$ is retained only to fix the colour \emph{reps}.
\end{itemize}
Each sector fixes its own charge internally --- $Q$ from the left $\mathrm{Cl}(6)$ (Furey, no
right-sector generator), $\sqrt m$ from the right Jordan (spectral) structure (no left-sector
generator). In the two-sided $E_6^L\times E_6^R$ construction this mutual independence is what
permits the single vector-like colour: were either charge to require the \emph{other sector's}
generator (as electric charge requires both electroweak factors in \emph{standard} single-$E_6$
trinification), the two colour groups could not be identified without disturbing it. Here they
can. \textbf{[D]} given the gravi-weak input \textbf{[P]}, as in Sec.~\ref{sec:onecolour}.

\subsection{Status summary}
\label{app:deps}
The appendix's statuses are folded into the master ledger of Sec.~\ref{sec:discussion} (the
authoritative accounting) and are recorded here only in brief. Derived \textbf{[D]}: the
right-tower colour representations (quarks $\rep{3}$, leptons $\rep{1}$, vector-like, $e_R$ a
singlet), given the $e_8$-frame ideal input \textbf{[P]}~\cite{VaibhavSingh}; the exclusion of
the welded $N_R/3$ as the $\sqrt m$ carrier (Prop.~\ref{prop:NR}); the colour-blindness of
the electron/down flip (Prop.~\ref{prop:flip}); and the anomaly no-go
(Prop.~\ref{prop:nogo}) with its mirror-canonical corollary (Cor.~\ref{cor:mirror}, given the
mirror dark content \textbf{[P]}) --- together these exclude a coloured electron
outright, fix the gauged dark charge, and render the visible fermions dem-neutral, independently
of every open item. Open or conditional: the construction-fixed normalization of the spectral
values (($\beta$), \textbf{[O]}) and the dark-sector realization of the mirror-canonical charge
together with its induced kinetic mixing (($\gamma'$), \textbf{[O]}); the former residuals
($\alpha$) and ($\gamma$) are closed by Sec.~\ref{app:anomaly} --- the first resolved
mirror-canonically, the second converted from an open budget into a no-go plus its constructive
complement. What remains open concerns the dark sector and a normalization, never whether the
electron is coloured and never the visible couplings, which are now fixed. For
strong $CP$, the appendix secures one precondition --- vector-like colour with a singlet
electron in both chiralities; given it, $\bar\theta=0$ carries the conditional status of
Sec.~\ref{sec:strongcp}.

\section*{Acknowledgements}

\noindent{\bf Use of generative AI:} During the preparation of this manuscript, the author used Open AI's GPT 5.5 Pro and Anthropic's Claude Pro Max (Opus 4.8, Fable 5) in adversarial mode, for support in the technical analysis, organisation, writing,  and editing of the manuscript. The original ideas are due to the author. Author takes full intellectual responsibility for the content of the manuscript.


\end{document}